\documentclass[useAMS,usedcolumn,usegraphicx,usenatbib]{mn2e}
\usepackage{amssymb,amstext,amsfonts} 
\usepackage[fleqn]{amsmath}
\usepackage{subfigure}
\usepackage{slashed}
\usepackage[lowtilde]{url}
\usepackage{mathrsfs}
\usepackage{dcolumn}
\usepackage{hyperref}

\newcommand{\eqnref}[1]{(\ref{eq:#1})}
\newcommand{\figref}[1]{Fig.~\ref{fig:#1}}
\newcommand{\Figref}[1]{Figure~\ref{fig:#1}}

\newcommand{\secref}[1]{Sec.~\ref{sec:#1}}

\newcommand{\apref}[1]{Appendix~\ref{ap:#1}}

\newcommand{\units}[1]{\ensuremath{~\mathrm{#1}}}

\DeclareMathOperator{\sinc}{sinc}

\newcommand{\sub}[1]{\ensuremath{_\mathrm{#1}}}
\newcommand{\super}[1]{\ensuremath{^\mathrm{#1}}}
\newcommand{\dd}{\ensuremath{\mathrm{d}}}
\newcommand{\diff}[2]{\ensuremath{\frac{\dd {#1}}{\dd {#2}}}}

\newcommand{\intd}[4]{\ensuremath{\int_{#1}^{#2}{#3}\,\dd{#4}}}
\newcommand{\recip}[1]{\ensuremath{\frac{1}{#1}}}

\newcommand{\order}[1]{\ensuremath{\mathcal{O}({#1})}}

\newcommand{\innerprod}[2]{\ensuremath{\left({#1}\middle|{#2}\right)}}

\newcommand{\Ibar}{{\declareslashed{}{\text{-}}{0.04}{-0.2}{I}\slashed{I}}}
\title[Observing the Galaxy's MBH with GW bursts]{Observing the Galaxy's massive black hole with gravitational wave bursts}
\author[C.\ P.\ L.\ Berry and J.\ R.\ Gair]{C.\ P.\ L.\ Berry$^{1}$\thanks{E-mail: cplb2@cam.ac.uk}  and J.\ R.\ Gair$^{1}$\\
$^{1}$Institute of Astronomy, University of Cambridge, Madingley Road, Cambridge, CB3 0HA}

\begin{document}

\date{\today}

\pagerange{\pageref{firstpage}--\pageref{lastpage}} \pubyear{2012}

\maketitle

\label{firstpage}

\begin{abstract}
An extreme-mass-ratio burst (EMRB) is a gravitational wave signal emitted when a compact object passes through periapsis on a highly eccentric orbit about a much more massive body, in our case a stellar mass object about a $10^6 M_\odot$ black hole. EMRBs are a relatively unexplored means of probing the spacetime of massive black holes (MBHs). We conduct an investigation of the properties of EMRBs and how they could allow us to constrain the parameters, such as spin, of the Galaxy's MBH. We find that if an EMRB event occurs in the Galaxy, it should be detectable for periapse distances $r\sub{p} < 65 r\sub{g}$ for a $\mu = 10 M_\odot$ orbiting object, where $r\sub{g} = GM_\bullet/c^2$ is the gravitational radius. The signal-to-noise ratio scales as $\log(\rho) \simeq -2.7\log(r\sub{p}/r\sub{g}) + \log(\mu/M_\odot) + 4.9$. For periapses $r\sub{p} \lesssim 10 r\sub{g}$, EMRBs can be informative, and provide good constraints on both the MBH's mass and spin. Closer orbits provide better constraints, with the best giving accuracies of better than one part in $10^4$ for both the mass and spin parameter.
\end{abstract}


\section{Background and introduction}\label{sec:Intro}

Many, if not all, galactic nuclei have harboured a massive black hole (MBH) during their evolution \citep{Lynden-Bell1971, Rees1984}. Observations show there exist well-defined correlations between the MBHs' masses and the properties of their host galaxies, such as bulge luminosity, mass, velocity dispersion and light concentration \citep[e.g.][]{Kormendy1995, Magorrian1998, Graham2001, Tremaine2002, Graham2011}. These suggest coeval evolution of the MBH and galaxy \citep{Peng2007, Jahnke2011}, possibly with feedback mechanisms coupling the two \citep{Haiman2004, Volonteri2009}. The MBH and the surrounding spheroidal component share a common history, such that the growth of one can inform us about the growth of the other.

The best opportunity to study MBHs comes from the compact object in our own Galactic Centre (GC), which is coincident with Sagittarius A* (Sgr A*). Through careful monitoring of stars orbiting the GC, this has been identified as an MBH of mass $M_\bullet = 4.31 \times 10^6 M_\odot$ at a distance of only $R_0 = 8.33\units{kpc}$ \citep{Gillessen2009}.

According to the no-hair theorem, the MBH should be described completely by just its mass $M_\bullet$ and spin $a$, since we expect the charge of an astrophysical black hole (BH) to be negligible \citep{Chandrasekhar1998}. The spin parameter $a$ is related to the BH's angular momentum $J$ by
\begin{equation}
J = M_\bullet ac;
\end{equation}
it is often convenient to use the dimensionless spin
\begin{equation}
a_\ast = \frac{cJ}{GM_\bullet^2}.
\end{equation}
As we have a good estimate of the mass, to gain a complete description of the MBH we have only to measure its spin; this shall give us insight into its history and role in the evolution of the Galaxy.

The spin of an MBH is determined by several competing processes. An MBH accumulates mass and angular momentum through accretion \citep{Volonteri2010}. Accretion from a gaseous disc shall spin up the MBH, potentially leading to high spin values \citep{Volonteri2005}; a series of randomly orientated accretion events leads to a low spin value: we expect an average value $|a_\ast| \sim 0.1$--$0.3$ \citep{King2006}. The MBH also grows through mergers \citep{Yu2002, Malbon2007}. Minor mergers with smaller BHs can decrease the spin \citep*{Hughes2003}, while a series of major mergers, between similar mass MBHs, would lead to a likely spin of $|a_\ast| \sim 0.69$ \citep{Berti2007, Gonzalez2007}. Measuring the spin of MBHs shall help us understand the relative importance of these processes, and perhaps gain a glimpse into their host galaxies' pasts.

Elliptical and spiral galaxies are believed to host MBHs of differing spins because of their different evolutions: we expect MBHs in elliptical galaxies to have on average higher spins than MBHs in spiral galaxies, where random, small accretion episodes have played a more important role \citep*{Volonteri2007, Sikora2007}.

It has been suggested that the spin of the Galaxy's MBH could be inferred from careful observation of the orbits of stars within a few milliparsecs of the GC \citep{Merritt2010}, although this is complicated because of perturbations due to other stars, or from observations of quasi-periodic oscillations in the luminosity of flares believed to originate from material orbiting close to the innermost stable orbits \citep{Genzel2003a, Hamaus2009}, though there are difficulties in interpreting these results \citep{Psaltis2008a}.

This latter method, combined with a disc-seismology model, has produced a value of the dimensionless spin of $a_\ast = 0.44 \pm 0.08$. To obtain this result \citet{Kato2010} have combined their observations of Sgr A* with observations of galactic X-ray sources containing solar mass BHs, to find a best-fit unique spin parameter for all BHs. It is not clear that all BHs should share the same spin parameter; especially as the BHs considered here differ in mass by six orders of magnitude. Even if BH spin is determined by a universal process, we expect some distribution of spin parameters \citep*{King2008, Berti2008}. Thus we cannot precisely determine the spin of the GC's MBH from an average.

The spins of MBHs in active galactic nuclei have been inferred using X-ray observations of $\mathrm{Fe}$ $\mathrm{K}$ emission lines \citep{Miller2007, McClintock2011}. This has been done for a handful of other galaxies' MBHs \citep{Brenneman2006, Miniutti2009, Schmoll2009, delaCallePerez2010, Zoghbi2010, Nardini2011,  Patrick2011}. Estimates for the spin cover a range up to the maximal value for an extremal Kerr BH. Typical results are in the intermediate range of $a_\ast \sim 0.7$ with an uncertainty of about $10\%$ on each measurement.

While we can use the spin of other BHs as a prior, to inform us of what we should expect for the Galaxy's MBH, it is desirable to have an independent observation, a direct measurement.

An exciting means of inferring information about the MBH is through gravitational waves (GWs) emitted when compact objects (COs), such as stellar mass BHs, neutron stars (NSs), white dwarfs (WDs) or low mass main sequence (MS) stars, pass close by \citep{Sathyaprakash2009}. A space-borne detector, such as the \textit{Laser Interferometer Space Antenna} (\textit{LISA}) or the \textit{evolved Laser Interferometer Space Antenna} (\textit{eLISA}), is designed to be able to detect GWs in the frequency range of interest for these encounters \citep{Bender1998, Danzmann2003, Jennrich2011, Amaro-Seoane2012a}.\footnote{The revised \textit{eLISA} concept shares the same descoped design as the \textit{New Gravitational-wave Observatory} (\textit{NGO}) submitted to the European Space Agency for their L1 mission selection.} The identification of waves requires a set of accurate waveform templates covering parameter space. Much work has already been done on the waveforms generated when companion objects inspiral towards an MBH \citep{Glampedakis2005, Barack2009}; as they orbit, GWs carry away energy and angular momentum, causing the orbit to shrink until eventually the CO plunges into the MBH. The initial orbits may be highly elliptical and a burst of radiation is emitted during each close encounter. These are extreme mass-ratio bursts (EMRBs; \citealt*{Rubbo2006}). Assuming the companion is not scattered, and does not plunge straight into the MBH, its orbit evolves, becoming more circular, and it begins to continuously emit significant gravitational radiation in the \textit{LISA}/\textit{eLISA} frequency range. The resulting signals are extreme mass-ratio inspirals (EMRIs; \citealt{Amaro-Seoane2007}).

Studies of these systems have usually focused upon the phase when the orbit is close to plunge and completes a large number of cycles in the detector's frequency band, allowing a high signal-to-noise ratio (SNR) to be accumulated. Here, we investigate high eccentricity orbits. These are the initial bursting orbits from which an EMRI may evolve, and are the consequence of scattering from two body encounters. The event rate for the detection of such EMRBs with \textit{LISA} has been estimated to be as high as $15\units{yr^{-1}}$ \citep{Rubbo2006}, although this has been subsequently revised downwards to the order of $1\units{yr^{-1}}$ \citep*{Hopman2007}. Even if only a single burst is detected during a mission, this is still an exciting possibility since the information carried by the GW gives an unparalleled probe of the spacetime of the GC. Exactly what can be inferred depends upon the orbit, which we investigate here. 

We make the simplifying assumption that all these orbits are marginally bound, or parabolic, since highly eccentric orbits appear almost indistinguishable from an appropriate parabolic orbit. Here ``parabolic'' and ``eccentricity'' refer to the energy of the geodesic and not to the geometric shape of the orbit.\footnote{Marginally bound Keplerian orbits in flat spacetime are parabolic in both senses.} Following such a trajectory an object may make just one pass of the MBH or, if the periapsis distance is small enough, it may complete a number of rotations. Such an orbit is referred to as zoom-whirl \citep{Glampedakis2002a}.

In order to compute the gravitational waveform produced in such a case, we integrate the geodesic equations for a parabolic orbit in Kerr spacetime. We assume the orbiting body is a test particle, such that it does not influence the underlying spacetime, and that the orbital parameters evolve negligibly during the orbit such that they may be held constant. We use this to construct an approximate numerical kludge (NK) waveform \citep{Babak2007}.

This paper is organised as follows. We begin in \secref{Geodesic} with the construction of the geodesic orbits; these trajectories are used for the NK waveforms as explained in \secref{Kludge}. In \secref{Signal} we establish what the \textit{LISA} detectors would measure and how the signal would be analysed. This includes a brief mention of window functions which is expanded in \apref{window}. Here we present a novel window function, the Planck-Bessel window, of use for signals with a large dynamic range. In \secref{Waveforms} we look at our NK waveforms. We give fiducial power-law fits for SNR as a function of periapse radius, useful for back-of-the-envelope estimates. We confirm the accuracy of the kludge waveforms in \secref{Energy} by comparing the energy flux to fluxes calculated using other approaches. The typical error introduced by the NK approximation may be a few percent, but this worsens as the periapsis approaches the last non-plunging orbit. We explain how to extract the information from the bursts in \secref{Estimation}. Results estimating the measurement precision are presented in \secref{Results}. We briefly mention the possibility of detecting bursts from extra-galactic sources in \secref{Extragal}, before concluding in \secref{End} with a summary of our results. EMRBs may be informative if the event rate is high enough for them to be a viable source.

There are currently no funded space-borne detector missions. The \textit{eLISA} mission concept remains an active field of study. It is hoped to submit this to the European Space Agency as a potential cornerstone mission. We use the classic \textit{LISA} design. This is done from historical affection in lieu of a definite alternative. Should funding for a space-borne detector be secured in the future it is hoped that it shall have comparable sensitivity to \textit{LISA}, and that studies using the \textit{LISA} design shall be a sensible benchmark for comparison. We find that to obtain good results the periapse radius must be $r\sub{p} \lesssim 10 r\sub{g}$, where $r\sub{g} = GM_\bullet / c^2$ is a gravitational radius; at this point the SNR is already high: for parameter estimation the orbit is more important that the signal strength, and so the exact detector performance should be of secondary importance.

We adopt a metric with signature $(+,-,-,-)$. Greek indices are used to represent spacetime indices $\mu = \{0,1,2,3\}$ and lowercase Latin indices from the middle of the alphabet are used for spatial indices $i = \{1,2,3\}$. Uppercase Latin indices from the beginning of the alphabet are used for the output of the two \textit{LISA} detector-arms $A = \{\mathrm{I}, \mathrm{II}\}$, and lowercase Latin indices from the beginning of the alphabet are used for parameter space. Summation over repeated indices is assumed unless explicitly noted otherwise. Geometric units with $G = c = 1$ are used where noted, but in general factors of $G$ and $c$ are retained.

\section{Parabolic orbits in Kerr spacetime}\label{sec:Geodesic}

Astrophysical BHs are described by the Kerr metric \citep{Kerr1963}. This is conveniently written in terms of Boyer-Lindquist coordinates $\{t, r, \theta, \phi\}$ (\citealt*{Boyer1967, Hobson2006}, section 13.7). For this section we shall work in natural units with $G = c = 1$.

Geodesics are parametrized by three conserved quantities (aside from the particle's mass $\mu$): energy (per unit mass) $E$, specific angular momentum about the symmetry axis (the $z$-axis) $L_z$, and Carter constant $Q$ (\citealt{Carter1968, Chandrasekhar1998}, section 62). For a parabolic orbit $E = 1$; the particle is at rest at infinity. This simplifies the geodesic equations. It also allows us to give a simple interpretation for the Carter constant: this is defined as
\begin{equation}
Q = L_\theta^2 + \cos^2\theta\left[a^2\left(1 - E^2\right) + \frac{L_z^2}{\sin^2\theta}\right],
\end{equation}
where $L_\theta$ is the (non-conserved) specific angular momentum in the $\theta$-direction. For $E = 1$
\begin{equation}
Q = L_\theta^2 + \cot^2\theta\, L_z^2 = L_\infty^2 - L_z^2;
\end{equation}
here $L_\infty$ is the total specific angular momentum at infinity, where the metric is asymptotically flat \citep{DeFelice1980}.\footnote{See \citet*{Rosquist2009} for a discussion of the interpretation of $Q$ in the limit $G \rightarrow 0$, corresponding to a flat spacetime.} This is as in Schwarzschild spacetime.

\subsection{Integration variables and turning points}

In integrating the geodesic equations, difficulties can arise because of the presence of turning points, when the sign of the $r$ or $\theta$ geodesic equation changes. The radial turning points are at the periapsis $r\sub{p}$ and at infinity. We may locate the periapsis by finding the roots of
\begin{align}
V_r = {} & 2M_\bullet r^3 - \left(L_z^2+Q\right)r^2 + 2M_\bullet\left[\left(L_z - a\right)^2 + Q\right]r - a^2 Q \nonumber \\
 = {} & 0.
\end{align}
This has three roots, which we denote $\{r_1, r_2, r\sub{p}\}$; the periapsis $r\sub{p}$ is the largest real root. We do not find the apoapsis as a (fourth) root as we have removed it by taking $E = 1$ before solving.\footnote{This turning point can be found by setting the unconstrained expression for $V_r$ equal to zero, and then solving for $E(r)$; taking the limit $r \rightarrow \infty$ gives $E \rightarrow 1$ \citep{Wilkins1972}.}

We avoid the difficulties associated with the turning point by introducing angular variables that always increase with proper time \citep{Drasco2004}: inspired by Keplerian orbits
\begin{equation}
r = \frac{p}{1+e\cos\psi},
\end{equation}
where $e = 1$ is the eccentricity and $p = 2r\sub{p}$ is the semilatus rectum. As $\psi$ covers its range from $-\pi$ to $\pi$, $r$ traces out a complete orbit. The geodesic equation for $\psi$ is
\begin{align}
\left(r^2 + a^2\cos^2\theta\right)\diff{\psi}{\tau} = {} & \left\{M_\bullet\left[2r\sub{p} - \left(r_1 + r_2\right)\left(1 + \cos\psi\right) \vphantom{\frac{r_1 r_2}{2r\sub{p}}} \right.\right. \nonumber \\*
 {} & + \left.\left. \frac{r_1 r_2}{2r\sub{p}}\left(1 + \cos\psi\right)^2\right]\right\}^{1/2},
\end{align}
for proper time $\tau$. Parametrizing an orbit by its periapsis and eccentricity has the additional benefit of allowing easier comparison with its flat-space equivalent \citep*{Gair2005}.

\begin{figure}
\begin{center}
\includegraphics[width=0.2\textwidth]{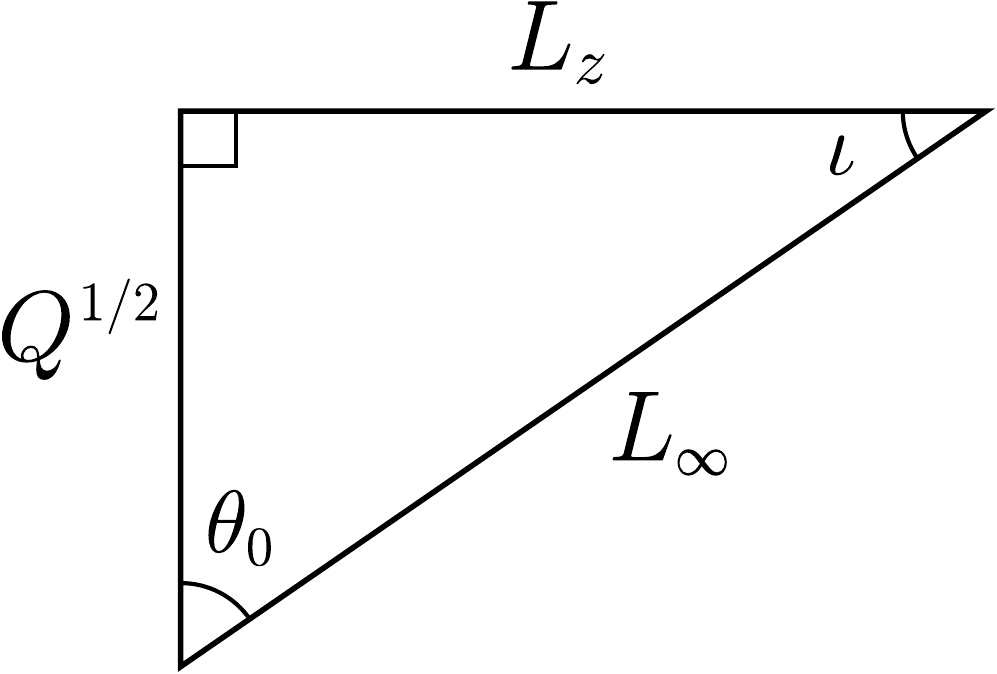}
    \caption{The angular momenta $L_\infty$, $L_z$ and $\sqrt{Q}$ define a right-angled triangle. The acute angles are $\theta_0$, the extremal value of the polar angle, and $\iota$, the orbital inclination \citep*{Glampedakis2002}.}
   \label{fig:L_triangle}
\end{center}
\end{figure}

The $\theta$ motion is usually bounded, with $\theta_0 \leq \theta \leq \pi - \theta_0$; in the event that $L_z = 0$ the particle follows a polar orbit and $\theta$ covers its full range \citep{Wilkins1972}. Turning points are given by
\begin{equation}
V_\theta = Q - \cot^2\theta\, L_z^2 = 0.
\end{equation}
Changing variable to $\xi = \cos^2\theta$, we have a maximum $\xi_0 = \cos^2\theta_0$ given by
\begin{equation}
\xi_0 = \frac{Q}{Q+L_z^2} = \frac{Q}{L_\infty^2}.
\label{eq:theta_0}
\end{equation}
See \figref{L_triangle} for a geometrical visualization.
Introducing a second angular variable \citep{Drasco2004}
\begin{equation}
\xi = \xi_0\cos^2\chi;
\end{equation}
over one $2\pi$ period of $\chi$, $\theta$ oscillates from its minimum to its maximum and back. The geodesic equation for $\chi$ is
\begin{equation}
\left(r^2 + a^2\cos^2\theta\right)\diff{\chi}{\tau} = \sqrt{Q + L_z^2}.
\end{equation}

\section{Waveform Construction}\label{sec:Kludge}

We can now calculate the geodesic trajectory. The orbiting body is assumed to follow this track exactly; we ignore evolution due to the radiation of energy and angular momentum, which should be negligible for EMRBs. From this trajectory we calculate the waveform using a semirelativistic approximation \citep{Ruffini1981}: we assume the particle moves along the Kerr geodesic, but radiates as if it were in flat spacetime. This quick-and-dirty technique is known as a numerical kludge (NK), and has been shown to approximate well results computed by more accurate methods \citep{Babak2007}. It is often compared to a bead travelling along a wire. The shape of the wire is set by the geodesic, but the bead moves along in flat space.

\subsection{Kludge approximation}

Numerical kludge approximations aim to encapsulate the main characteristics of a waveform by using the exact particle trajectory (ignoring inaccuracies from radiative effects and from the particle's self-force), whilst saving on computational time by using approximate waveform generation techniques.

We build an equivalent flat-space trajectory by identifying the Boyer-Lindquist coordinates with a set of flat-space coordinates. We consider two choices:
\begin{enumerate}
\item Identify the Boyer-Lindquist coordinates with flat-space spherical polars $\{r\sub{BL},$ $\theta\sub{BL},$ $\phi\sub{BL}\} \rightarrow \{r\sub{sph}, \theta\sub{sph}, \phi\sub{sph}\}$, then define Cartesian coordinates \citep{Gair2005, Babak2007}
\begin{equation}
\boldsymbol{x} = \begin{pmatrix}
r\sub{sph} \sin\theta\sub{sph}\cos\phi\sub{sph} \\
r\sub{sph} \sin\theta\sub{sph}\sin\phi\sub{sph} \\
r\sub{sph} \cos\theta\sub{sph}
\end{pmatrix}.
\end{equation}
\item Identify the Boyer-Lindquist coordinates with flat-space oblate-spheroidal coordinates $\{r\sub{BL}, \theta\sub{BL}, \phi\sub{BL}\} \rightarrow \{r\sub{ob}, \theta\sub{ob}, \phi\sub{ob}\}$ so that the Cartesian coordinates are
\begin{equation}
\boldsymbol{x} = \begin{pmatrix}
\sqrt{{r\sub{ob}}^2 + a^2} \sin\theta\sub{ob}\cos\phi\sub{ob} \\
\sqrt{{r\sub{ob}}^2 + a^2} \sin\theta\sub{ob}\sin\phi\sub{ob} \\
r\sub{ob} \cos\theta\sub{ob}
\end{pmatrix}.
\end{equation}
These are appealing because in the limit that $G \rightarrow 0$, where the gravitating mass goes to zero, the Kerr metric in Boyer-Lindquist coordinates reduces to the Minkowski metric in oblate-spheroidal coordinates.
\end{enumerate}
The two coincide for $a \rightarrow 0$ or $r \rightarrow \infty$.

There is no well motivated argument that either coordinate system must yield an accurate GW; their use is justified {\it post facto} by comparison with results obtained from more accurate, and computationally intensive, methods \citep{Gair2005, Babak2007}. The ambiguity in assigning flat-space coordinates reflects the inconsistency of the semirelativistic approximation: the geodesic trajectory was calculated for the Kerr geometry; by moving to flat spacetime we lose the reason for its existence. This should not be regarded as a major problem; it is an artifact of the basic assumption that the shape of the trajectory is important for determining the character of the radiation, but the curvature of the spacetime in the vicinity of the source is not. By binding the particle to the exact geodesic, we ensure that the waveform has spectral components at the correct frequencies, but by assuming flat spacetime for generation of GWs they shall not have the correct amplitudes.

\subsection{Quadrupole-octupole formula}

Now we have a flat-space particle trajectory $x\sub{P}^\mu(\tau)$, we may apply a flat-space wave generation formula. We use the quadrupole-octupole formula to calculate the gravitational strain \citep{Bekenstein1973, Press1977, Yunes2008}
\begin{equation}
h^{jk}(t, \boldsymbol{x}) = -\frac{2G}{c^6r}\left(\ddot{I}^{jk} - 2n_i\ddot{S}^{ijk} + n_i\dddot{M}^{ijk}\right)_{t'\, =\, t - r/c},
\label{eq:octupole}
\end{equation}
where an over-dot represents differentiation with respect to time $t$, $t'$ is the retarded time, $r = \left|\boldsymbol{x} - \boldsymbol{x}\sub{P}\right|$ is the radial distance, $\boldsymbol{n}$ is the radial unit vector, and ${I}^{jk}$, ${S}^{ijk}$ and ${M}^{ijk}$ are the mass quadrupole, current quadrupole and mass octupole respectively. This is correct for a slowly moving source. It is the familiar quadrupole formula (\citealt*[section 36.10]{Misner1973}; \citealt[section 17.9]{Hobson2006}), derived from linearized theory, plus the next order terms. 

\section{Signal detection and analysis}\label{sec:Signal}

\subsection{The \textit{LISA} detector}\label{sec:Detector}

The classic \textit{LISA} design is a three arm, space-borne laser interferometer \citep{Bender1998, Danzmann2003}. The arms form an equilateral triangle that rotates as the system's centre of mass follows a circular, heliocentric orbit, trailing $20^{\circ}$ behind the Earth. \textit{eLISA} has a similar design, but  has only two arms, which are shorter in length, and trails $9^{\circ}$ behind the Earth \citep{Jennrich2011}.

\begin{figure}
\begin{center}
 \includegraphics[width=0.35\textwidth]{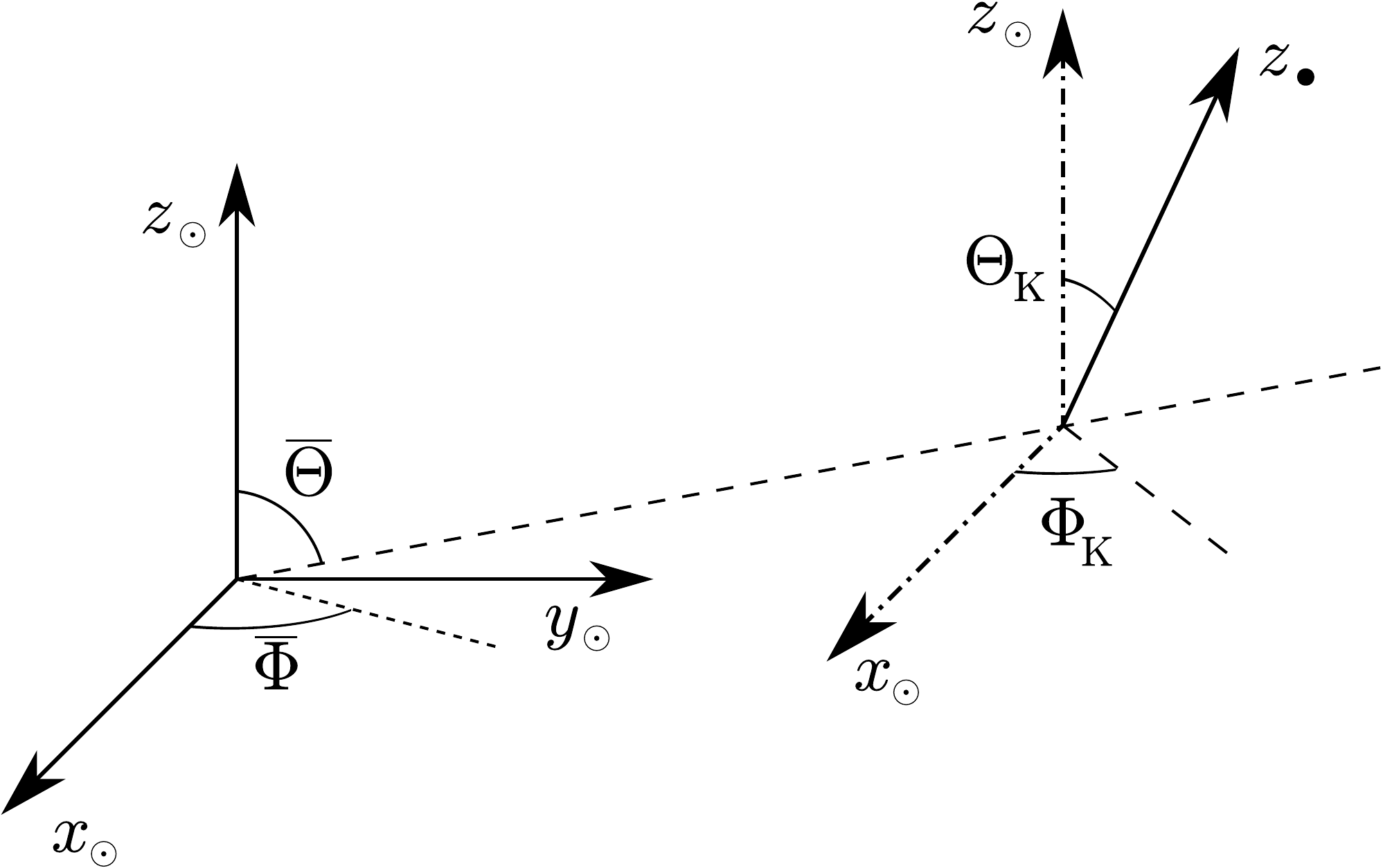}
    \caption{The relationship between the MBH's coordinate system $x_\bullet^i$ and the SS coordinate system $x_\odot^i$. The MBH's spin axis is aligned with the $z_\bullet$-axis. The orientation of the MBH's $x$- and $y$-axes is arbitrary. We choose $x_\bullet$ to be orthogonal to the direction to the SS.}
   \label{fig:BH_SS}
\end{center}
\end{figure}

\begin{figure}
\begin{center}
 \includegraphics[width=0.27\textwidth]{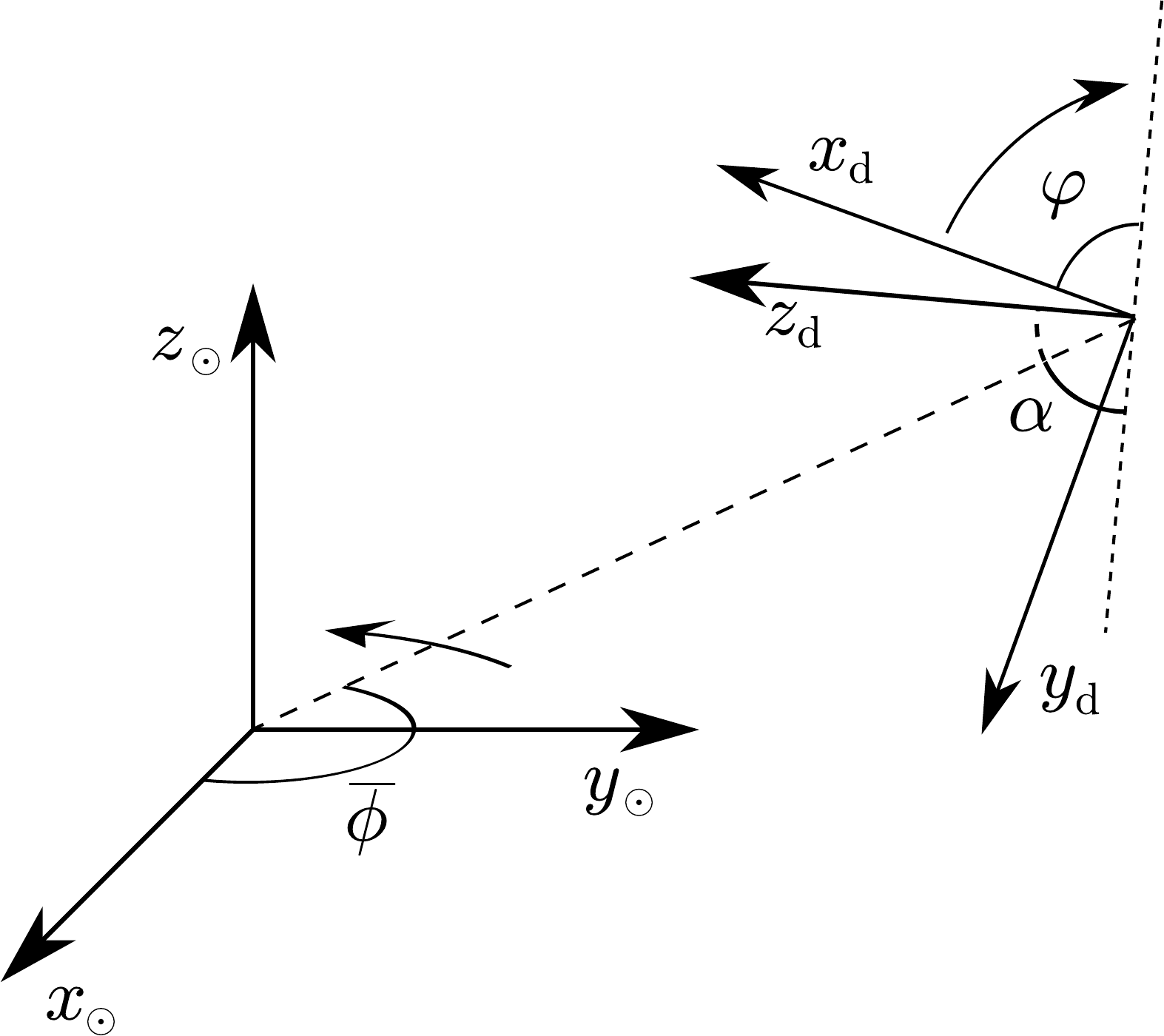}
    \caption{The relationship between the detector coordinates $x\sub{d}^i$ and the ecliptic coordinates of the SS $x_\odot^i$ \citep{Bender1998, Jennrich2011}. The detector inclination is $\alpha = 60^{\circ}$.}
   \label{fig:SS_LISA}
\end{center}
\end{figure}

To describe the detector configuration, and to transform from the MBH coordinate system to those of the detector, we use three coordinate systems: those of the BH at the GC $x_\bullet^i$; ecliptic coordinates centred at the solar system (SS) barycentre $x_\odot^i$, and coordinates that co-rotate with the detector $x\sub{d}^i$. The MBH's coordinate system and the SS coordinate system are depicted in \figref{BH_SS}.
The mission geometry for \textit{LISA}/\textit{eLISA} is shown in \figref{SS_LISA}.
We define the detector coordinates such that the detector-arms lie in the $x\sub{d}$-$y\sub{d}$ plane as in \citet{Cutler1998}. We have computed the waveforms in the MBH's coordinates, but it is simplest to describe the measured signal using the detector's coordinates.

The strains measured in the three arms can be combined such that \textit{LISA} behaves as a pair of $90^{\circ}$ interferometers at $45^{\circ}$ to each other, with signals scaled by ${\sqrt{3}}/{2}$ \citep{Cutler1998}. We denote the two detectors as I and II and use vector notation $\boldsymbol{h}(t) = \left(h\sub{I}(t), h\sub{II}(t)\right) = \left\{h_A(t)\right\}$ to represent signals from both detectors.

\subsection{Frequency domain formalism}

Having constructed the GW $\boldsymbol{h}(t)$ that shall be incident upon the detector, we may consider how to analyse the waveform and extract the information it contains. We briefly recap GW signal analysis, with application to \textit{LISA}. A more complete discussion can be found in \citet{Finn1992} and \citet{Cutler1994}. Adaption for \textit{eLISA} requires a substitution of the noise distribution, and the removal of the sum over data channels, since it would only have one.

The measured strain $\boldsymbol{s}(t)$ is the combination of the signal and the detector noise
\begin{equation}
\boldsymbol{s}(t) = \boldsymbol{h}(t) + \boldsymbol{n}(t);
\end{equation}
we assume the noise $n_A(t)$ is stationary and Gaussian, and that noise in the two detectors is uncorrelated, but shares the same characterisation \citep{Cutler1998}.

The properties of the noise allow us to define a natural inner product and associated distance on the space of signals \citep{Cutler1994}
\begin{equation}
\innerprod{\boldsymbol{g}}{\boldsymbol{k}} = 2\intd{0}{\infty}{\frac{\tilde{g}_A^\ast(f)\tilde{k}_A(f) + \tilde{g}_A(f)\tilde{k}_A^\ast(f)}{S\sub{n}(f)}}{f},
\label{eq:inner}
\end{equation}
introducing Fourier transforms
\begin{equation}
\tilde{g}(f) = \mathscr{F}\{g(t)\} = \intd{-\infty}{\infty}{g(t)\exp(2\pi i ft)}{t},
\end{equation}
and $S\sub{n}(f)$ is the noise spectral density. The signal-to-noise ratio is approximately
\begin{equation}
\rho[\boldsymbol{h}] = \innerprod{\boldsymbol{h}}{\boldsymbol{h}}^{1/2}.
\label{eq:SNR}
\end{equation}
The probability of a particular realization of noise $\boldsymbol{n}(t) = \boldsymbol{n}_0(t)$ is
\begin{equation}
p(\boldsymbol{n}(t) = \boldsymbol{n}_0(t)) \propto \exp\left[-\recip{2}\innerprod{\boldsymbol{n}_0}{\boldsymbol{n}_0}\right].
\end{equation}
Thus, if the incident waveform is $\boldsymbol{h}(t)$, the probability of measuring signal $\boldsymbol{s}(t)$ is
\begin{equation}
p(\boldsymbol{s}(t)|\boldsymbol{h}(t)) \propto \exp\left[-\recip{2}\innerprod{\boldsymbol{s}-\boldsymbol{h}}{\boldsymbol{s}-\boldsymbol{h}}\right].
\label{eq:sig_prob}
\end{equation}

\subsection{Noise curve}\label{sec:Noise}

\textit{LISA}'s noise has two sources: instrumental noise and confusion noise, primarily from white dwarf binaries. The latter may be divided into contributions from galactic and extragalactic binaries. We use the noise model of \citet{Barack2004}. The instrumental noise dominates at both high and low frequencies. The confusion noise is important at intermediate frequencies, and is responsible for the cusp around $10^{-3}\units{Hz}$. \textit{eLISA} shares the same sources of noise, but is less affected by confusion. Its sensitivity regime is shifted to higher frequencies because of a shorter arm length.

\subsection{Window functions}

There is one remaining complication regarding signal analysis: as we are Fourier transforming a finite signal we encounter spectral leakage; a contribution from large amplitude spectral components leaks into surrounding components (sidelobes), obscuring and distorting the spectrum at these frequencies \citep{Harris1978}. This is an inherent problem with finite signals; it shall be as much of a problem when analysing signals from an actual mission as it is here. To mitigate, but unfortunately not eliminate, these effects, the time-domain signal can be multiplied by a window function. These are discussed in detail in \apref{window}. We adopt the Nuttall four-term window with continuous first derivative \citep{Nuttall1981} for our results.

\section{Waveforms and detectability}\label{sec:Waveforms}

\subsection{Model parameters}

The shape of the waveform depends on parameters defining the MBH; the companion object on its orbit, and the detector. Let us define $\boldsymbol{\lambda} = \left\{\lambda^1, \lambda^2, \ldots, \lambda^Z\right\}$ as the set of $Z$ parameters which specify the GW. For our model, these are:
\begin{enumerate}
\item[(1)] The MBH's mass $M_\bullet$. This is well constrained by the observation of stellar orbits about Sgr A* \citep{Ghez2008, Gillessen2009}, with best estimate $M_\bullet = (4.31 \pm 0.36) \times 10^6 M_\odot$. This depends upon the GC distance $R_0$ as $M_\bullet = (3.95 \pm 0.06|\sub{stat} \pm 0.18|_{R_0, \, \mathrm{stat}} \pm  0.31|_{R_0, \, \mathrm{sys}}) \times 10^6 M_\odot (R_0 / 8\units{kpc})^{2.19}$, where the errors are statistical, independent of $R_0$; statistical from the determination of $R_0$, and systematic from $R_0$ respectively.
\item[(2)] The spin parameter $a_\ast$. Naively this could be anywhere in the range $|a_\ast| < 1$; however, it is possible to place an upper bound by contemplating spin-up mechanisms. Considering the torque from radiation emitted by an accretion disc, and swallowed by the BH, it can be shown that $|a_\ast| \lesssim 0.998$ \citep{Thorne1974}. Magnetohydrodynamical simulations of accretion discs produce a smaller maximum value of $|a_\ast| \sim 0.95$ \citep*{Gammie2004}. The actual spin value could be much lower than this upper bound, depending upon the MBH's evolution.
\item[(3, 4)] The orientation angles for the MBH $\Theta\sub{K}$ and $\Phi\sub{K}$.
\item[(5)] The ratio of the SS-GC distance $R_0$ and the CO mass $\mu$, which we denote as $\zeta = R_0/\mu$. This scales the amplitude of the waveform. Bursts, unlike inspirals, do not undergo orbital evolution, hence we cannot break the degeneracy in $R_0$ and $\mu$. The distance, like $M_\bullet$, is constrained by stellar orbits, the best estimate being $R_0 = 8.33 \pm 0.35\units{kpc}$ \citep{Gillessen2009}. The mass of the orbiting particle depends upon the type of object: whether it is an MS star, WD, NS or BH. Since we shall not know the $\mu$ precisely, we shall not be able to infer anything more about the distance.
\item[(6, 7)] The angular momentum of the CO. This can be described using either $\{L_z, Q\}$ or $\{L_\infty, \iota\}$. We employ the latter, as the total angular momentum and inclination are less tightly correlated. Assuming spherical symmetry, we expect $\cos \iota$ to be uniformly distributed.
\item[(8--10)] A set of coordinates to specify the trajectory. We use the angular phases at periapse, $\phi\sub{p}$ and $\chi\sub{p}$ (which determines $\theta\sub{p}$), as well as the time of periapse $t\sub{p}$.
\item[(11, 12)] The coordinates of the MBH from the SS barycentre $\overline{\Theta}$ and $\overline{\Phi}$. These may be taken as the coordinates of Sgr A*, as the radio source is expected to be within $20 r\sub{g}$ of the MBH \citep{Reid2003,Doeleman2008}. We use the J2000.0 coordinates \citep{Reid1999, Yusef-Zadeh1999}. These change with time due to the rotation of the SS about the GC; the proper motion is about $6\units{mas\,yr^{-1}}$, mostly in the plane of the Galaxy \citep{Backer1999, Reid2003}. The position is already determined to high accuracy: an EMRB can only give weak constraints on source position.\footnote{For comparison, an EMRI, which should be more informative, can only give sky localisation to $\sim10^{-3}~\mathrm{steradians}$ \citep{Barack2004, Huerta2009}.} We take it as known and do not try to infer it.
\item[(13, 14)] The orbital position of the detector given by $\overline{\phi}$ and $\varphi$. We assume the initial positions are chosen such that $\overline{\phi} = 0$ when $\varphi = 0$ \citep{Cutler1998}; this choice does not qualitatively influence our results. The orbital position should be known, so need not be inferred.\\
\end{enumerate}
We therefore have a $14$ dimensional parameter space, of which we are interested in inferring $d = 10$ parameters.

\subsection{Waveforms}

\begin{figure}
  \begin{center}
   \subfigure[{Waveform for $a_\ast \simeq 0.12$, $r\sub{p} \simeq 15.6 r\sub{g}$ and $\iota \simeq 2.1$. The SNR for the spherical polar kludge is $\rho[\boldsymbol{h}\sub{sph}] \simeq 451$, for the oblate-spheroidal kludge $\rho[\boldsymbol{h}\sub{ob}] \simeq 451$ (agreement to $0.01\%$).}]{\label{fig:Orbit_233} \includegraphics[width=0.43\textwidth]{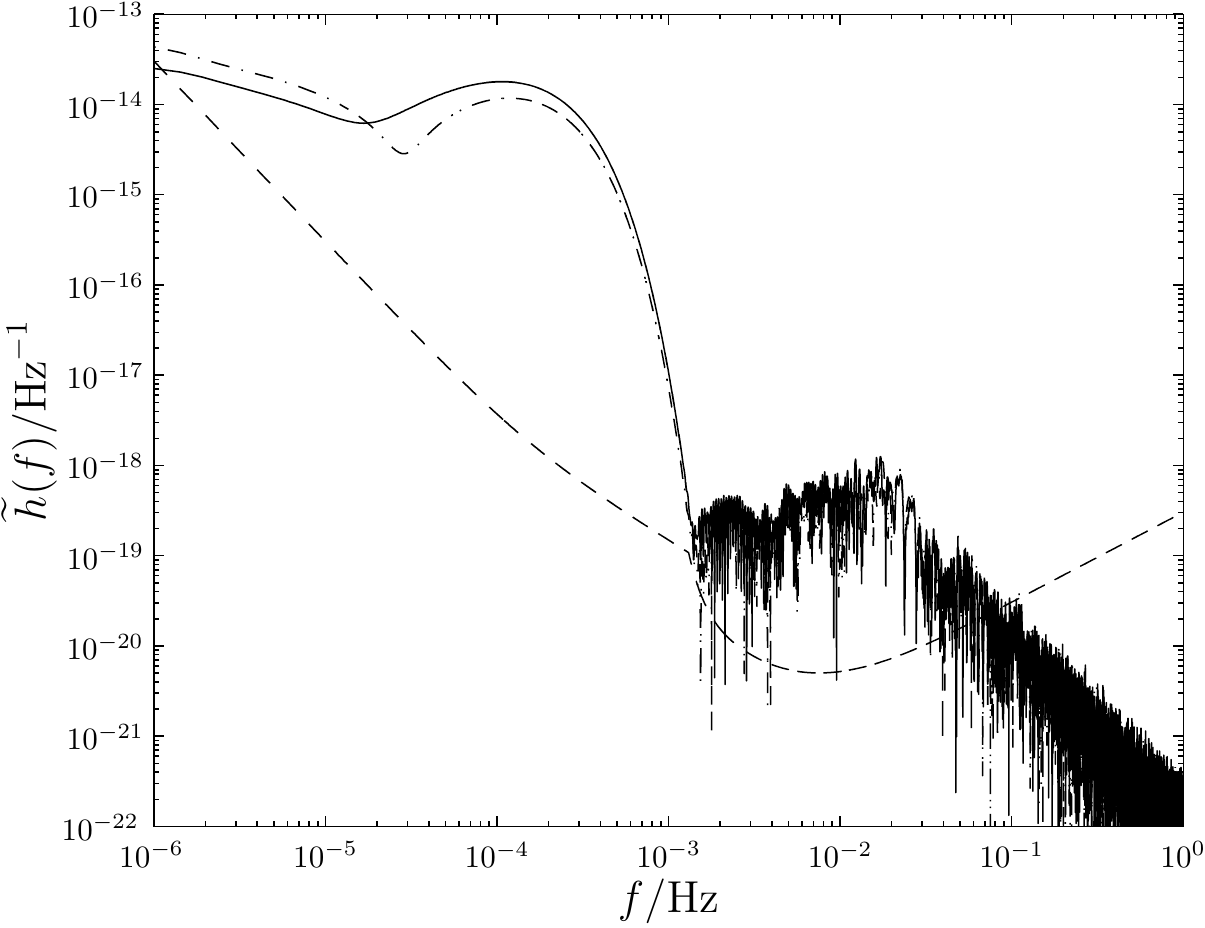}} \\
   \subfigure[{Waveform for $a_\ast \simeq 0.48$, $r\sub{p} \simeq 8.8 r\sub{g}$ and $\iota \simeq 2.0$. The SNR for the spherical polar kludge is $\rho[\boldsymbol{h}\sub{sph}] \simeq 2300$, for the oblate-spheroidal kludge $\rho[\boldsymbol{h}\sub{ob}] \simeq 2310$.}]{\label{fig:Orbit_254} \includegraphics[width=0.43\textwidth]{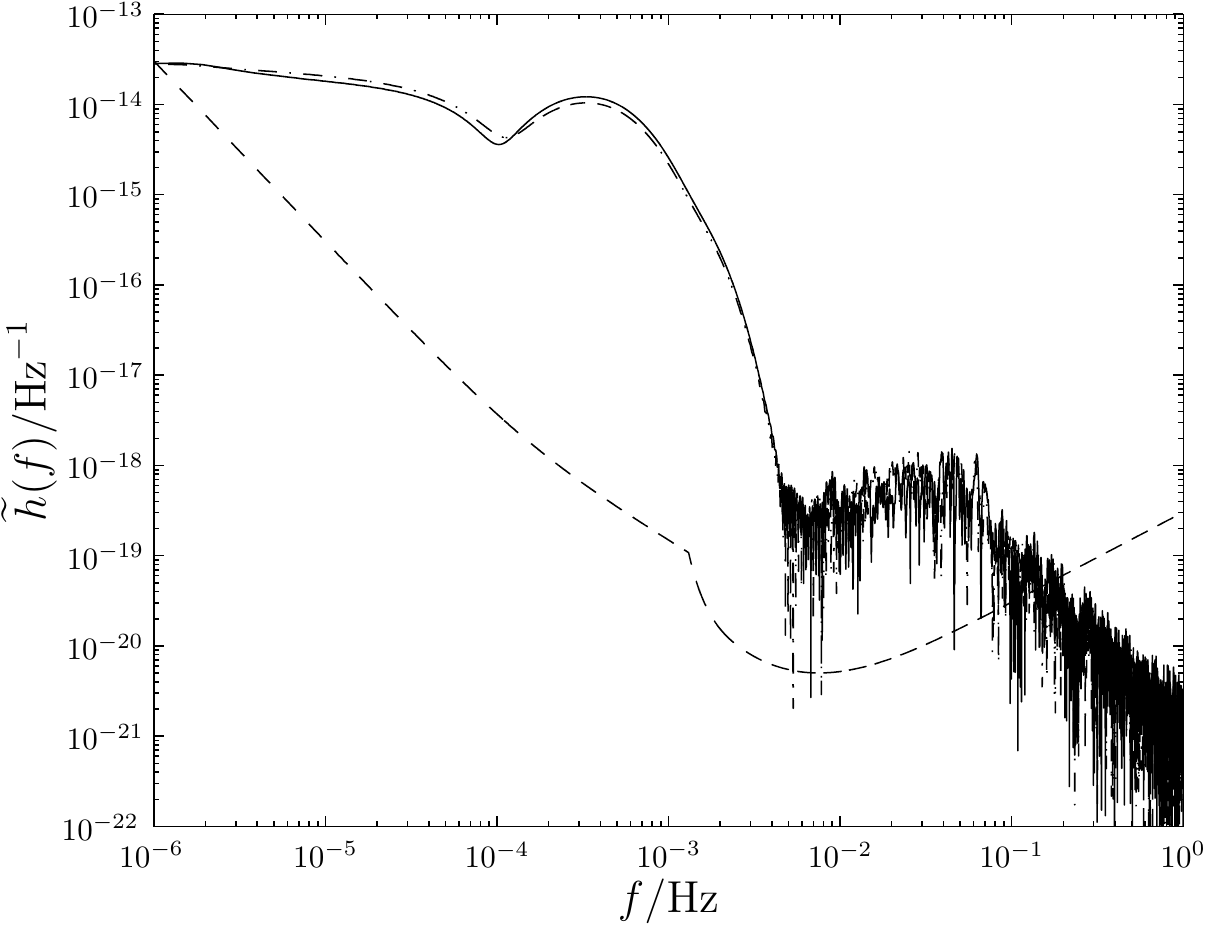}} \\
   \subfigure[{Waveform for $a_\ast \simeq 0.74$, $r\sub{p} \simeq 3.2 r\sub{g}$ and $\iota \simeq 1.2$. The SNR for the spherical polar waveform is $\rho[\boldsymbol{h}\sub{sph}] \simeq 70600$, for the oblate-spheroidal kludge $\rho[\boldsymbol{h}\sub{ob}] \simeq 74900$.}]{\label{fig:Orbit_135} \includegraphics[width=0.43\textwidth]{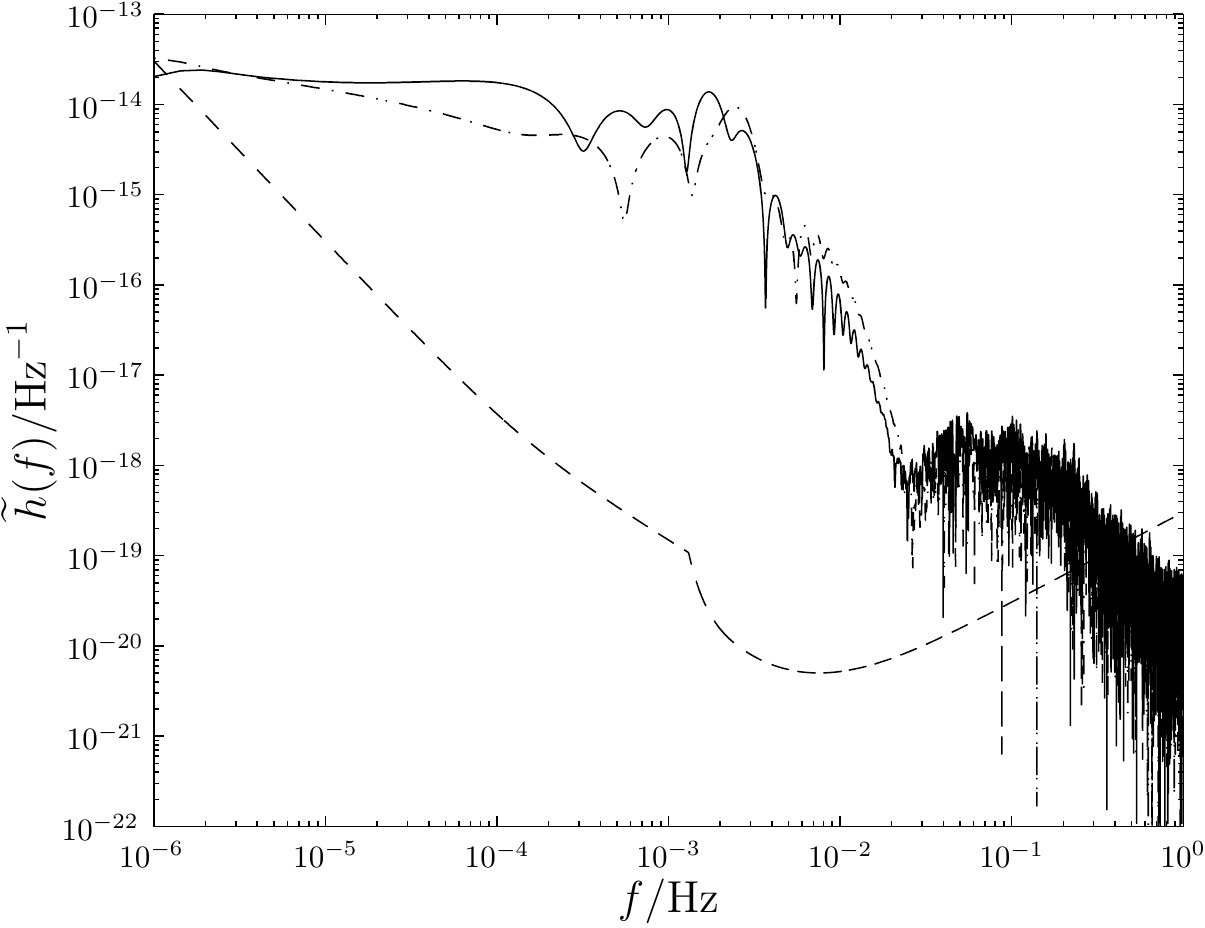}}  
\caption{Example burst waveforms from the GC. The strain $\widetilde{h}\sub{I}(f)$ is indicated by the solid line, $\widetilde{h}\sub{II}(f)$ by the dot-dashed line, and the noise curve by the dashed line. The kludge has been formulated using spherical polar coordinates.\label{fig:Examples}}
  \end{center}
\end{figure}

\Figref{Examples} shows example waveforms to demonstrate some possible variations in the signal. All assume the standard MBH mass and position as well as a $\mu = 10 M_\odot$ CO; other (randomly chosen) orbital parameters are specified in the captions. Radii are given in gravitational radii $r\sub{g} = GM_\bullet / c^2$.

The plotted waveforms use the spherical polar coordinate system for the NK. Using oblate-spheroidal coordinates makes a small difference: on the scale shown the only discernible difference would be in \figref{Orbit_135}; the maximum difference in the waveform (outside the high-frequency tail) is $\sim 10\%$. In the other cases the difference is entirely negligible (except in the high-frequency tail, which is not of physical significance). This behaviour is typical; for the closest orbits, with the most extreme spin parameters, the maximum difference in the waveforms may be $\sim30\%$. The difference is largely confined to the higher frequency components, which are most sensitive to the parts of the trajectory closer to the MBH: the change in flat-space radius for the same Boyer-Lindquist radial coordinate causes a slight shift in the shape of the spectrum. Enforcing the same flat-space periapse radius gives worse agreement across the spectrum.

\begin{figure}
\begin{center}
 \includegraphics[width=0.43\textwidth]{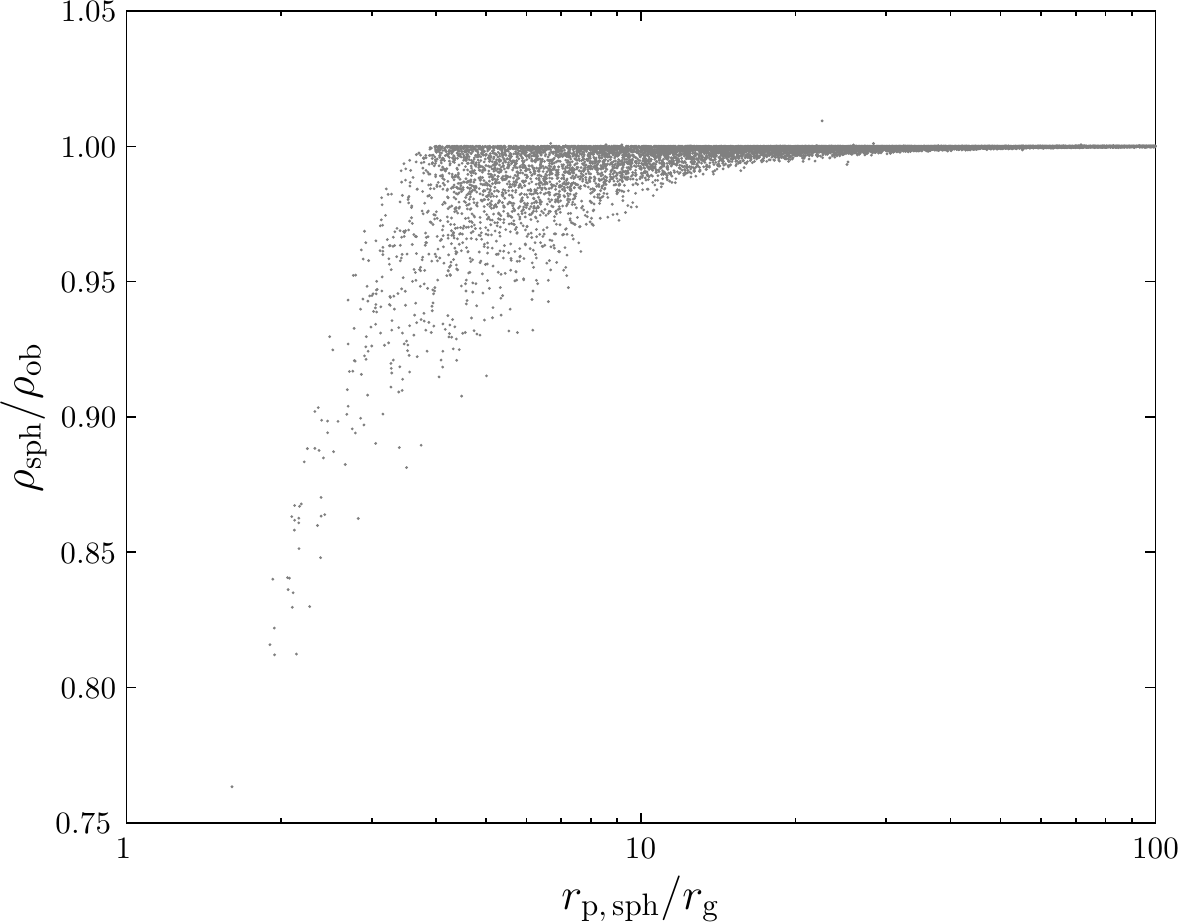}
 \caption{Ratio of the SNR for a waveform calculated using spherical polar coordinates to that for a waveform using oblate-spheroidal coordinates.\label{fig:Oblate_sphere}}
   \end{center}
\end{figure}

To examine the effect of the coordinate choice, we compare SNRs calculated using the alternative schemes. The MBH parameters were fixed as for the GC, the orbital parameters were chosen such that periapse distance was drawn from a logarithmic distribution (down to the innermost stable orbit), and other parameters were drawn from appropriate uniform distributions. The ratio of the two SNRs is shown in \figref{Oblate_sphere}.
The difference from the coordinate systems is only apparent for orbits with very small periapses. There is agreement to $10\%$ down to $r\sub{p} \simeq 4 r\sub{g}$; the maximal difference may be expected to be $\sim 20\%$, this is for periapses that are only obtainable for high spin values.

Since the deviation in the two waveforms is only apparent for small periapses, when the kludge approximation is least applicable, we conclude that the choice of coordinates is unimportant. The potential error of order $10\%$ is no greater than that inherent in the NK approximation (see \secref{Energy}). Without an accurate waveform template to compare against, we do not know if there is a preferable choice of coordinates. We adopt spherical coordinates for easier comparison with existing work.

\subsection{Signal-to-noise ratios}

The detectability of a burst depends upon its SNR. To characterise the variation of $\rho$ we considered a range of orbits, each for the standard MBH mass and position. These bursts were calculated for a $1 M_\odot$ CO. From \eqnref{octupole}, the amplitude of the waveform is proportional to the CO mass $\mu$ and so $\rho$ is also proportional to $\mu$; a $10 M_\odot$ object is ten times louder on the same orbit. To make results mass independent, we use a mass-normalised SNR
\begin{equation}
\hat{\rho}[\boldsymbol{h}] = \left(\frac{\mu}{M_\odot}\right)^{-1}\rho[\boldsymbol{h}].
\end{equation}

The spin of the MBH and the orbital inclination were randomly chosen, and the periapse distance was drawn from a logarithmic distribution down the inner-most stable orbit. For each set of these extrinsic parameters, the periapse position, orientation of the MBH, and orbital position of the detector were varied: five random combinations of these intrinsic parameters (each being drawn from a separate uniform distribution) were used for each point. We take the mean of $\ln \rho$ for each set of randomised intrinsic parameters.\footnote{The logarithm is a better quantity to work with since the SNR is a positive-definite quantity that may be distributed over a range of magnitudes \citep[sections 22.1, 23.3]{MacKay2003}. Using median values yields results that are quantitatively similar.}

\begin{figure}
  \begin{center}
  \includegraphics[width=0.43\textwidth]{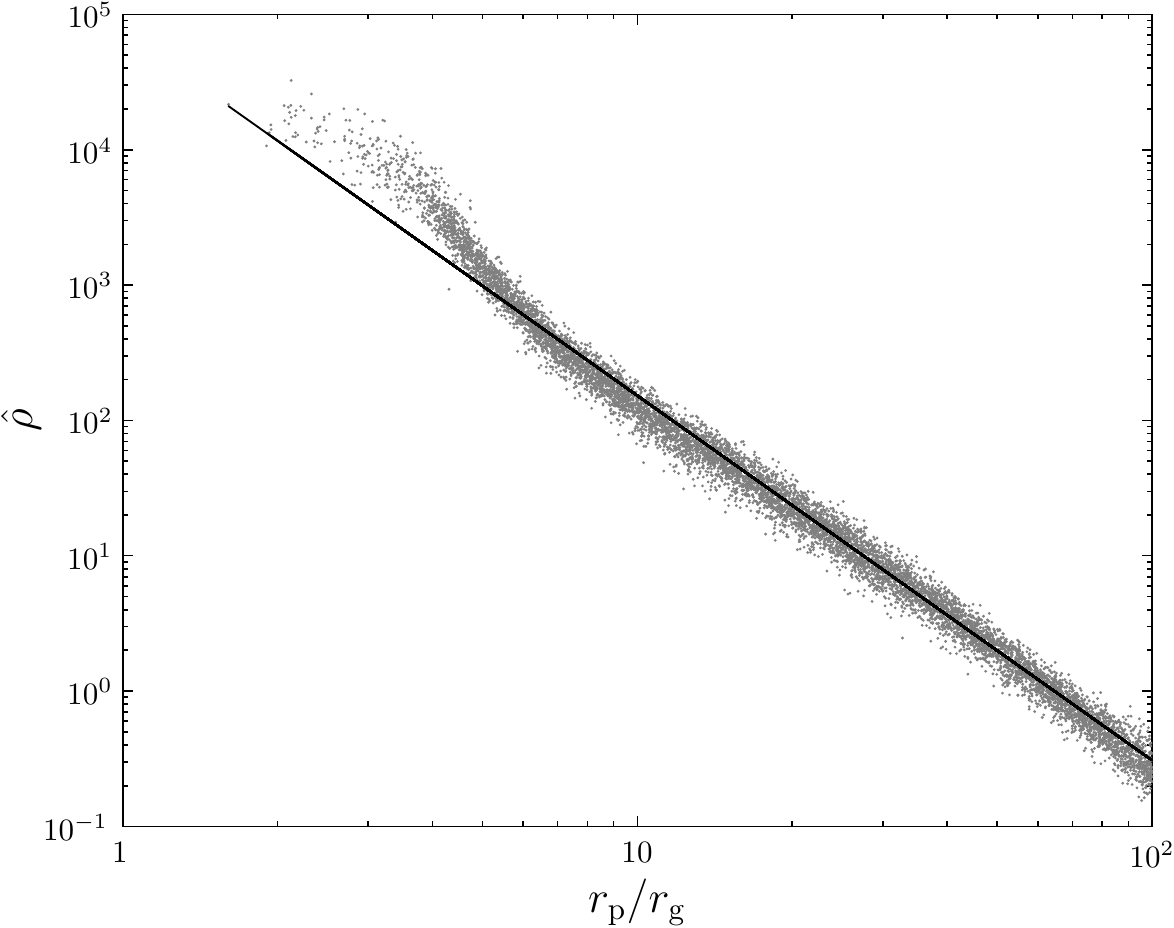}
    \caption{Mass-normalised SNR as a function of periapse radius. The plotted points are the values obtained by averaging over each set of intrinsic parameters. The best fit line is $\log(\hat{\rho}) = -2.69\log(r\sub{p}/r\sub{g}) + 4.88$. This is fitted to orbits with $r\sub{p} >  13.0 r\sub{g}$ and has a reduced chi-squared value of $\chi^2/\nu = 1.73$.\label{fig:SNR}}
      \end{center}
\end{figure}

There exists a correlation between the periapse radius and SNR, as shown in \figref{SNR}.
Closer orbits produce louder bursts. To reflect this trend, we have fitted a simple fiducial power law, indicated by the straight line.\footnote{Using oblate-spheroidal coordinates instead of spherical polars gives a fit consistent to within $0.1\%$ as we have excluded the closest orbits.} This was done by maximising the likelihood, assuming $\ln \rho$ has a Gaussian distribution with standard deviation derived from the scatter because of variation in the intrinsic parameters. The power law is a good fit only for larger periapses. The shape is predominately determined by the noise curve. The change in the trend reflects the transition as from approximately power law behaviour to the bucket of the noise curve. Hence, we fit a power law to orbits with a characteristic frequency of $f_\ast = \sqrt{GM_\bullet/r\sub{p}} < 1 \times 10^{-3}\units{Hz}$, to avoid spilling into the bucket. Changing the cut-off within a plausible region alters the fit coefficients by around $0.1$.\footnote{The power law exponent $-2.7$ is inconsistent with $-13/4$ as predicted by the approximate model of \citet{Hopman2007}. This is the result of their approximate waveform model.}

The SNR shows no clear correlation with the other parameters (excluding $\mu$). However, the SNR is sensitive to the intrinsic parameters, in particular the initial position, and may vary by an order of magnitude.

Setting a threshold of $\rho = 10$, a $1 M_\odot$ ($10 M_\odot$) object would be expected to be detectable if the periapse distance is less than $27 r\sub{g}$ ($65 r\sub{g}$). \citet{Hopman2007}, assuming a threshold of $\rho = 5$, used an approximate form for the SNR based upon the quadrupole component of a circular orbit; their model, with updated parameters for the MBH, predicts bursts would be detectable out to $66 r\sub{g}$ ($135 r\sub{g}$). This is overly optimistic.

\section{Energy spectra}\label{sec:Energy}

To check the NK waveforms, we compare the energy spectra calculated from these with those obtained from the classic treatment of \citet{Peters1963} and \citet{Peters1964}. This calculates GW emission for Keplerian orbits in flat spacetime, assuming only quadrupole radiation. The spectrum produced should be similar to that obtained from the NK in weak fields, that is for large periapses; we do not expect an exact match because of the differing input physics and varying approximations.

In addition to using the energy spectrum, we also use the total energy flux. This contains less information than the spectrum; however, results have been calculated for parabolic orbits in Schwarzschild spacetime using time-domain black hole perturbation theory \citep{Martel2004}. These should be more accurate than results calculated using the Peters and Mathews formalism.

We do not intend to use the kludge waveforms to calculate an accurate energy flux: this would be inconsistent as we assume the orbits do not evolve with time. We only calculate the energy flux as a sanity check, to confirm that the kludge approximation is consistent with other approaches.

\subsection{Kludge spectrum}

A gravitational wave in the TT gauge has an effective energy-momentum tensor (\citealt{Misner1973}, section 35.15)
\begin{equation}
T_{\mu\nu} = \frac{c^4}{32\pi G}\left\langle\partial_\mu h_{ij} \partial_\nu h^{ij}\right\rangle,
\end{equation}
where $\langle\ldots\rangle$ indicates averaging over several wavelengths or periods. The energy flux through a sphere of radius $R$ is
\begin{equation}
\diff{E}{t} = \frac{c^3}{32\pi G} R^2 \int{\dd\Omega}\left\langle\diff{h_{ij}}{t}\diff{h^{ij}}{t}\right\rangle,
\end{equation}
with $\int{\dd\Omega}$ representing integration over all solid angles. From \eqnref{octupole} the waves have a $1/{r}$ dependence; if we define
\begin{equation}
h_{ij} = \frac{H_{ij}}{r},
\end{equation}
we see, the flux is independent of $R$, as required for energy conservation, and
\begin{equation}
\diff{E}{t} = \frac{c^3}{32\pi G} \int{\dd\Omega}\left\langle\diff{H_{ij}}{t}\diff{H^{ij}}{t}\right\rangle.
\end{equation}
Integrating to find the total energy emitted we obtain
\begin{equation}
E = \frac{c^3}{32\pi G} \int{\dd\Omega}\int_{-\infty}^{\infty}{\dd t} \, \diff{H_{ij}}{t}\diff{H^{ij}}{t}.
\label{eq:integrate_E}
\end{equation}
Since we are considering all time, the localization of the energy is no longer of importance and it is unnecessary to average over several periods. Switching to Fourier representation $\widetilde{H}_{ij}(f) = \mathscr{F}\left\{H_{ij}(t)\right\}$,
\begin{equation}
E = \frac{\pi c^3}{4 G} \int{\dd\Omega}\int_{0}^{\infty}{\dd f} \, f^2 \widetilde{H}^{ij}(f)\widetilde{H}_{ij}^*(f),
\label{eq:total_E}
\end{equation}
using $\widetilde{H}_{ij}^*(f) = \widetilde{H}_{ij}(-f)$ as the signal is real. From this we identify the energy spectrum as
\begin{align}
\diff{E}{f} = \frac{\pi c^3}{4 G} \intd{}{}{}{\Omega} \, f^2 \widetilde{H}^{ij}(f)\widetilde{H}_{ij}^*(f).
\label{eq:NK_dEdf}
\end{align}

\subsection{Peters and Mathews spectrum}

To calculate the Peters and Mathews energy spectrum for a parabolic orbit, we use the limiting result of \citet{Turner1977}. This result should be accurate to $\sim10\%$ for orbits with periapse radii larger than $\sim20r\sub{g}$ \citep{Berry2010}.

\subsection{Comparison}

\begin{figure*}
  \begin{center}
   \subfigure[$r\sub{p} = 15.0 r\sub{g}$, log-log plot.]{\includegraphics[width=0.43\textwidth]{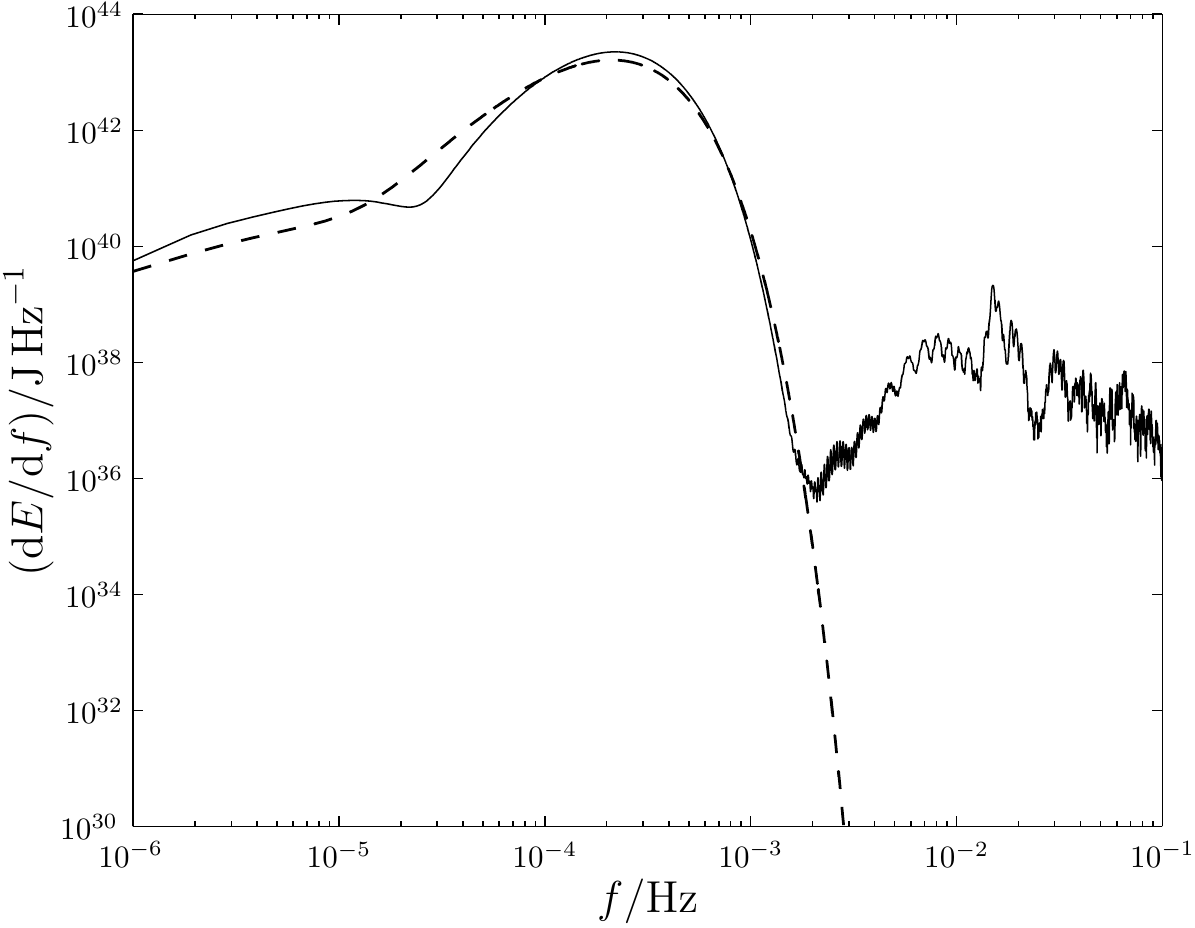}} \quad
   \subfigure[$r\sub{p} = 15.0 r\sub{g}$, log-linear plot.]{\includegraphics[width=0.43\textwidth]{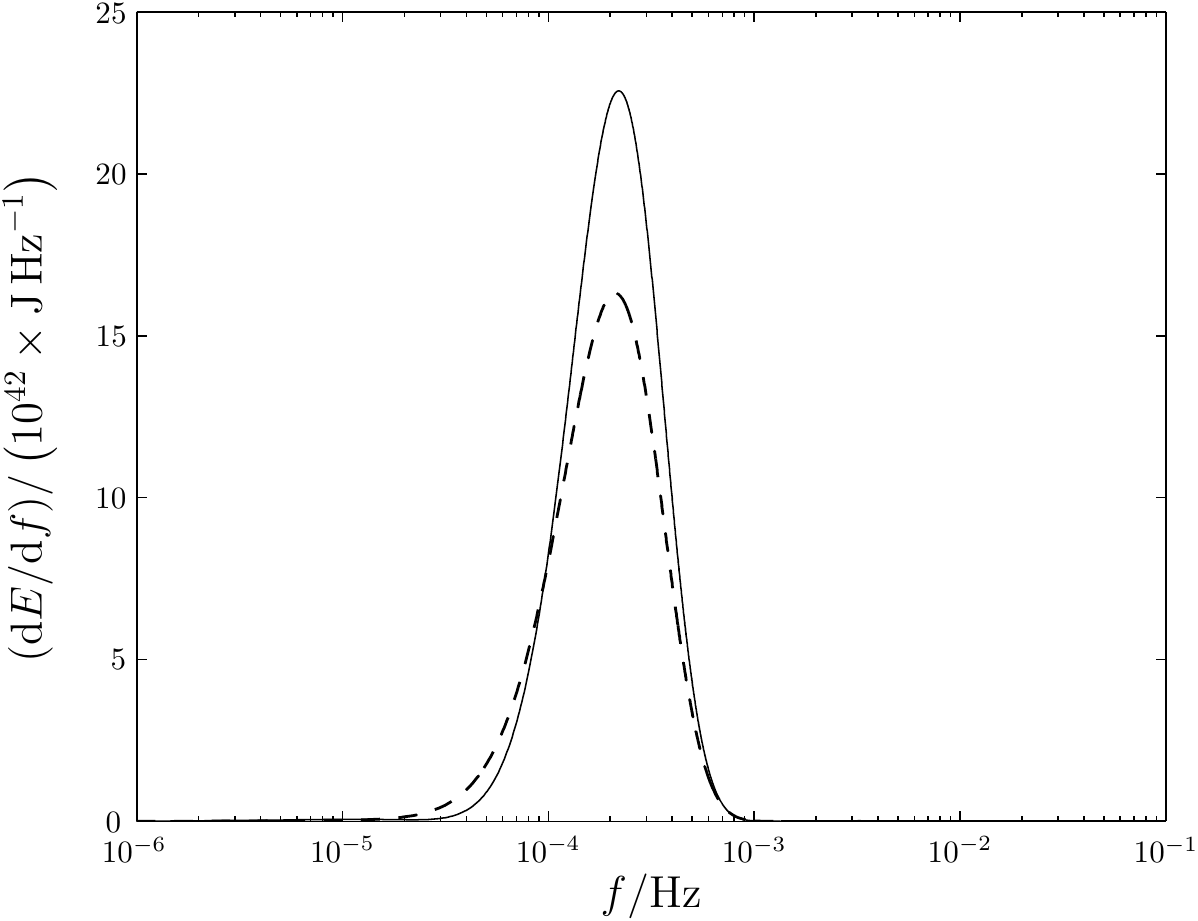}} \\
   \subfigure[$r\sub{p} = 30.0 r\sub{g}$, log-log plot.]{\includegraphics[width=0.43\textwidth]{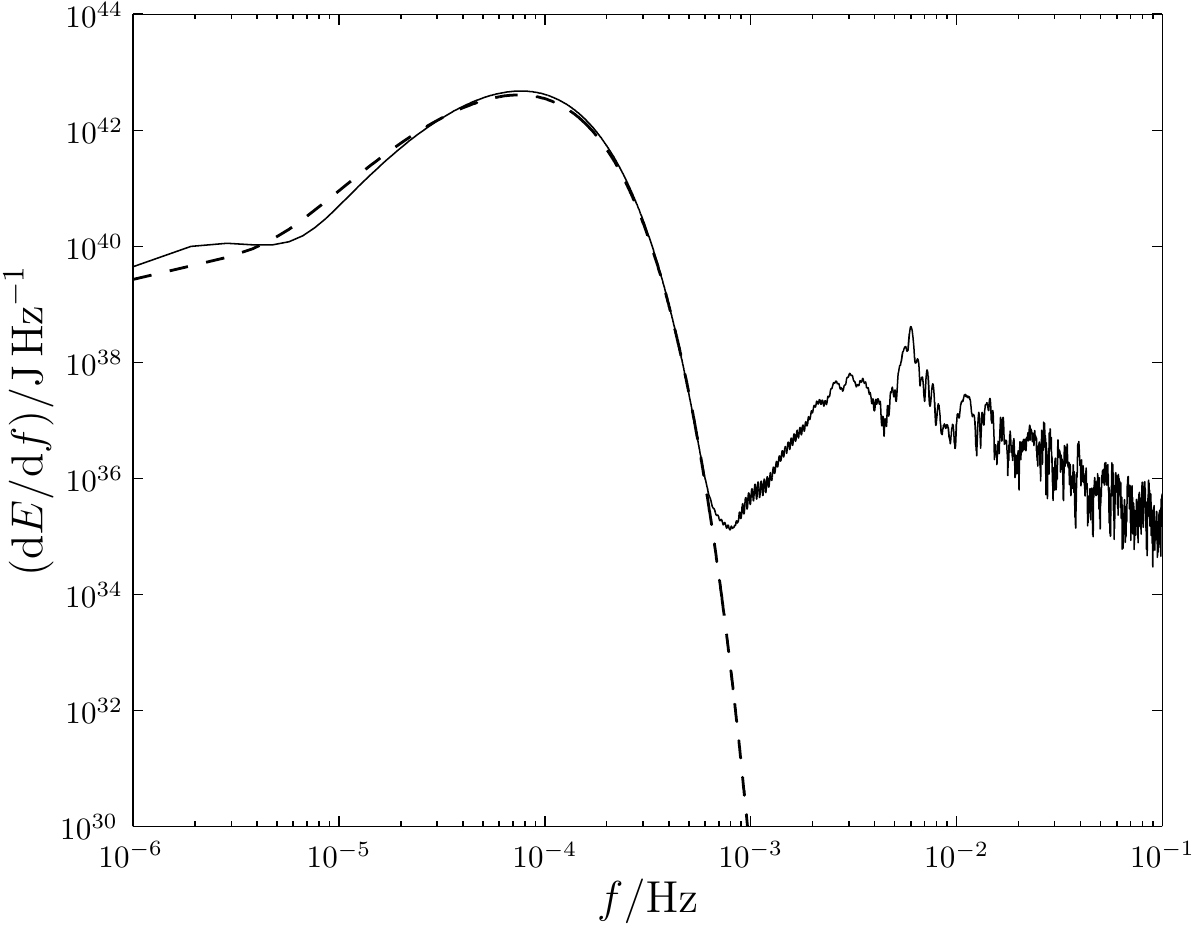}} \quad
   \subfigure[$r\sub{p} = 30.0 r\sub{g}$, log-linear plot.]{\includegraphics[width=0.43\textwidth]{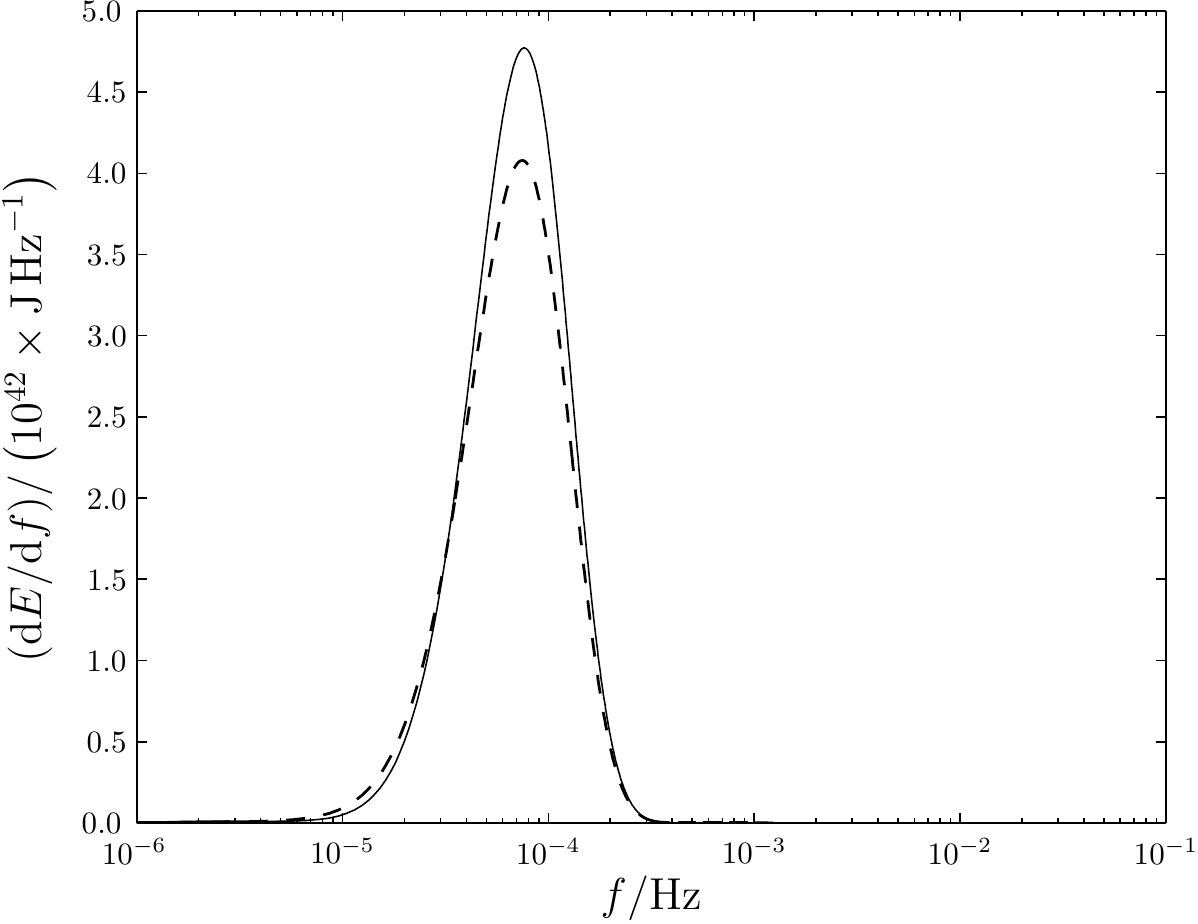}} \\
   \subfigure[$r\sub{p} = 60.0 r\sub{g}$, log-log plot.]{\includegraphics[width=0.43\textwidth]{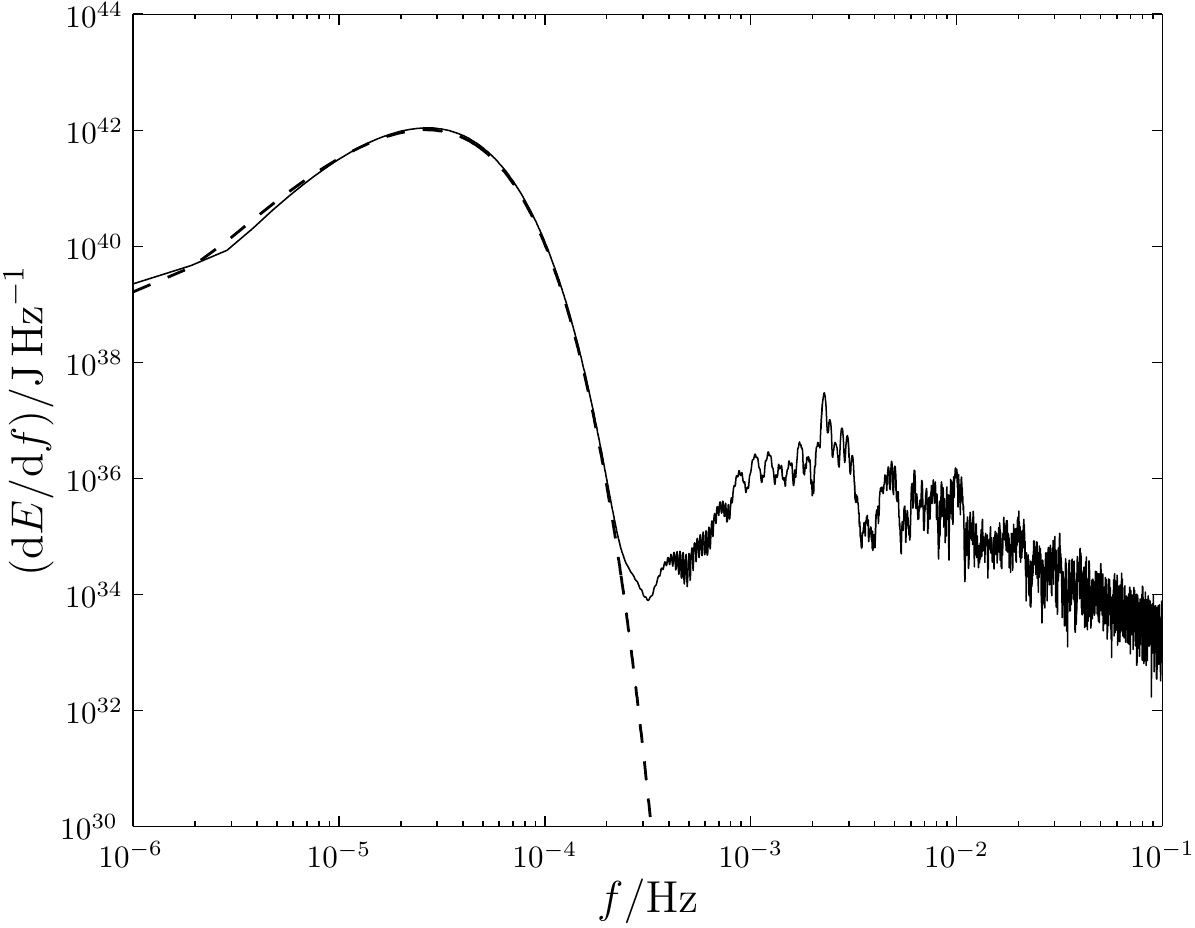}} \quad
   \subfigure[$r\sub{p} = 60.0 r\sub{g}$, log-linear plot.]{\includegraphics[width=0.43\textwidth]{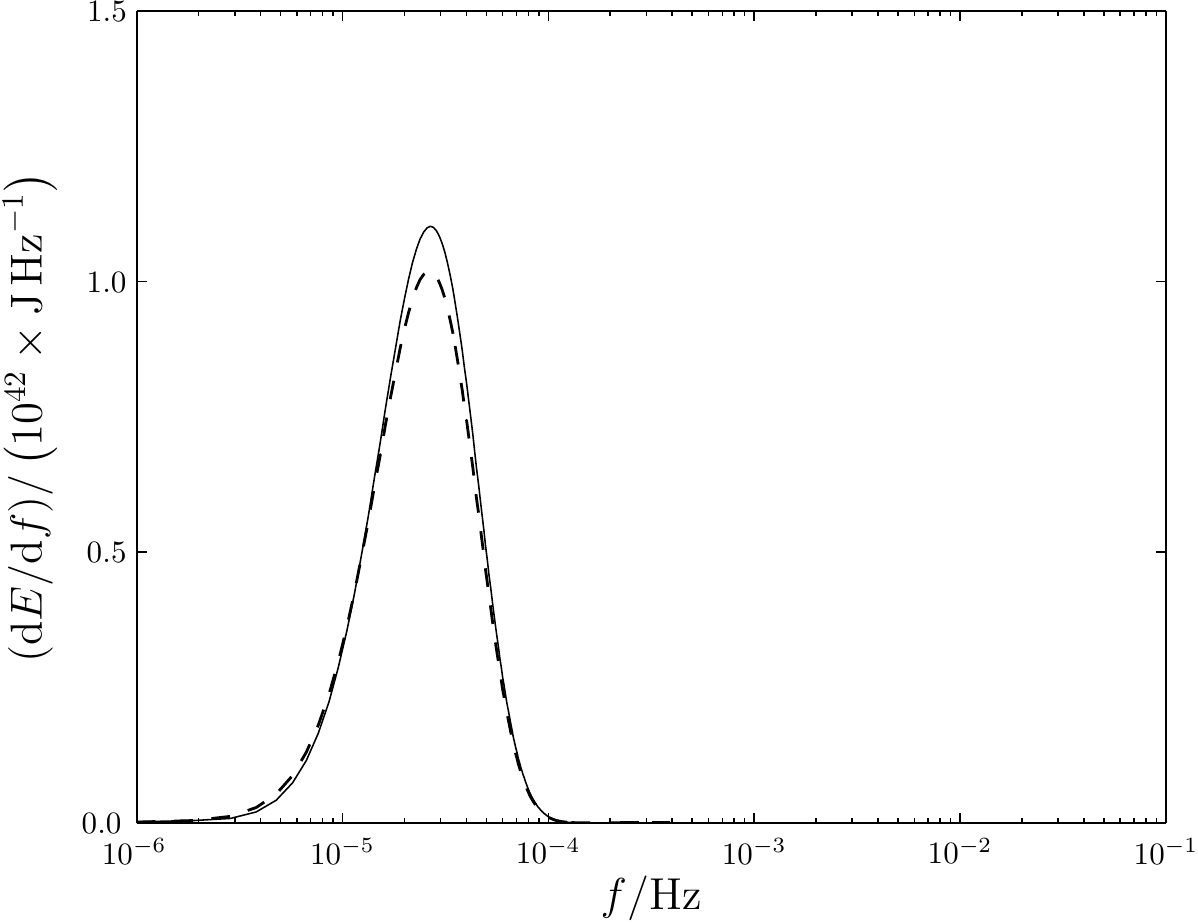}}
    \caption{Energy spectra for a parabolic orbit of a $\mu = 10 M_\odot$ object about a Schwarzschild BH with $M_\bullet = 4.31 \times 10^6 M_\odot$. The spectra calculated from the NK waveform are shown by the solid line and the Peters and Mathews flux is indicated by the dashed line. The NK waveform includes current quadrupole and mass octupole contributions. The high frequency tail is the result of spectral leakage.\label{fig:Energy}}
  \end{center}
\end{figure*}

Two energy spectra are plotted in \figref{Energy} for orbits with periapses of $r\sub{p} = 15.0 r\sub{g}$, $30.0 r\sub{g}$ and $60.0 r\sub{g}$.
The two spectra appear to be in good agreement, showing the same general shape in the weak-field limit. The NK spectrum is more tightly peaked, but is always within a factor of $2$ at the apex. The peak of the spectrum is shifted to a marginally higher frequency in the NK spectrum primarily because of the addition of the higher order terms.

\begin{figure*}
  \begin{center}
   \subfigure[Versus periapsis]{\includegraphics[width=0.43\textwidth]{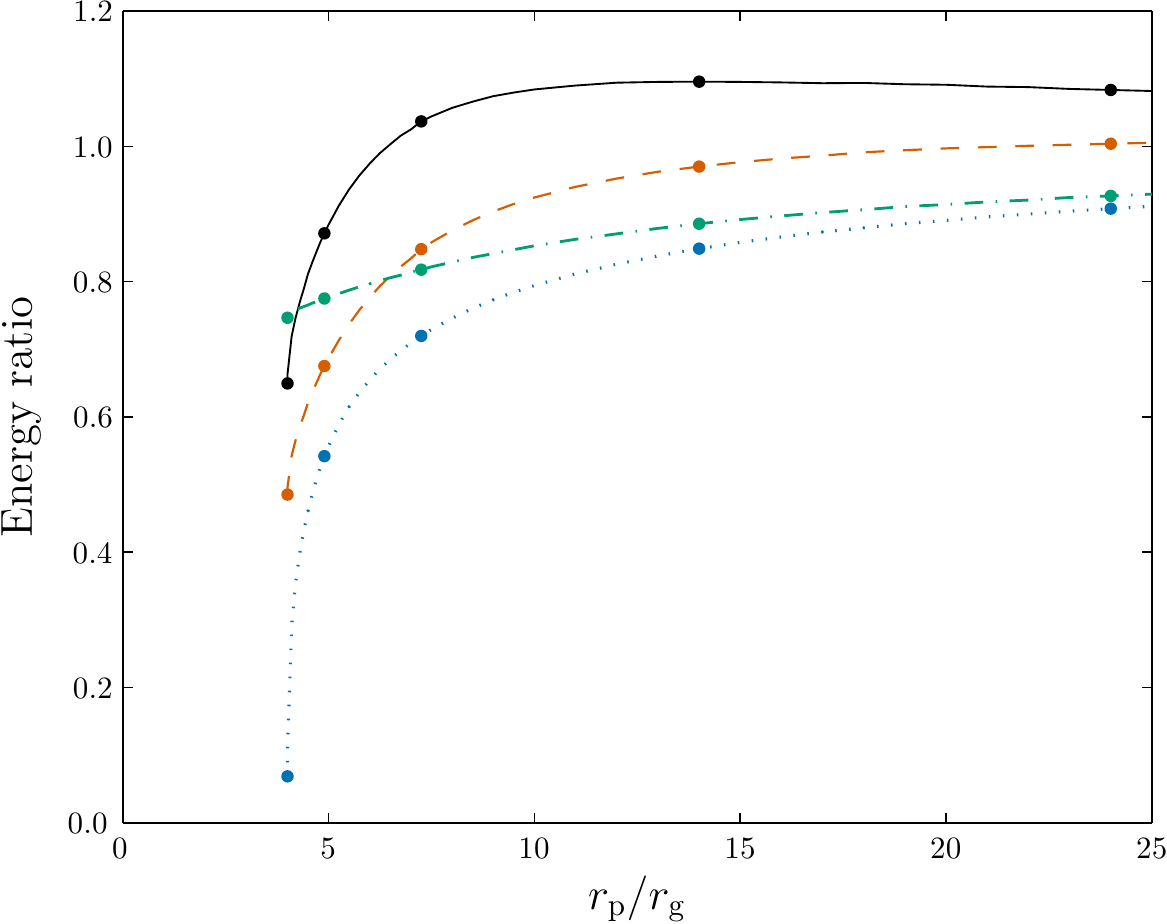}} \quad
   \subfigure[Versus amount of rotation]{\includegraphics[width=0.43\textwidth]{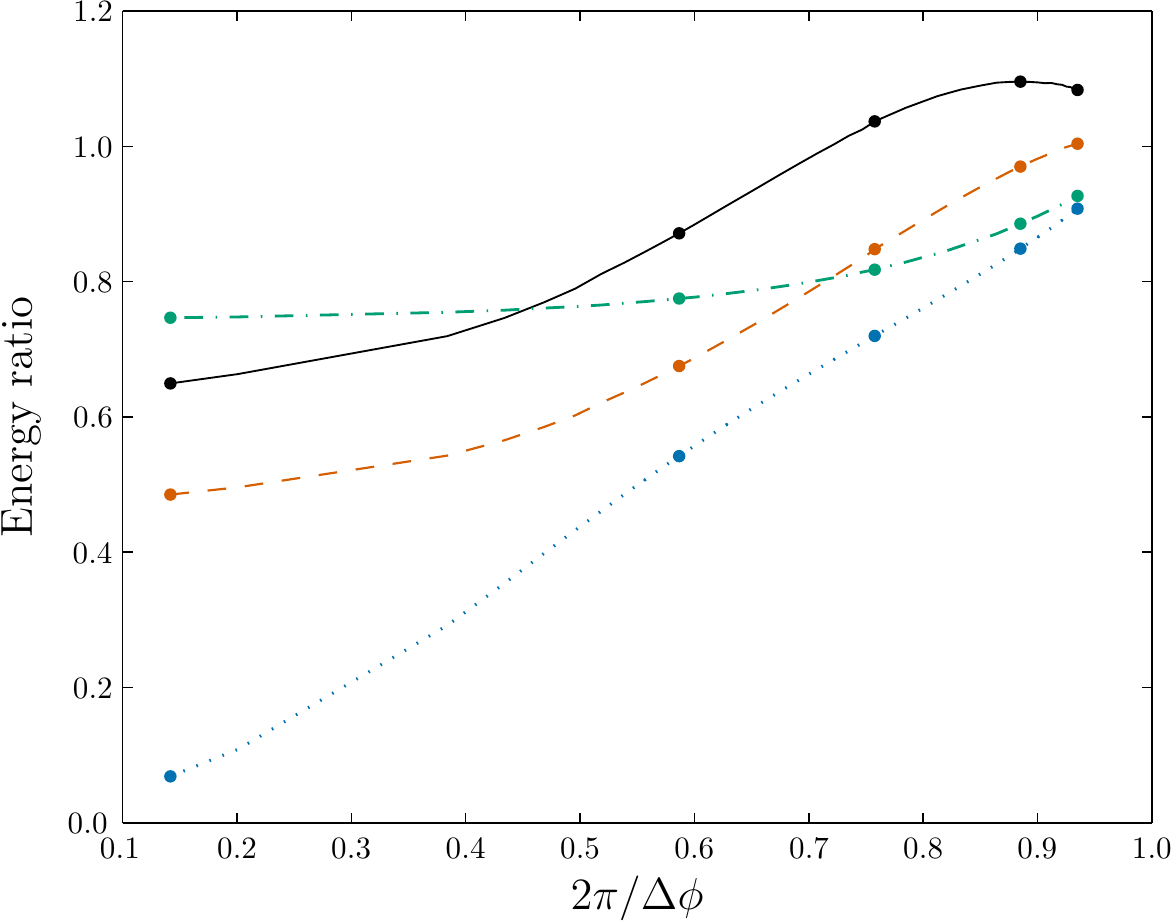}}
    \caption{Ratios of energies as a function of periapsis $r\sub{p}$ and $2\pi$ divided by the total angle of rotation in one orbit $\Delta\phi$ ($2\pi/\Delta\phi = 1$ for a Keplerian orbit). The solid line shows the ratio of the numerical kludge and Martel energies $E\sub{NK}/E\sub{M}$; the dashed line shows the ratio of the NK energy calculated using only the mass quadrupole term and the Martel energy $E\sub{NK(Q)}/E\sub{M}$; the dot-dashed line shows the ratio of the quadrupole and quadrupole-octupole NK energies $E\sub{NK(Q)}/E\sub{NK}$, and the dotted line shows the ratio of the Peters and Mathews and quadrupole NK energies $E\sub{PM}/E\sub{NK(Q)}$. The spots show the mapping between the two abscissa scales. Compare with figure 4 of \citet{Gair2005}.\label{fig:Energy_ratio}}
  \end{center}
\end{figure*}

Comparing the total energy fluxes, ratios of the various energies are plotted in \figref{Energy_ratio}.
We introduce an additional energy, the quadrupole NK energy $E\sub{NK(Q)}$. This allows easier comparison with the Peters and Mathews energy which includes only quadrupole radiation. It can be calculated in three ways:
\begin{enumerate}
\item Inserting the waveform $\widetilde{h}(f)$ generated including only the mass quadrupole term in \eqnref{octupole} into \eqnref{total_E} and integrating. This is equivalent to the method used to calculate $E\sub{NK}$.
\item Numerically integrating the quadrupole GW luminosity (\citealt{Misner1973}, section 36.7; \citealt{Hobson2006}, section 18.7)
\begin{equation}
E = \frac{G}{5c^9}\intd{}{}{\dddot{\Ibar}_{ij}\dddot{\Ibar}^{ij}}{t},
\label{eq:E_quad}
\end{equation}
where $\Ibar_{ij} = I_{ij} - (1/3)I\delta_{ij}$ is the reduced mass quadrupole tensor. We can obtain this from \eqnref{integrate_E}, by integrating over all angles when the waveform only contains the mass quadrupole component. This has the advantage of avoiding the effects of spectral leakage or the influence of window functions.
\item Using the analytic expressions for the integral \eqnref{E_quad} given in appendix A of \citet{Gair2005}.
\end{enumerate}
All three agree to within computational error. No difference is visible on the scale plotted in \figref{Energy_ratio}. This demonstrates the validity of the code, and shows that the use of a window function does not significantly distort the waveform.

The ratios all tend towards one in the weak field, as required, but differences become more pronounced in the strong field. The NK energy is larger than the Peters and Mathews result $E\sub{PM}$. This behaviour has been seen before for high eccentricity orbits about a non-spinning BH \citep{Gair2005}. It may be explained by considering the total path length for the different orbits: the Peters and Mathews spectrum assumes a Keplerian orbit, the orbit in Kerr geometry rotates more than this. The greater path length leads to increased emission of gravitational waves and a larger energy flux \citep{Berry2010}. Our bead must travel further along its wire. A good proxy for the path length is the angle of rotation $\Delta\phi$; this is $2\pi$ for a Keplerian orbit, in Kerr the angle should be $2\pi$ in the limit of an infinite periapsis, whereas for a periapsis small enough that the orbit shows zoom-whirl behaviour, the total angle may be many times $2\pi$. There is a reasonable correlation between the amount of rotation $2\pi/\Delta\phi$ and the ratio of energies.

Error in the NK energy compared with the time-domain black hole perturbation theory results of Martel comes from two sources: the neglecting of higher order multipole contributions and the ignoring of background curvature. The contribution of the former can be estimated by looking at the difference in the NK energy by including the current quadrupole and mass octupole terms. From \figref{Energy_ratio} we see these terms give a negligible contribution in the weak field, but the difference is $\sim20\%$ in the strong field. This explains why the Martel energy $E\sub{M}$ is greater in the strong field: it includes contributions from all multipoles. Neglecting the background curvature increases the NK energy relative to $E\sub{M}$. This partially cancels out the error introduced by not including higher order terms: this accidentally leads to $E\sub{NK(Q)}$ being more accurate than $E\sub{NK}$ for $r\sub{p} \gtrsim 10 r\sub{g}$ \citep{Tanaka1993}.

From the level of agreement we may be confident that the NK waveforms are a reasonable approximation. The difference in energy flux is only greater than $10\%$ for very strong fields $r\sub{p} \simeq 4 r\sub{g}$; since this is dependent on the square of the waveform, typical accuracy in the waveform may be $\sim 5\%$ \citep{Gair2005, Tanaka1993}. This is more significant than the variation in waveforms we generally found using the two alternative coordinate systems for the NK (in this case the two coincide because $a_\ast = 0$).

\section{Parameter estimation}\label{sec:Estimation}

Having detected a signal, we are interested in what we can learn about the source. We have an inference problem that can be solved by  application of Bayes' Theorem (\citealt{Jaynes2003}, chapter 4): the probability distribution for our parameters given that we have detected the signal $\boldsymbol{s}(t)$ is given by the posterior
\begin{equation}
p(\boldsymbol{\lambda}|\boldsymbol{s}(t)) = \frac{p(\boldsymbol{s}(t)|\boldsymbol{\lambda})p(\boldsymbol{\lambda})}{p(\boldsymbol{s}(t))}.
\end{equation}
Here $p(\boldsymbol{s}(t)|\boldsymbol{\lambda})$ is the likelihood of the parameters, $p(\boldsymbol{\lambda})$ is the prior probability distribution for the parameters, and the evidence $p(\boldsymbol{s}(t)) = \intd{}{}{p(\boldsymbol{s}(t)|\boldsymbol{\lambda})}{^d \lambda}$ is, for our purposes, a normalising constant. The likelihood depends upon the realization of noise. If parameters $\boldsymbol{\lambda}_0$ define a waveform $\boldsymbol{h}_0(t) = \boldsymbol{h}(t; \boldsymbol{\lambda}_0)$, the probability that we observe signal $\boldsymbol{s}(t)$ GW is given by \eqnref{sig_prob}, so the likelihood is
\begin{equation}
p(\boldsymbol{s}(t)|\boldsymbol{\lambda}_0) \propto \exp\left[-\recip{2}\innerprod{\boldsymbol{s}-\boldsymbol{h}_0}{\boldsymbol{s}-\boldsymbol{h}_0}\right].
\label{eq:likelihood}
\end{equation}
If we were to define this as a probability distribution for the parameters $\boldsymbol{\lambda}$, the modal values are the maximum-likelihood (ML) parameters $\boldsymbol{\lambda}\sub{ML}$. The waveform $\boldsymbol{h}(t; \boldsymbol{\lambda}\sub{ML})$ is the signal closest to $\boldsymbol{s}(t)$, where distance is defined using the inner product \eqnref{inner} \citep{Cutler1994}.

\subsection{Fisher matrices}

In the limit of a high SNR, we may approximate \citep{Vallisneri2008}
\begin{equation}
p(\boldsymbol{s}(t)|\boldsymbol{\lambda}_0) \propto \exp\left[-\recip{2}\innerprod{\partial_a\boldsymbol{h}}{\partial_b\boldsymbol{h}}\left(\lambda^a - \langle\lambda^a\rangle_\ell\right)\left(\lambda^b - \left\langle\lambda^b\right\rangle_\ell\right)\right],
\label{eq:LSA}
\end{equation}
where the mean is defined as
\begin{equation}
\langle\lambda^a\rangle_\ell = \frac{\intd{}{}{\lambda^a p(\boldsymbol{s}(t)|\boldsymbol{\lambda})}{^d \lambda}}{\intd{}{}{p(\boldsymbol{s}(t)|\boldsymbol{\lambda})}{^d \lambda}}.
\end{equation}
In the high SNR limit, this is the ML value $\langle\lambda^a\rangle_\ell = \lambda^a\sub{ML}$. The quantity
\begin{equation}
\Gamma_{ab} = \innerprod{\partial_a\boldsymbol{h}}{\partial_b\boldsymbol{h}}
\label{eq:Fisher}
\end{equation}
is the Fisher information matrix (FIM). It controls the variance of the likelihood distribution.

The form of the posterior distribution depends upon the nature of the prior information. If we have an uninformative prior, such that $p(\boldsymbol{\lambda})$ is a constant, the posterior distribution is determined by the likelihood. In the high SNR limit, we obtain a Gaussian with variance-covariance matrix
\begin{equation}
\boldsymbol{\Sigma} = \boldsymbol{\Gamma}^{-1}.
\label{eq:InvFisher}
\end{equation}
The FIM therefore gives the uncertainty associated with the inferred parameters, in this case the ML values.

If the prior restricts the allowed range for a parameter, as is the case for the spin $a_\ast$, then the posterior is a truncated Gaussian, and $\boldsymbol{\Gamma}^{-1}$ may no longer represent the variance-covariance.

If the prior is approximately Gaussian with variance-covariance matrix $\boldsymbol{\Sigma}_0$, the posterior is also Gaussian.\footnote{If we only know the typical value and spread of a parameter, a Gaussian is the maximum entropy prior (\citealt{Jaynes2003}, section 7.11): the prior that is least informative given what we know.} The posterior variance-covariance is \citep{Cutler1994, Vallisneri2008}
\begin{equation}
\boldsymbol{\Sigma} = \left(\boldsymbol{\Gamma} + \boldsymbol{\Sigma}_0^{-1}\right)^{-1}.
\label{eq:Posterior_variance}
\end{equation}
From this the inverse FIM $\boldsymbol{\Gamma}^{-1}$ is an upper bound on the size of the posterior covariance matrix.\footnote{It is also the Cram\'{e}r-Rao bound on the error covariance of an unbiased estimator \citep{Cutler1994, Vallisneri2008}. Thus it represents the frequentist error: the lower bound on the covariance for an unbiased parameter estimator $\boldsymbol{\lambda}\sub{est}$ calculated from an infinite set of experiments with the same signal $\boldsymbol{h}(t)$ but different realisations of the noise $\boldsymbol{n}(t)$.}

The FIM gives a quick way of estimating the range of the posterior. It is widely used because of this. However, it is only appropriate when the approximation of \eqnref{LSA} holds. This is known as the linearised-signal approximation (LSA), where higher order derivatives are neglected. To assess the validity of this, \citet{Vallisneri2008} recommends use of the maximum-mismatch (MM) criterion
\begin{equation}
\ln r = -\recip{2}\innerprod{\Delta\lambda^a\partial_a\boldsymbol{h}\sub{ML} - \Delta\boldsymbol{h}}{\Delta\lambda^b\partial_b\boldsymbol{h}\sub{ML} - \Delta\boldsymbol{h}}.
\end{equation}
Here $\Delta \boldsymbol{\lambda}$ is the displacement to some point on the $1\sigma$ surface
\begin{equation}
\Delta \boldsymbol{\lambda} = \boldsymbol{\lambda}\sub{1\sigma} - \boldsymbol{\lambda}\sub{ML},
\end{equation}
and $\Delta \boldsymbol{h}$ is the corresponding change in the waveform
\begin{equation}
\Delta \boldsymbol{h} = \boldsymbol{h}(\boldsymbol{\lambda}\sub{1\sigma}) - \boldsymbol{h}(\boldsymbol{\lambda}\sub{ML}).
\end{equation}
The $1\sigma$ surface is defined from the inverse of the FIM. If higher order terms are indeed negligible, the MM criterion is small. We check this by picking a random selection of points on the $1\sigma$ surface and evaluating $|\ln r|$. If this is smaller than a fiducial value ($|\ln r| = 0.1$) over the majority ($90\%$) of the surface we consider the LSA sufficiently justified.

We calculated FIMs for a wide range of orbits and checked the MM criterion. We found that for the overwhelming majority the test failed: the LSA is not appropriate. This behaviour was seen even for orbits with $\rho \sim 10^3$--$10^4$.\footnote{In this study, to increase $\rho$ we must reduce the periapse distance; this also reduces the region where the LSA is valid as parameter dependencies become more non-linear. If we had the luxury of increasing $\rho$ by moving the GC closer, things could be different.} Higher order terms are important, and cannot be neglected.

EMRBs have a short duration and accordingly are not the most informative of signals. Therefore, the $1\sigma$ surface as defined by considering only the LSA terms is large. Taking such a step in parameter space moves the signal beyond the region of linear changes.

We hope that this shall serve as an example to others. What constitutes high SNR depends upon the signal; it is not enough for $\rho > 1$. As stressed by \citet{Vallisneri2008}, it is essential to check the MM criterion for individual waveforms: the threshold for the LSA to become applicable could be much greater than naively thought.

As we cannot be confident in FIM results, we abandon this approach in favour of using Markov chain Monte Carlo simulations to explore constraints from different regions of parameter space. These are computationally more expensive, but do not rely on any approximations.

\subsection{Markov chain Monte Carlo methods}

Markov chain Monte Carlo (MCMC) methods are widely used for inference problems; they are a family of algorithms for integrating over complicated distributions and are efficient for high-dimensional problems~\citep[chapter 29]{MacKay2003}. Parameter space is explored by constructing a chain of $N$ samples. The distribution of points visited by the chain maps out the underlying distribution; this becomes asymptotically exact as $N \rightarrow \infty$. Samples are added sequentially, if the current state is $\boldsymbol{\lambda}_n$ a new point $\boldsymbol{\lambda}^\ast$ is drawn and accepted with probability
\begin{equation}
\mathcal{A} = \min\left\{\frac{\pi(\boldsymbol{\lambda}^\ast)\mathcal{L}(\boldsymbol{\lambda}^\ast)\mathcal{Q}(\boldsymbol{\lambda}_n;\,\boldsymbol{\lambda}^\ast)}{\pi(\boldsymbol{\lambda}_n)\mathcal{L}(\boldsymbol{\lambda}_n)\mathcal{Q}(\boldsymbol{\lambda}_n;\,\boldsymbol{\lambda}^\ast)}, 1\right\},
\end{equation}
setting $\boldsymbol{\lambda}_{n + 1} = \boldsymbol{\lambda}^\ast$, where $\mathcal{L}(\boldsymbol{\lambda})$ is the likelihood, in our case from \eqnref{likelihood}; $\pi(\boldsymbol{\lambda})$ is the prior, and $\mathcal{Q}$ is a proposal distribution. If the move is not accepted  $\boldsymbol{\lambda}_{n + 1} = \boldsymbol{\lambda}_n$. This is the Metropolis-Hastings algorithm~\citep{Metropolis1953,Hastings1970}.

Waiting long enough yields an exact posterior, but it is desirable for the MCMC to converge quickly. This requires a suitable choice for the proposal distribution, which can be difficult, since we do not yet know the shape of the target distribution.

One method to define the proposal is to use the previous results in the chain and refine $\mathcal{Q}$ by learning from these. Such approaches are known as adaptive methods. Updating using previous points means that the chain is no longer Markovian. Care must be taken to ensure that ergodicity is preserved and convergence obtained~\citep{Roberts2007,Andrieu2008}. To avoid this complication, we follow \citet{Haario1999}, and use the adapting method as a burn in phase. We have an initial phase where the proposal is updated based upon accepted points. After this we fix the proposal and proceed as for a standard MCMC. By only using samples from the second part, we guarantee that the chain is Markovian and ergodic, whilst still enjoying the benefits of a tailor-made proposal. After only a finite number of samples we cannot assess the optimality of this~\citep{Andrieu2008}, but the method is still effective.

To tune $\mathcal{Q}$, we use an approach based upon the adaptive Metropolis algorithm~\citep{Haario2001}. The proposal is taken to be a multivariate normal distribution centred upon the current point, the covariance of which is
\begin{equation}
\boldsymbol{C} = s \left(\boldsymbol{V}_n + \varepsilon\boldsymbol{C}_0\right),
\end{equation}
where $\boldsymbol{V}_n$ is the covariance of the accepted points $\{\boldsymbol{\lambda}_1,\ldots,\boldsymbol{\lambda}_n\}$, $s$ is a scaling factor that controls the step size, $\varepsilon$ is a small positive constant (typically $0.0025$), and $\boldsymbol{C}_0$ is a constant matrix included to ensure ergodicity.

Our adaptation is run in three phases. The initial phase is to get the chain moving. For this $\boldsymbol{C}_0\super{init}$ is a diagonal matrix with elements calibrated from initial one dimensional MCMCs. This finishes after $N\sub{init}$ accepted points.

For the second phase, we use the proposal covariance from the initial phase $\boldsymbol{C}\super{init}$ for $\boldsymbol{C}_0\super{main}$. We reset the covariance of the accepted points so that it only includes points from this phase. This is the main adaptation phase and lasts until $N\sub{main}$ points have been accepted.

In the final adaptation phase we restart the chain at the true parameter values. We no longer update the shape of the covariance ($\boldsymbol{V}_n$ remains fixed), but adjust the step size $s$ to tune the acceptance rate; it is then fixed, along with everything else, for the final MCMC.

Throughout the adaptation, we update the step size $s$ after every $100$ trial points (whether or not they are accepted). While updating, the covariance $\boldsymbol{V}_n$ changes after every $1000$ trial points. We set $N\sub{init} = 50000$ and $N\sub{main} = 450000$.

We initially aimed for an acceptance rate of $0.234$; this is optimal for a random walk Metropolis algorithm with some specific high-dimensional target distributions~\citep{Roberts1997,Roberts2001}. In many cases we found better convergence when aiming for a lower acceptance rate, say $0.1$. This is not unexpected: the optimal rate may be lower than $0.234$ when the parameters are not independent and identically distributed~\citep{Bedard2007, Bedard2008, Bedard2008a}. In practice, the final acceptance rate is (almost always) lower than the target rate as the use of a multivariate Gaussian for the proposal distribution is rarely a good fit at the edges of the posterior. Consequently, the precise choice for the target acceptance rate is unimportant as long as it is of the correct magnitude. Final rates are typically within a factor of $2$ of the target value. As an initial choice, we set $s = 2.38^2/d$, which is the optimal choice if $\boldsymbol{C}$ was the true target covariance for a high dimensional target of independent and identically distributed parameters~\citep{Gelman1996,Roberts1997,Roberts2001,Haario2001}.\footnote{Reasonably good results may be obtained by fixing $s$ at this value, and not adjusting to fine tune the acceptance rate.}

To assess the convergence of the MCMC we check the trace plot (the parameters' values throughout the run) for proper mixing, that the one and two dimensional posterior plots fill out to a smooth distribution, and that the distribution widths tend towards consistent values.

\section{Results}\label{sec:Results}

\subsection{Data set}

To investigate the information contained in EMRBs, we again considered a range of orbits. The MBH was assumed to have the standard mass and position. The CO was chosen to be $10 M_\odot$, as the most promising candidates for EMRBs would be BHs: they are massive and hence produce higher SNR bursts, they are more likely to be on close orbits as a consequence of mass segregation \citep{Bahcall1977, Alexander2009, Preto2010}, and they cannot be tidally disrupted.

Other potential source candidates would be NSs and low mass MS stars. Like BHs, NSs are not tidally disrupted in the GC, and benefit (albeit weakly) from mass segregation. MS stars have the great advantage of being more numerous than stellar remnants. Most MS stars would be tidally disrupted at periapses of interest, but the lowest mass stars could be viable sources \citep{Freitag2003}.\footnote{The optimal density to resist disruption occurs at $\mu \simeq 0.07 M_\odot$ \citep{Chabrier2000}. Bursts for these objects should be detectable if $r\sub{p} \lesssim 10 r\sub{g}$.} These could produce more intricate waveforms, since they are extended bodies, subject to tidal distortion, rather than point masses. We stick to BHs as the most credible and simplest source. Modelling extended bodies is left for future work.

Orbits were chosen with periapses uniformly distributed in logarithmic space between the the inner-most orbit and $16 r\sub{g}$. The other parameters were chosen randomly from appropriate uniform distributions. 

\begin{figure*}
\begin{center}
   \includegraphics[width=0.92\textwidth]{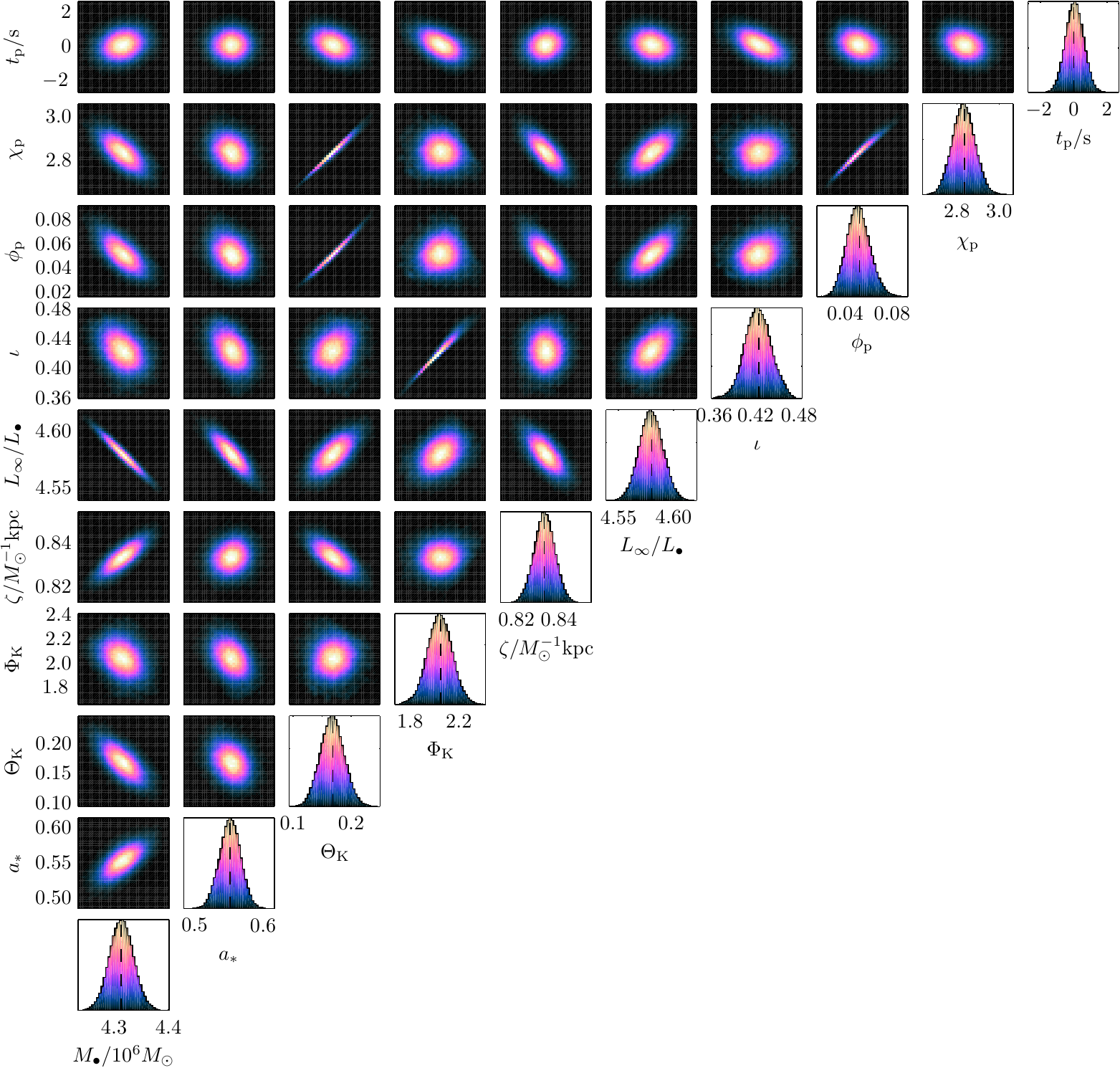}
\caption{Marginalised one and two dimensional posteriors. The scales are identical in both sets of plots. The dotted line indicates the true value. These distributions are fairly cromulent and well converged. Angular momentum is in units of $L_\bullet = GM_\bullet c^{-1}$. The input orbit has $r\sub{p} \simeq 8.54 r\sub{g}$ and $\rho \simeq 916$.\label{fig:MCMC-1}}
\end{center}
\end{figure*}

\begin{figure*}
\begin{center}
   \includegraphics[width=0.92\textwidth]{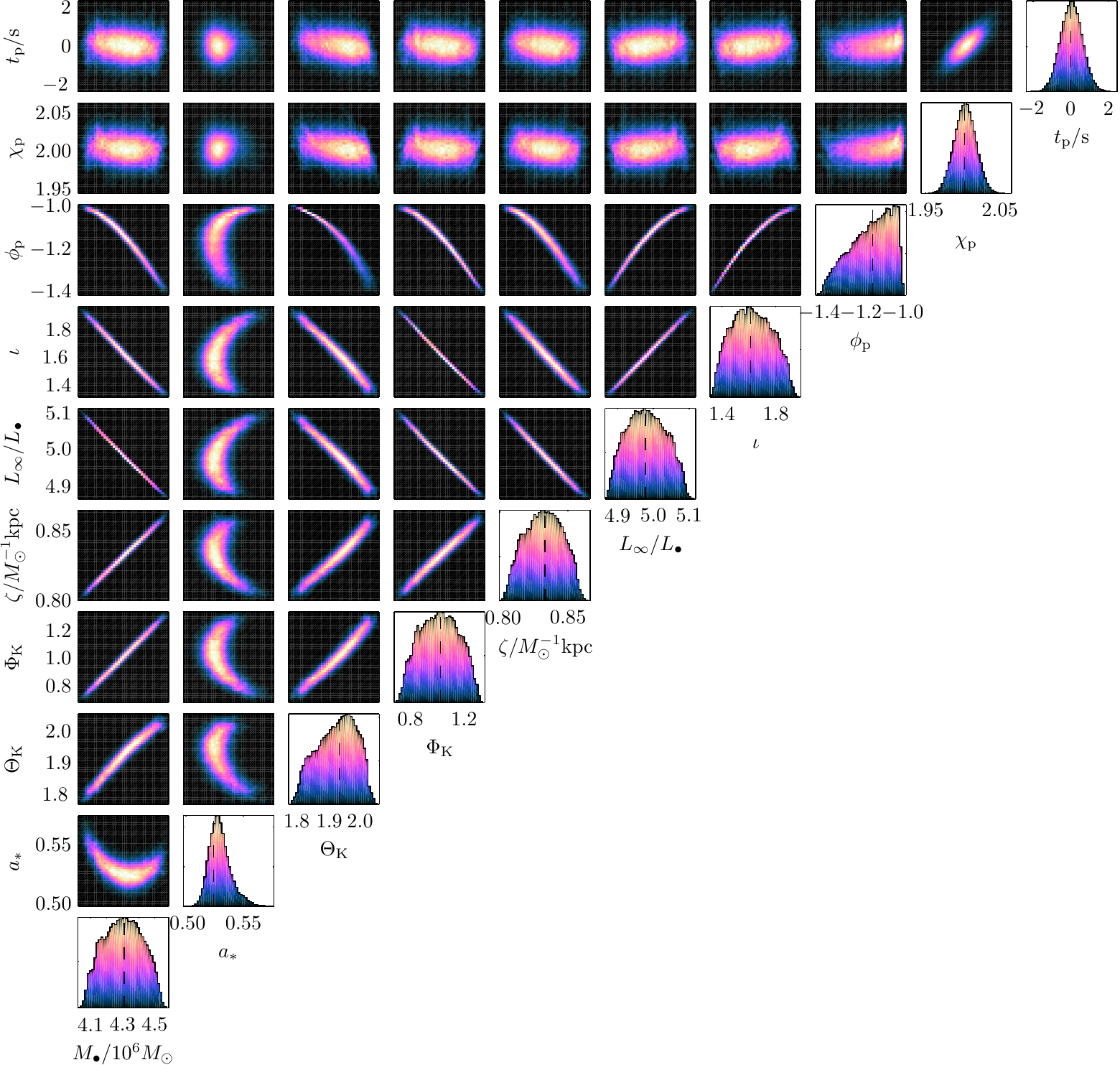}
\caption{Marginalised one and two dimensional posteriors. The scales are identical in both sets of plots. The dotted line indicates the true value. These distributions show definite non-gaussianity. The input orbit has $r\sub{p} \simeq 9.86 r\sub{g}$ and $\rho \simeq 1790$.\label{fig:MCMC-2}}
\end{center}
\end{figure*}

\begin{figure*}
\begin{center}
   \includegraphics[width=0.92\textwidth]{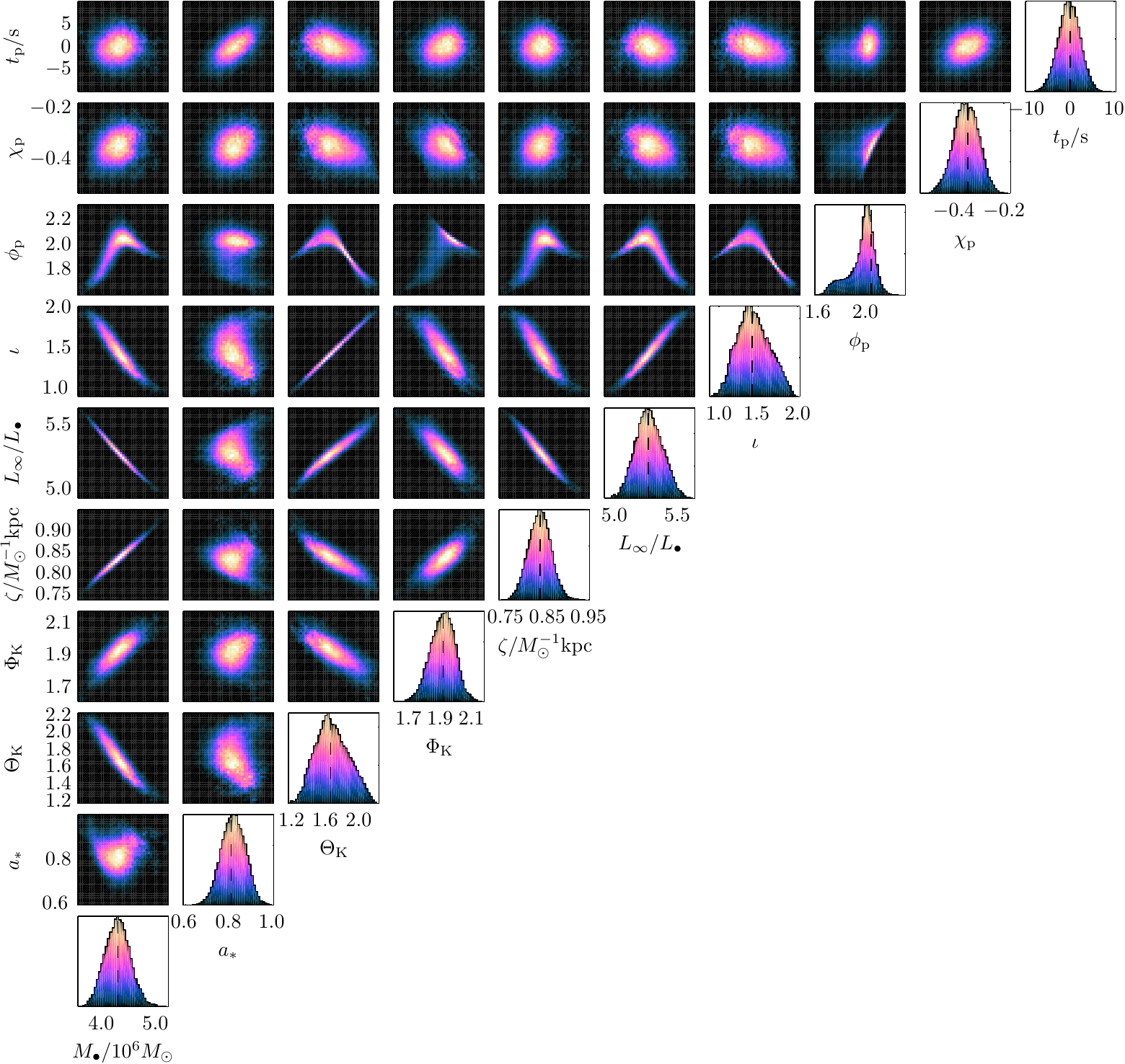}
\caption{Marginalised one and two dimensional posteriors. The scales are identical in both sets of plots. The dotted line indicates the true value. These distributions show complicated degeneracies. The input orbit has $r\sub{p} \simeq 11.60 r\sub{g}$ and $\rho \simeq 590$.}
\label{fig:MCMC-3}
\end{center}
\end{figure*}

The results of the MCMC runs illustrate why the FIM approach was insufficient. There are strong and complex parameter dependencies. For many sets of parameters the posteriors are far from Gaussian as assumed in the LSA. Some example results are shown in \figref{MCMC-1}, \ref{fig:MCMC-2} and \ref{fig:MCMC-3}.

The first is well-behaved. It is almost Gaussian, but we see some asymmetries and imperfections. There are also strong degeneracies, indicated by needle-like distributions. This is a fairly standard example: there are runs which are closer to being Gaussian (especially at higher SNR), and equally there are tighter degeneracies. The lenticular $M_\bullet$--$L_\infty$ degeneracy is common.

The second shows banana-like degeneracies. These are not uncommon; there are varying degrees of curvature. The more complicated shape makes it harder for the MCMC to converge, so the final distribution is not as smooth as for the first example. The curving degeneracies also bias the one dimensional marginalisations away from the true values.

The third shows more intricate behaviour. This is more rare, but indicates the variety of shapes that is obtainable. Again the convergence is more difficult, so the distributions are rougher around the edges; there is also some biasing due to the curving degeneracies.

These results do not incorporate any priors (save to keep them within realistic ranges); we have not folded in the existing information we have, for example, about the MBH's mass. Therefore, the resulting distributions characterise what we could learn from EMRBs alone. By the time a space-borne GW detector finally flies, we will have much better constraints on some parameters.

It is possible to place good constraints from the closest orbits. These can provide sufficient information to give beautifully behaved posteriors although significant correlation between parameters persists.

\subsection{Distribution widths}

Characteristic distribution widths are shown in \figref{sigmas}. Plotted are the standard deviation $\sigma\sub{SD}$; a scaled $50$-percentile range $\sigma_{50} = W_{50}/1.34898$, where $W_{50}$ is the range that contains the median $50\%$ of points, and a scaled $95$-percentile range $\sigma_{95} = W_{95}/3.919928$, where $W_{95}$ is the $95\%$ range. These widths are equal for a normal distribution. Filled circles are used for runs that appear to have converged. Open circles are for those yet to converge, but which appear to be approaching an equilibrium state; widths should be accurate to within a factor of a few.
\begin{figure*}
\begin{center}
\subfigure[MBH mass $M_\bullet$ versus periapsis.]{\includegraphics[width=0.42\textwidth]{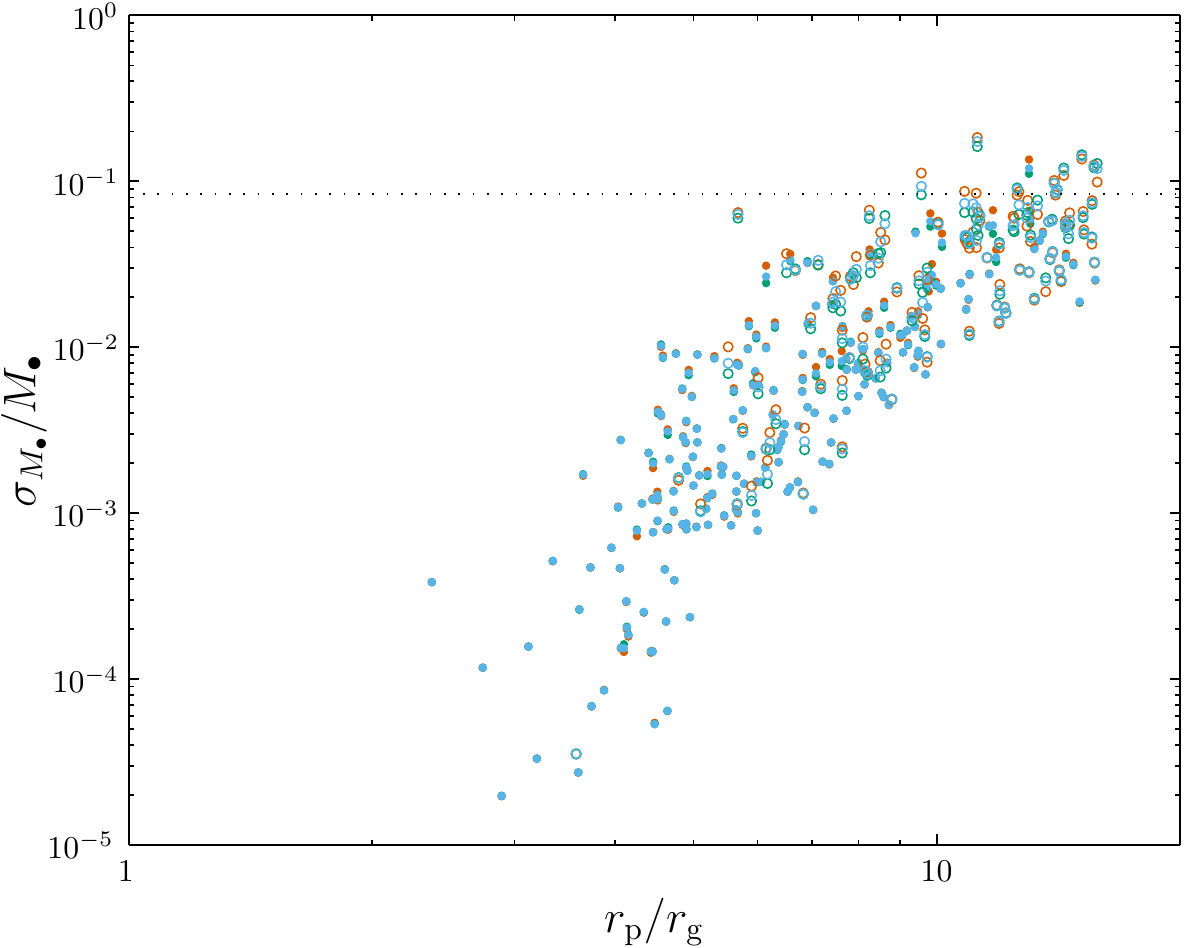}} \quad
\subfigure[MBH mass $M_\bullet$ versus SNR.]{\includegraphics[width=0.43\textwidth]{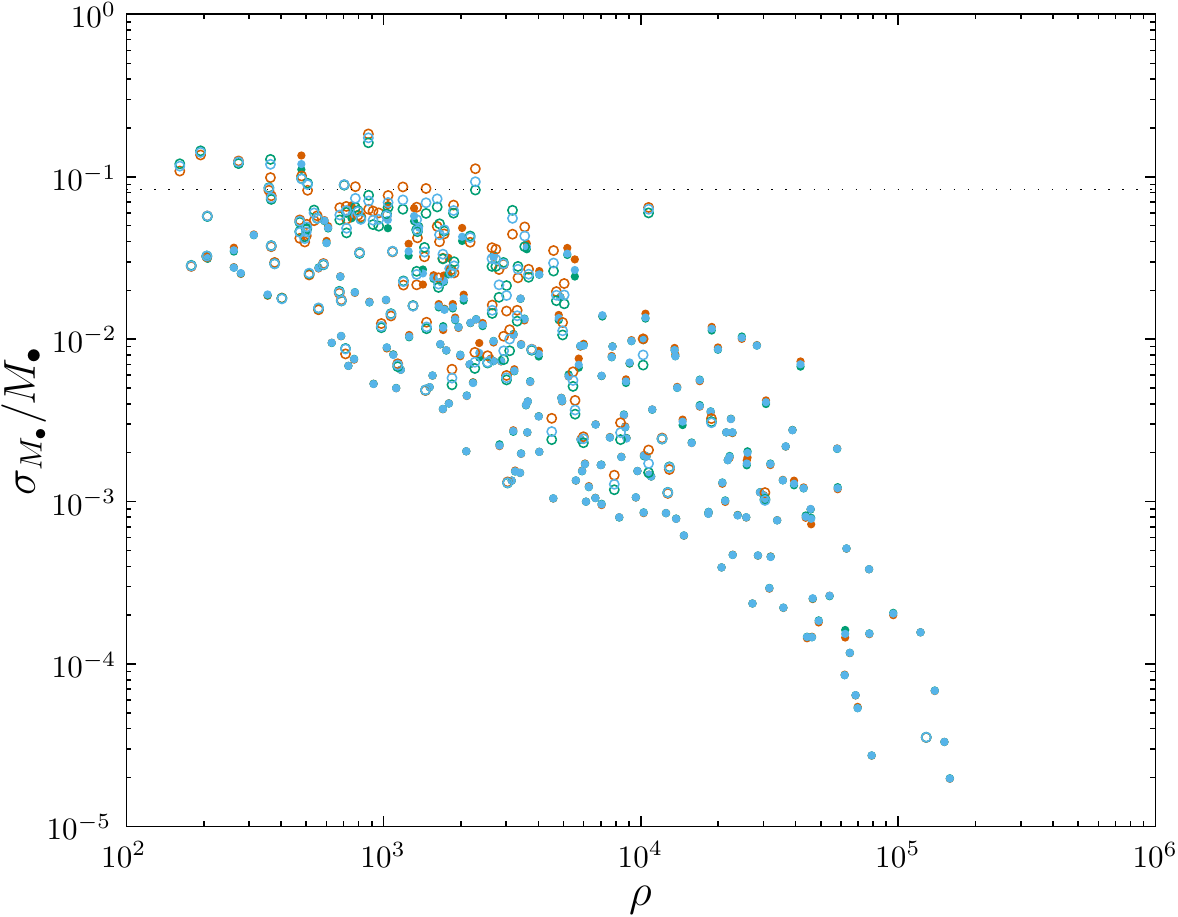}} \\
\subfigure[MBH spin $a_\ast$ versus periapsis.]{\includegraphics[width=0.42\textwidth]{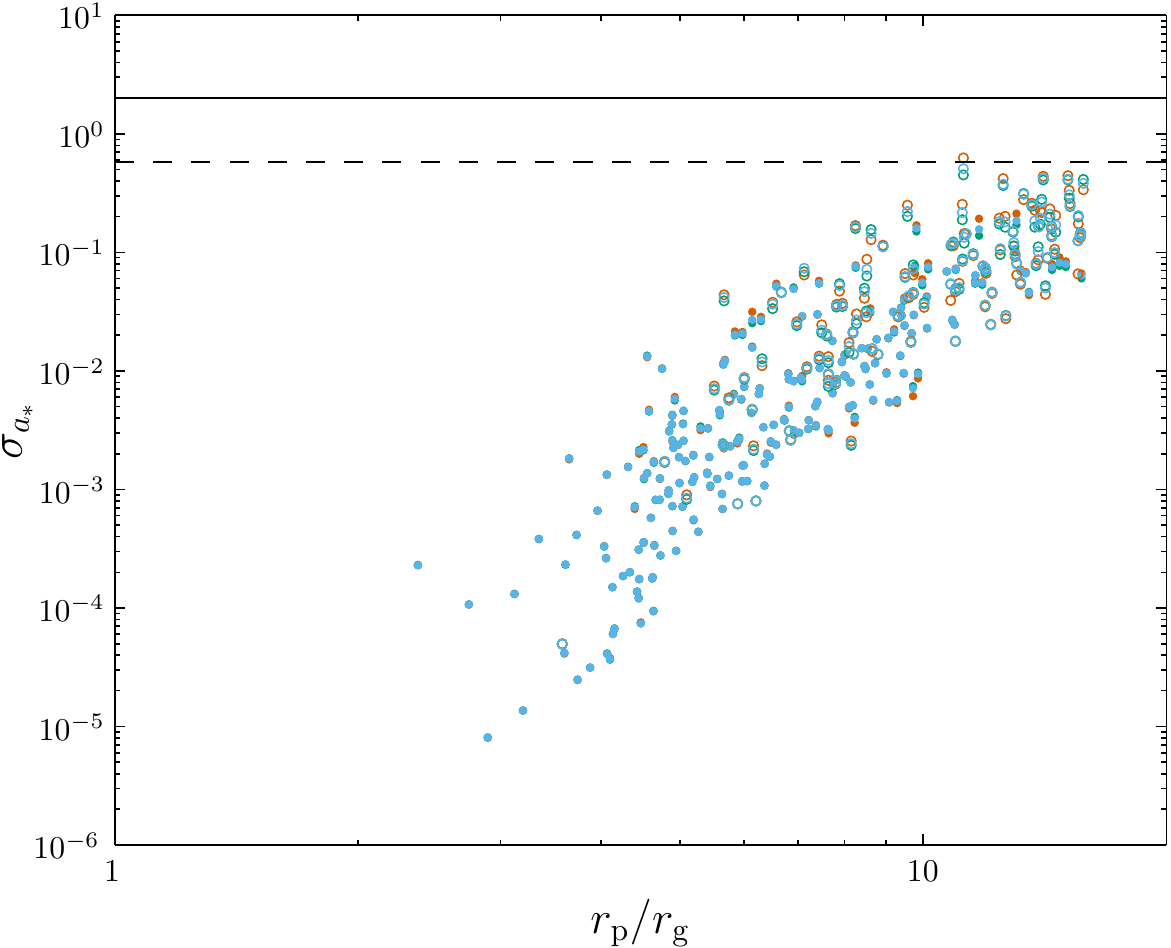}} \quad
\subfigure[MBH spin $a_\ast$ versus SNR.]{\includegraphics[width=0.43\textwidth]{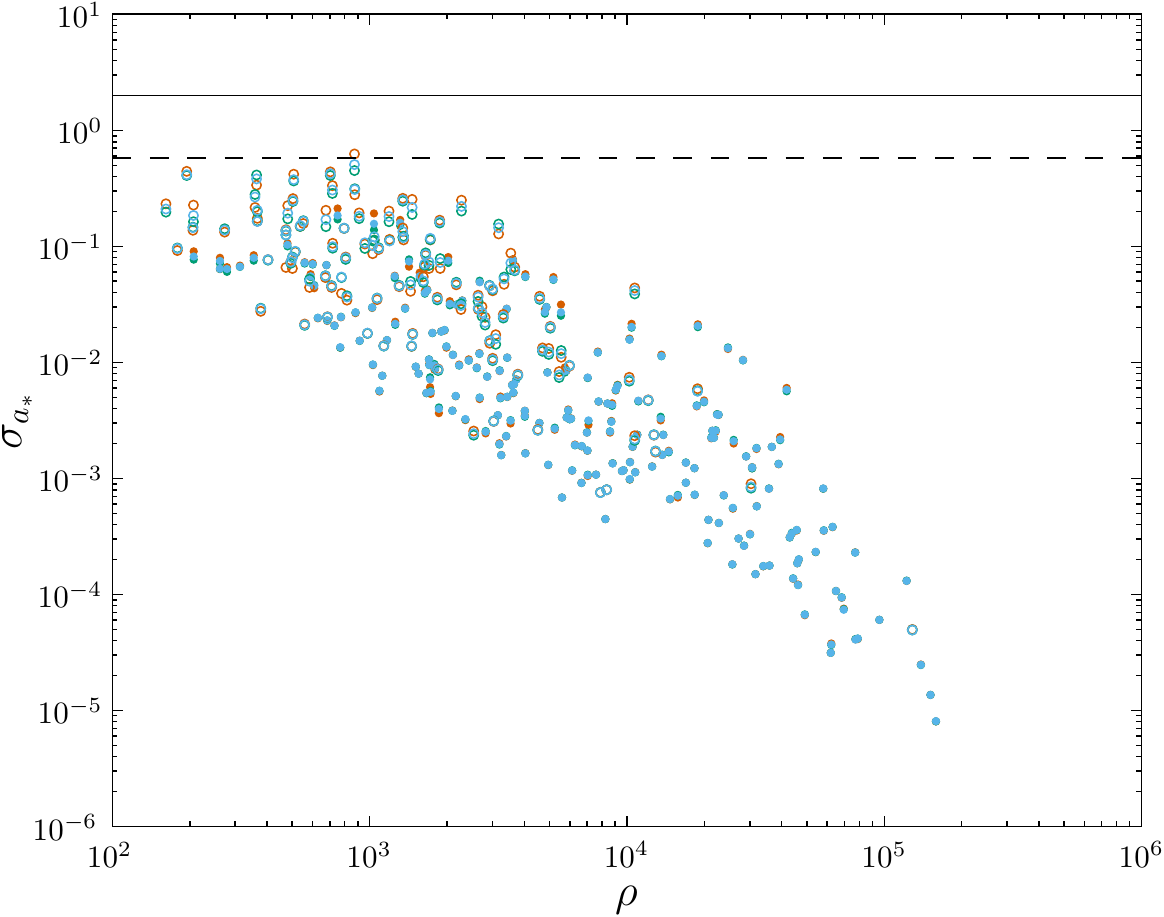}} \\
\subfigure[Orientation angle $\Theta\sub{K}$ versus periapsis.]{\includegraphics[width=0.42\textwidth]{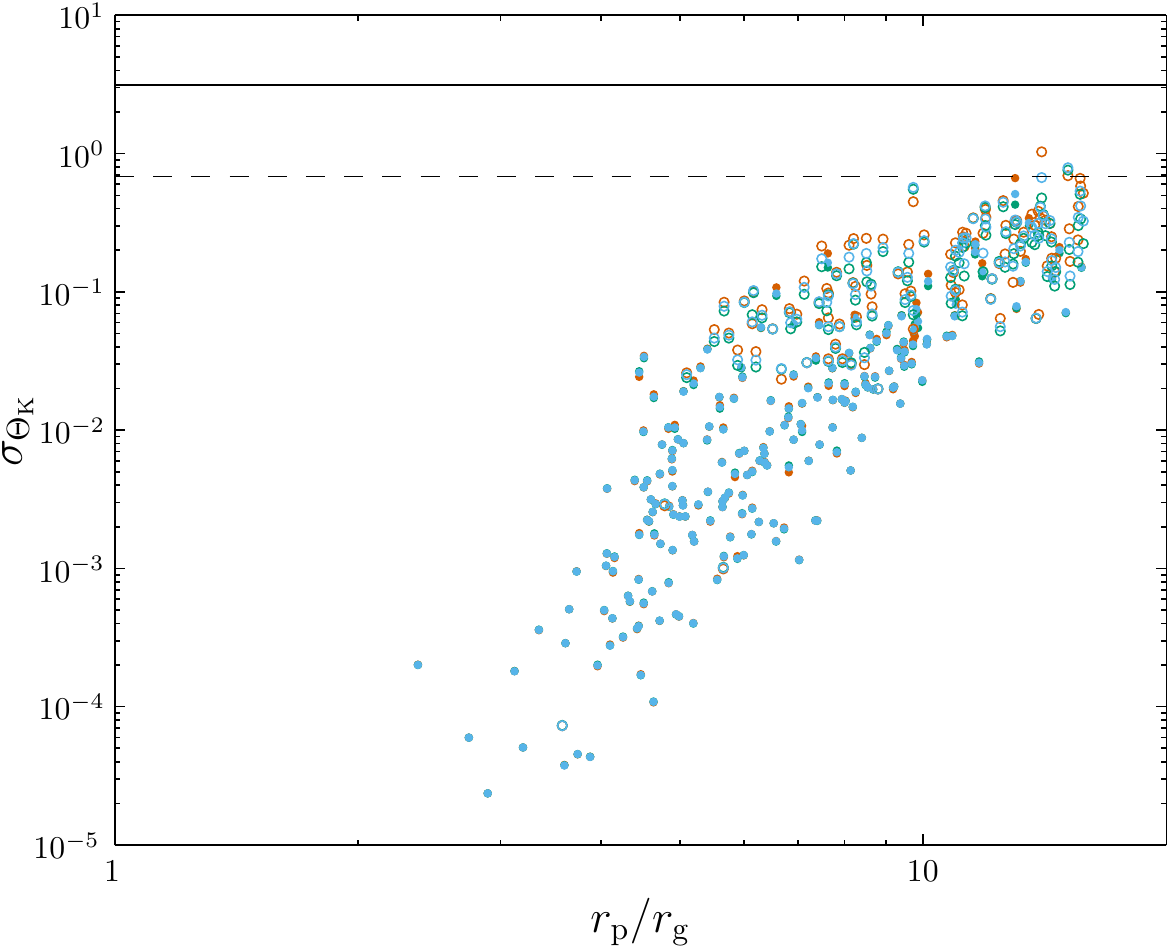}} \quad
\subfigure[Orientation angle $\Theta\sub{K}$ versus SNR.]{\includegraphics[width=0.43\textwidth]{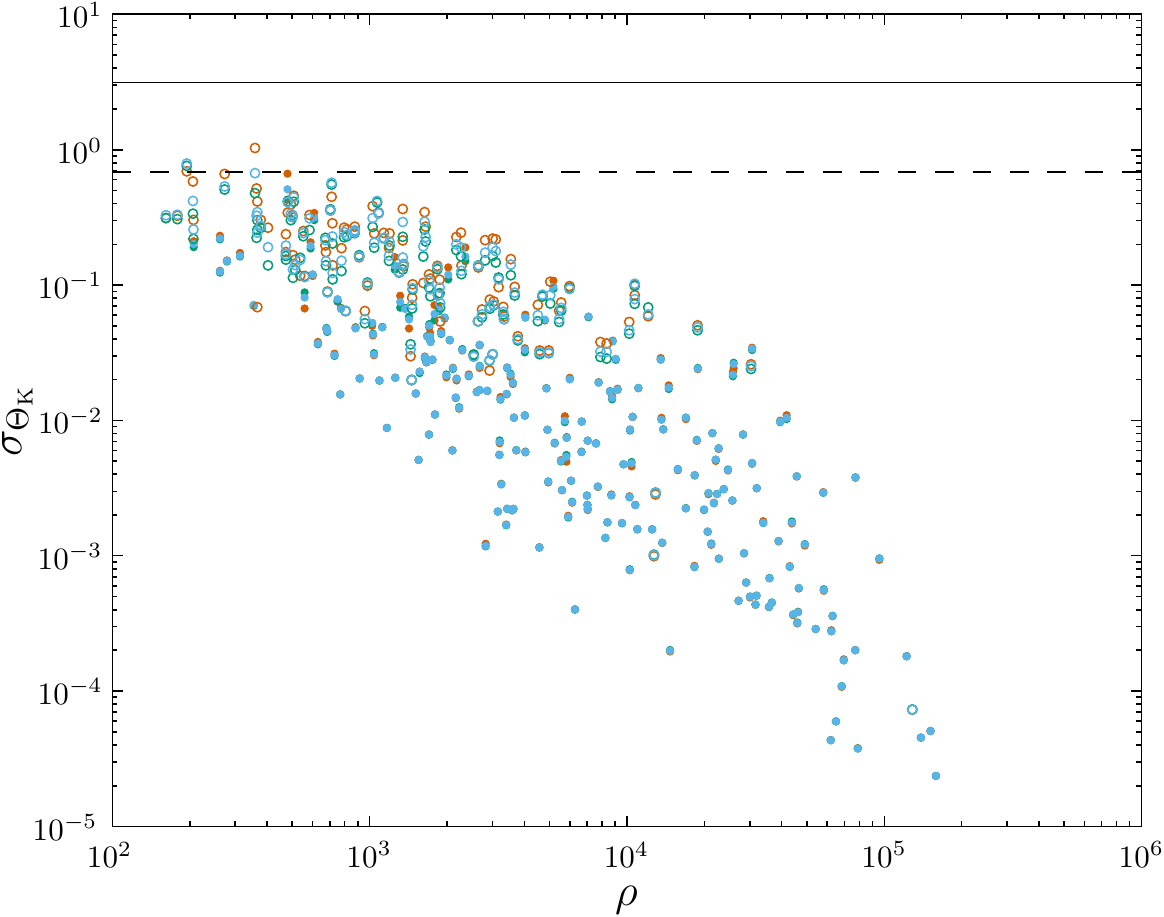}}
\caption{Distribution widths as functions of periapse $r\sub{p}$ and SNR $\rho$. Light blue is used for the standard deviation, red is the scaled $50$-percentile range and green is the scaled $95$-percentile range: all three coincide for a normal distribution. Filled circles are used for converged runs, open circles for those yet to converge. The dotted line indicates the current uncertainty for $M_\bullet$; the dashed lines the standard deviation for an uninformative prior, and the solid lines the total prior range.\label{fig:sigmas}}
\end{center}
\end{figure*}
\begin{figure*}
\setcounter{subfigure}{6}
\begin{center}
\subfigure[Orientation angle $\Phi\sub{K}$ versus periapsis.]{\includegraphics[width=0.42\textwidth]{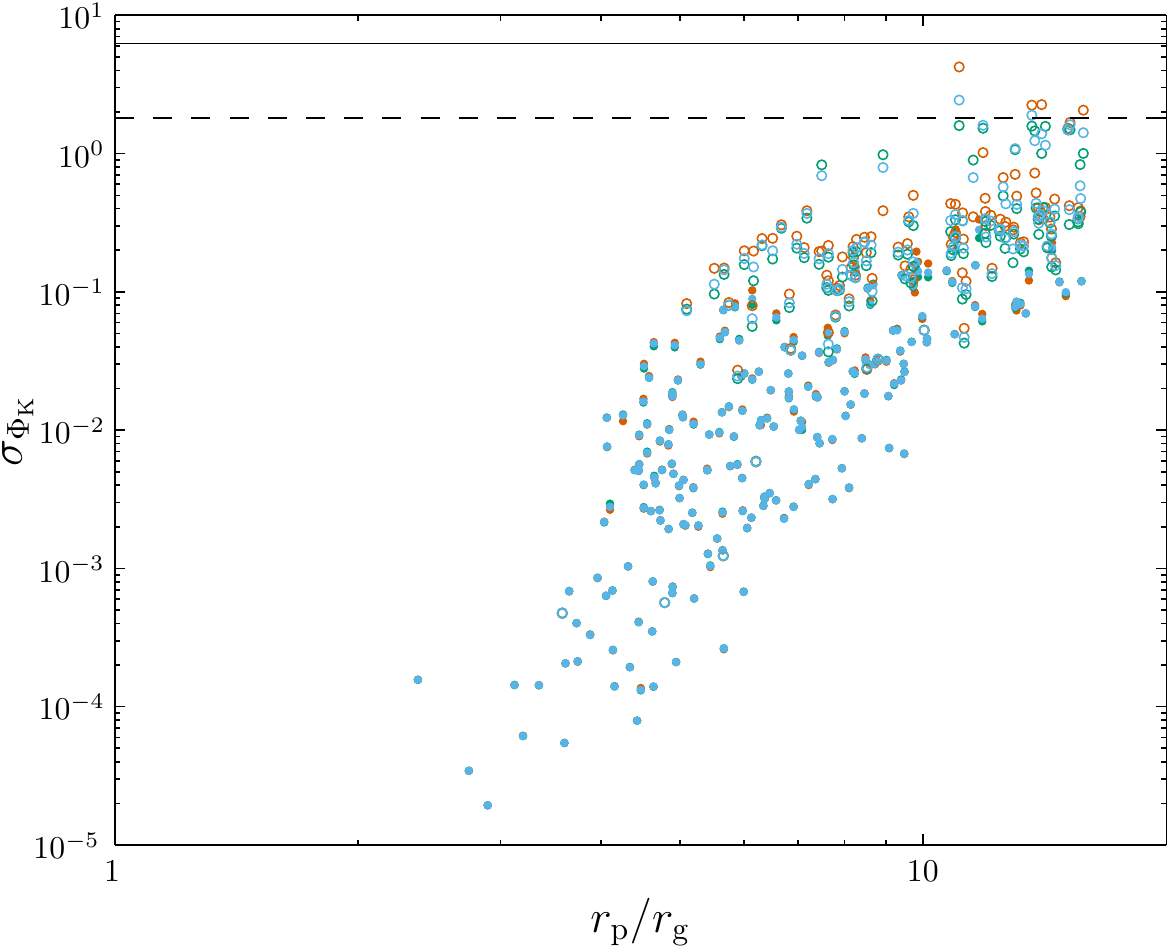}} \quad
\subfigure[Orientation angle $\Phi\sub{K}$ versus SNR.]{\includegraphics[width=0.43\textwidth]{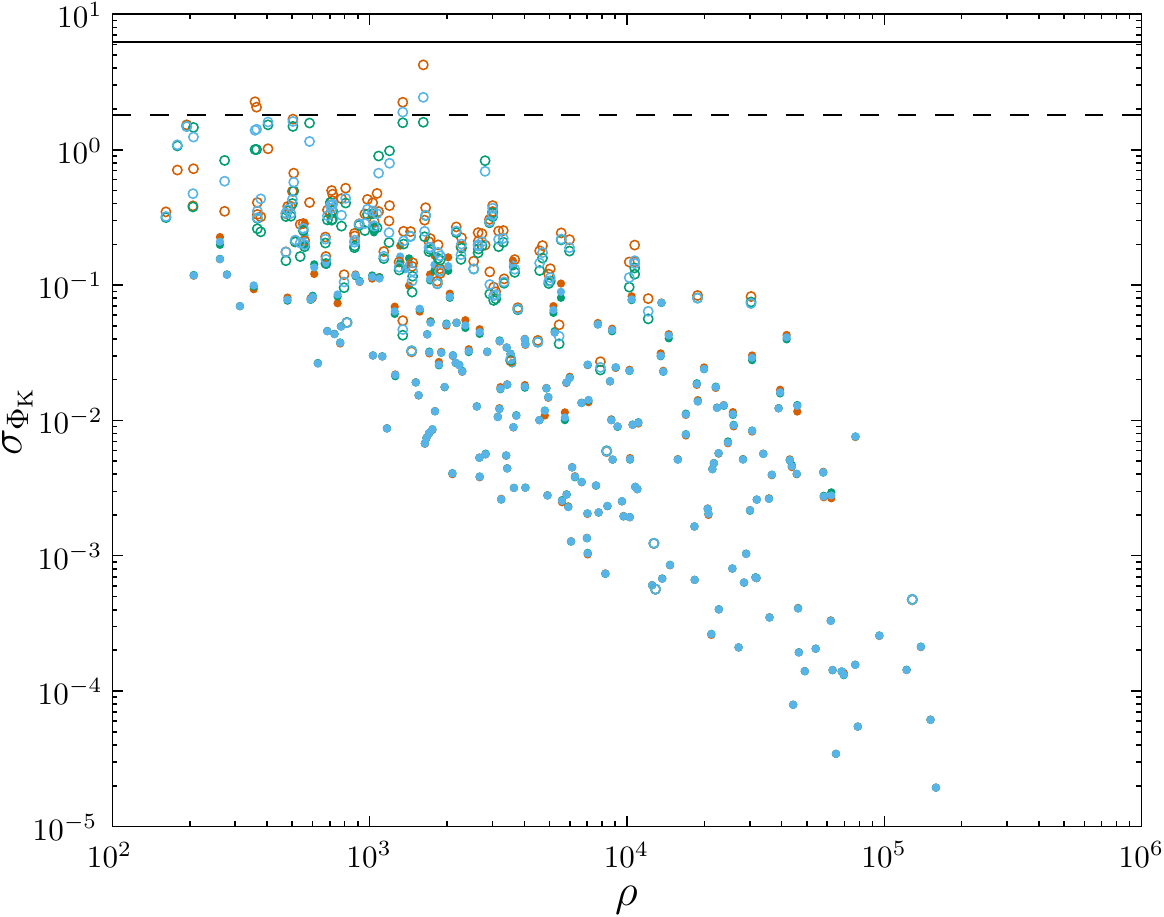}} \\
\subfigure[Scaled distance $\zeta$ versus periapsis.]{\includegraphics[width=0.42\textwidth]{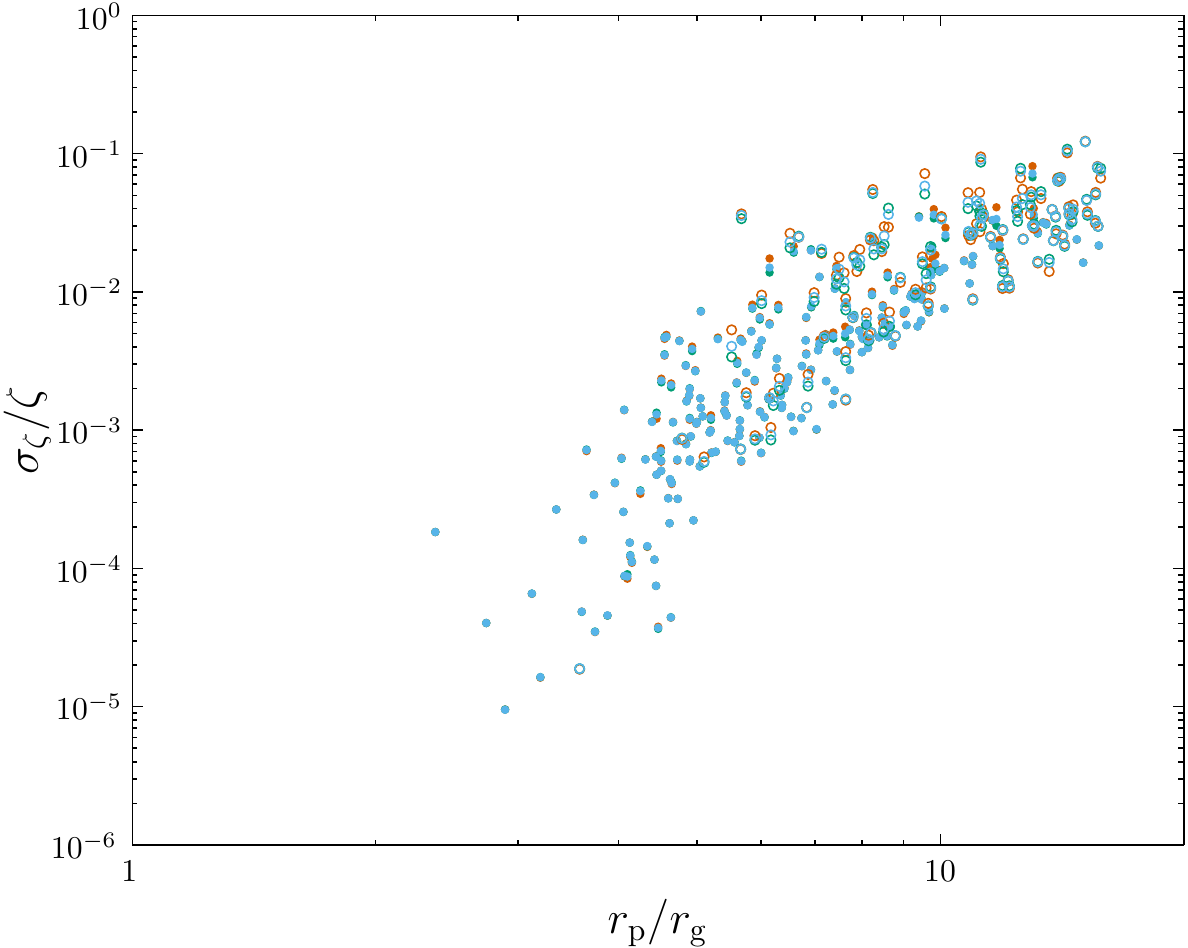}} \quad
\subfigure[Scaled distance $\zeta$ versus SNR.]{\includegraphics[width=0.43\textwidth]{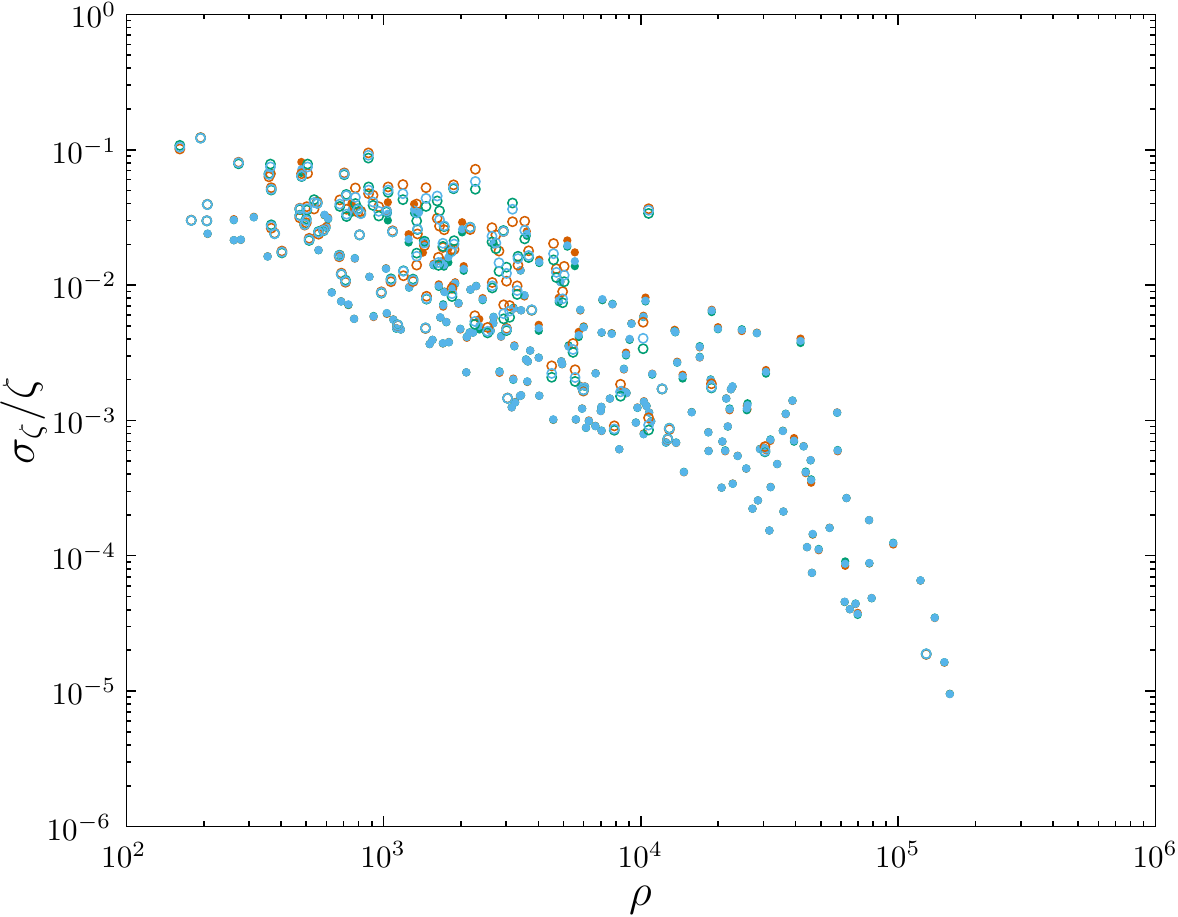}} \\
\subfigure[Angular momentum $L_\infty$ versus periapsis.]{\includegraphics[width=0.42\textwidth]{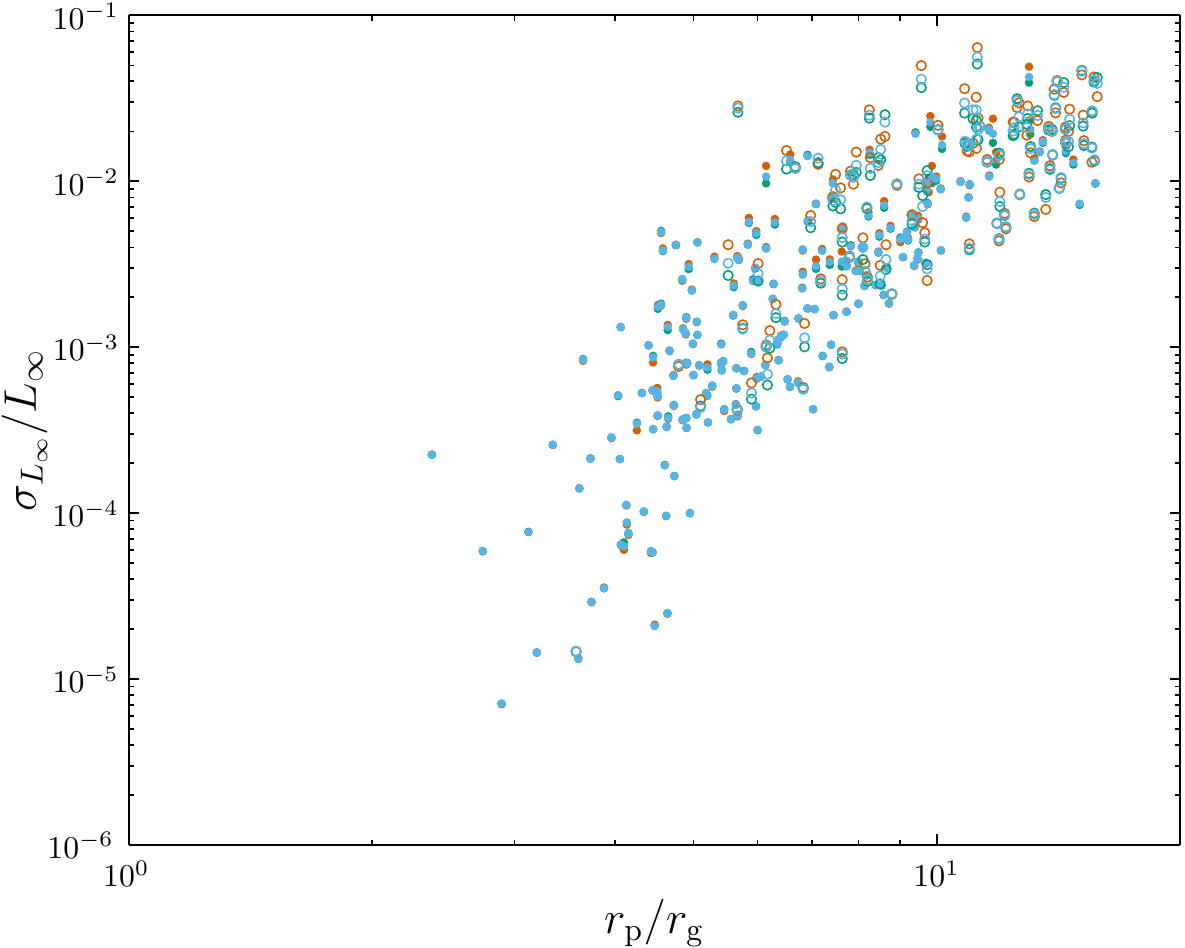}} \quad
\subfigure[Angular momentum $L_\infty$ versus SNR.]{\includegraphics[width=0.43\textwidth]{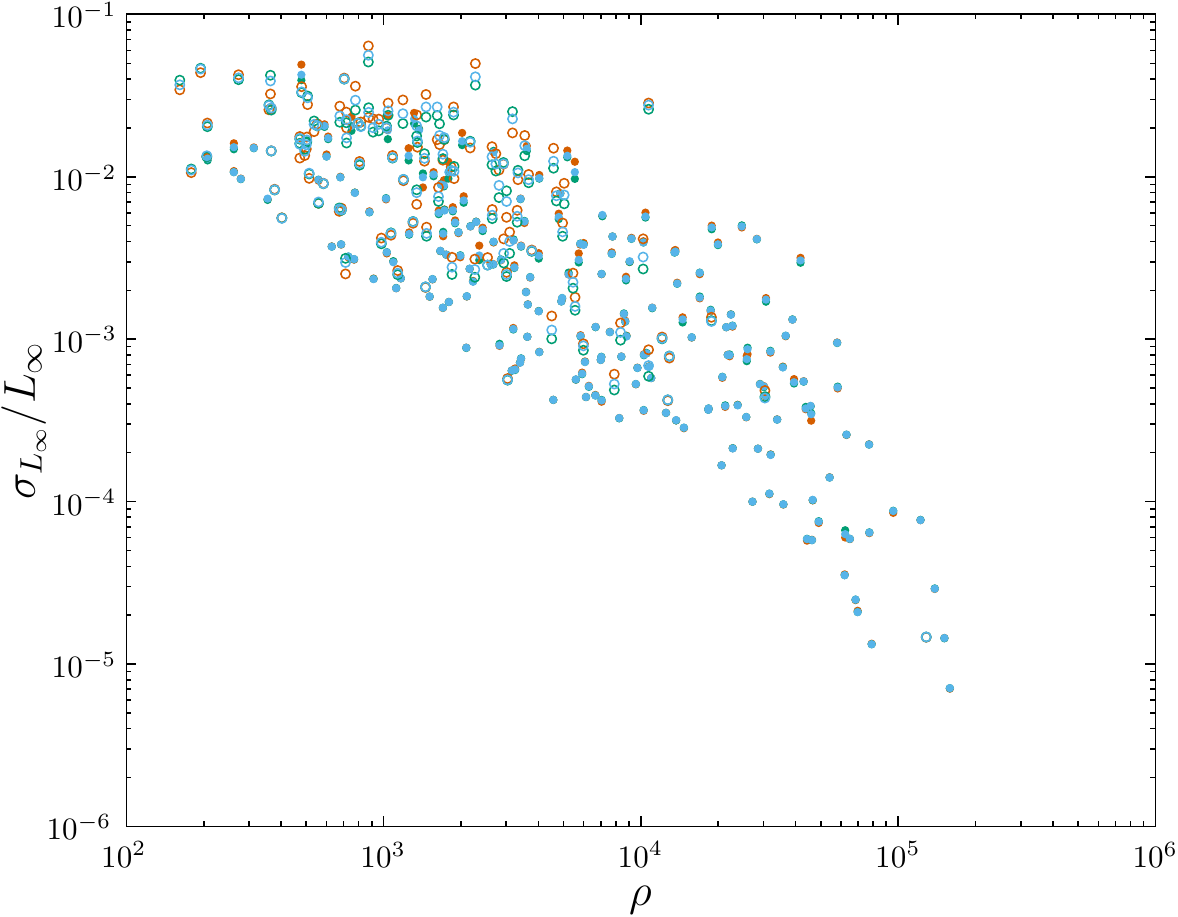}} \\
\contcaption{Distribution widths as functions of periapse $r\sub{p}$ and SNR $\rho$. Light blue is used for the standard deviation, red is the scaled $50$-percentile range and green is the scaled $95$-percentile range: all three coincide for a normal distribution. Filled circles are used for converged runs, open circles for those yet to converge. The dotted line indicates the current uncertainty for $M_\bullet$; the dashed lines the standard deviation for an uninformative prior, and the solid lines the total prior range.}
\end{center}
\end{figure*}
\begin{figure*}
\setcounter{subfigure}{12}
\begin{center}
\subfigure[Orbital inclination $\iota$ versus periapsis.]{\includegraphics[width=0.42\textwidth]{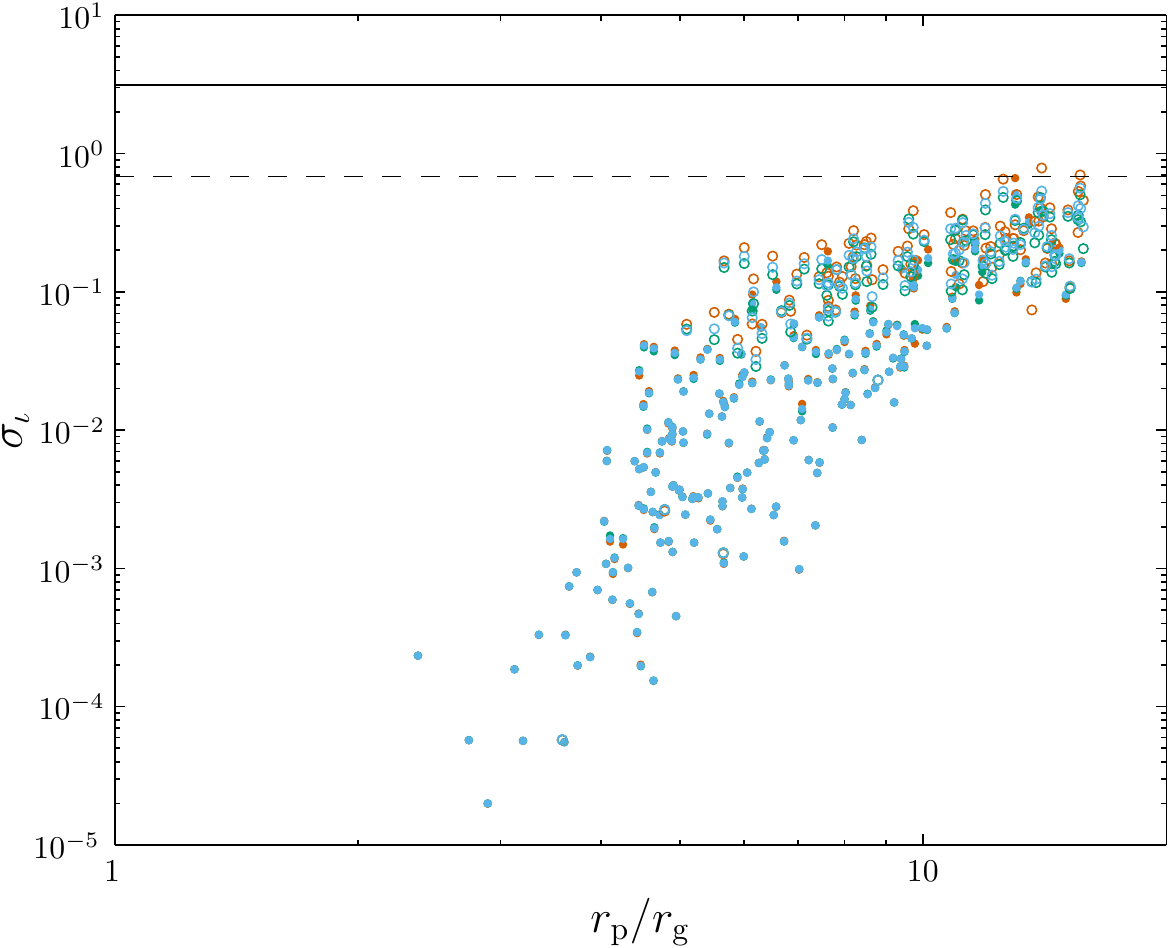}} \quad
\subfigure[Orbital inclination $\iota$ versus SNR.]{\includegraphics[width=0.43\textwidth]{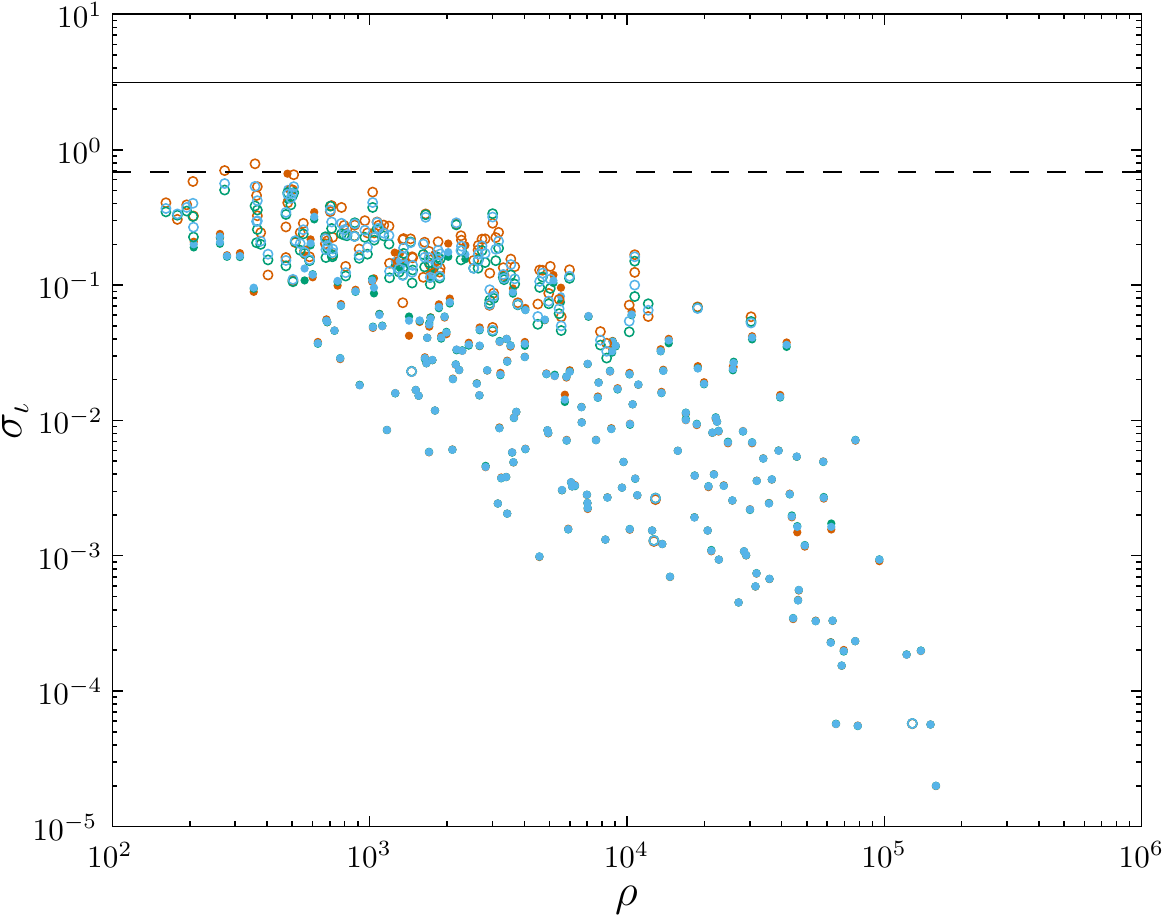}} \\
\subfigure[Periapse azimuthal phase $\phi\sub{p}$ versus periapsis.]{\includegraphics[width=0.42\textwidth]{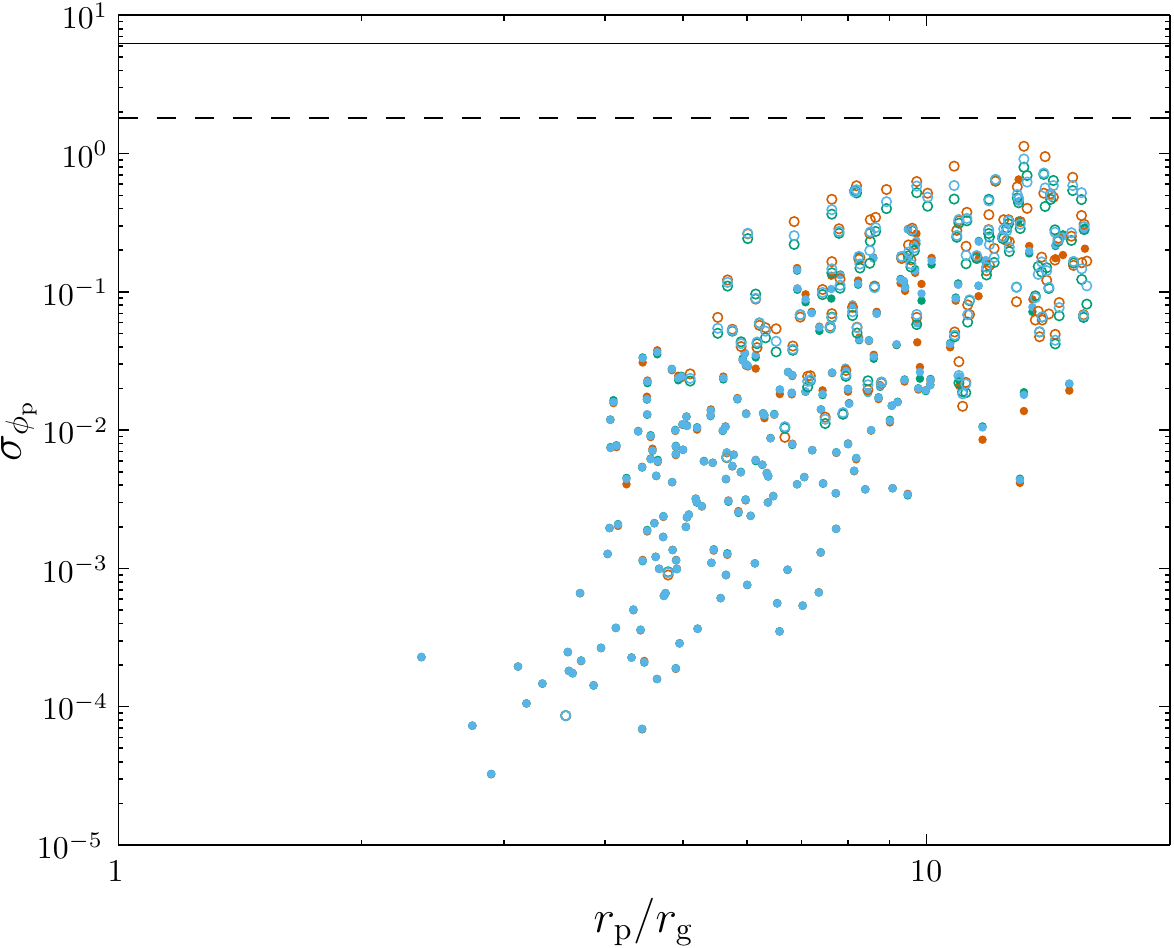}} \quad
\subfigure[Periapse azimuthal phase $\phi\sub{p}$ versus SNR.]{\includegraphics[width=0.43\textwidth]{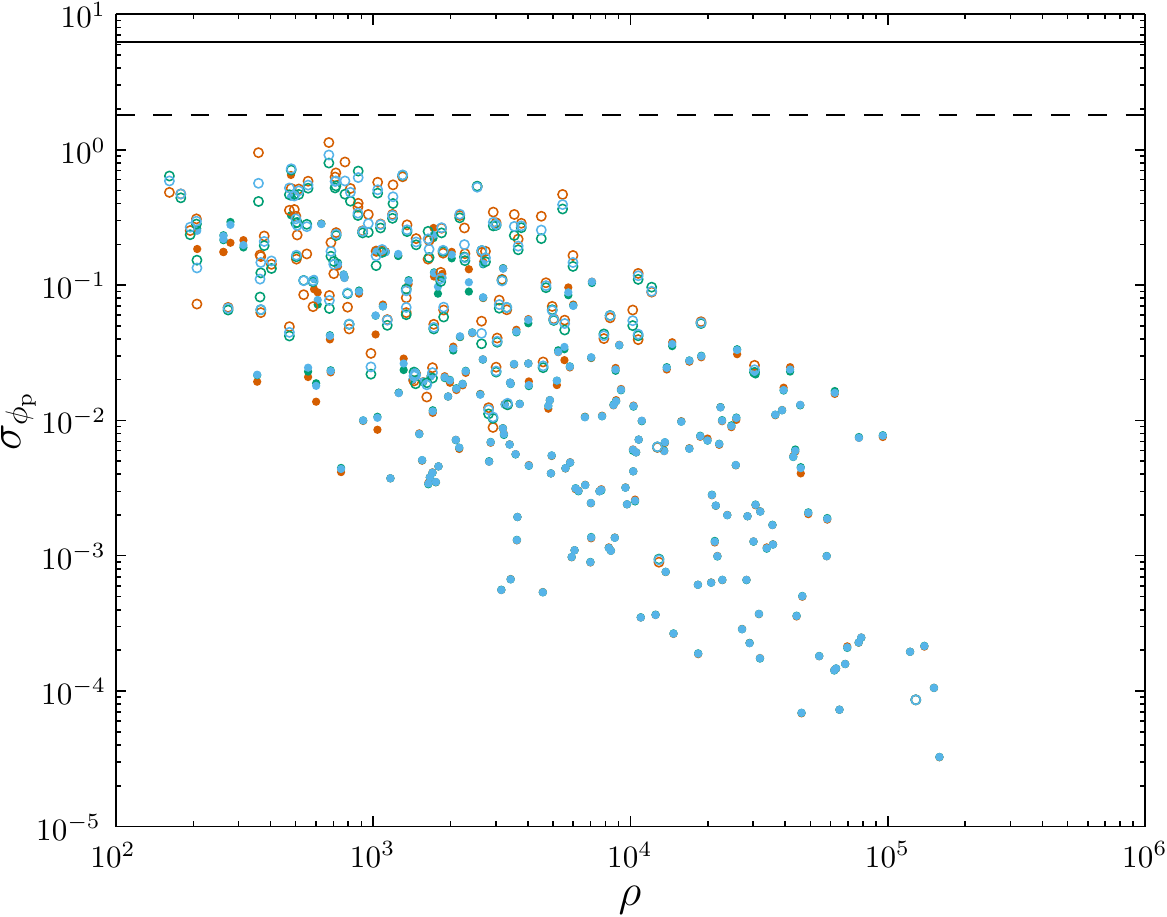}} \\
\subfigure[Periapse polar phase $\chi\sub{p}$ versus periapsis.]{\includegraphics[width=0.42\textwidth]{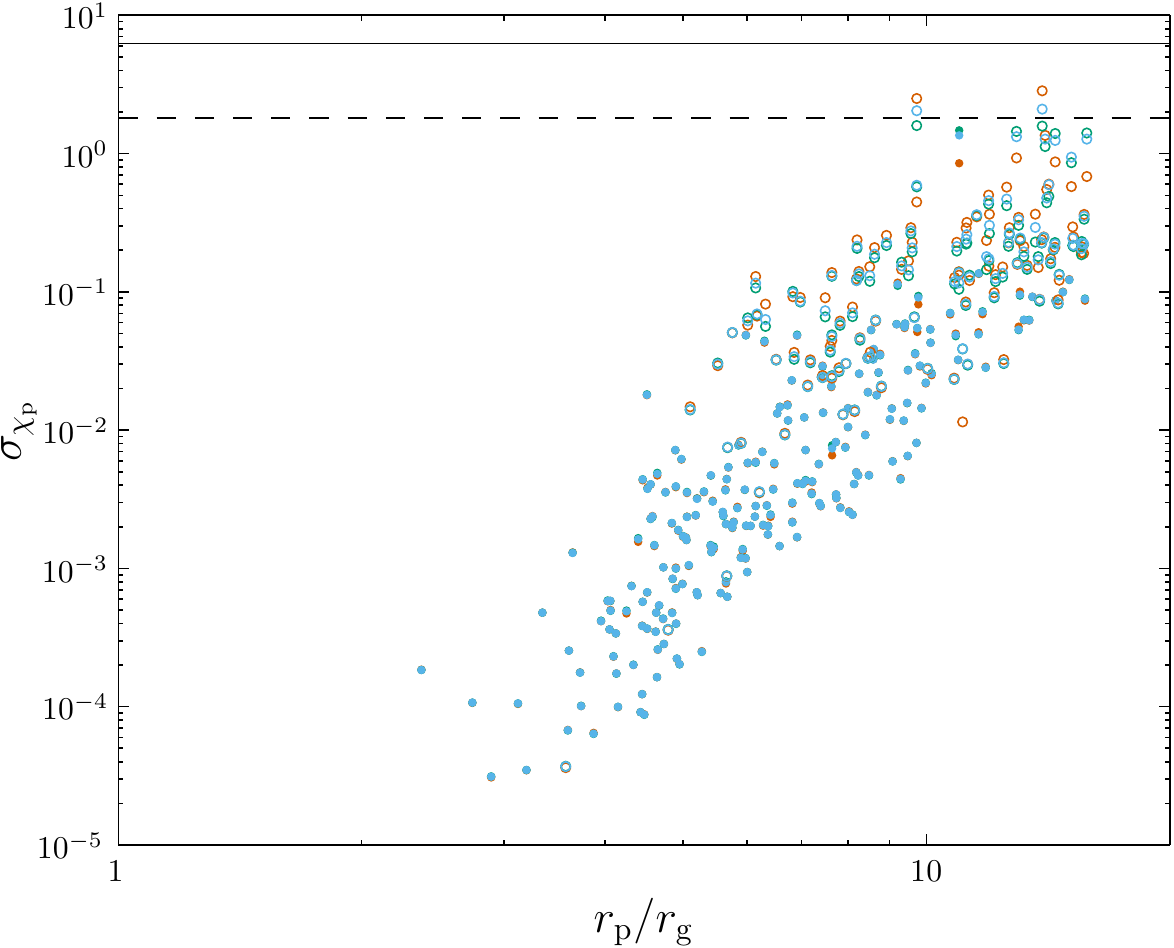}} \quad
\subfigure[Periapse polar phase $\chi\sub{p}$ versus SNR.]{\includegraphics[width=0.43\textwidth]{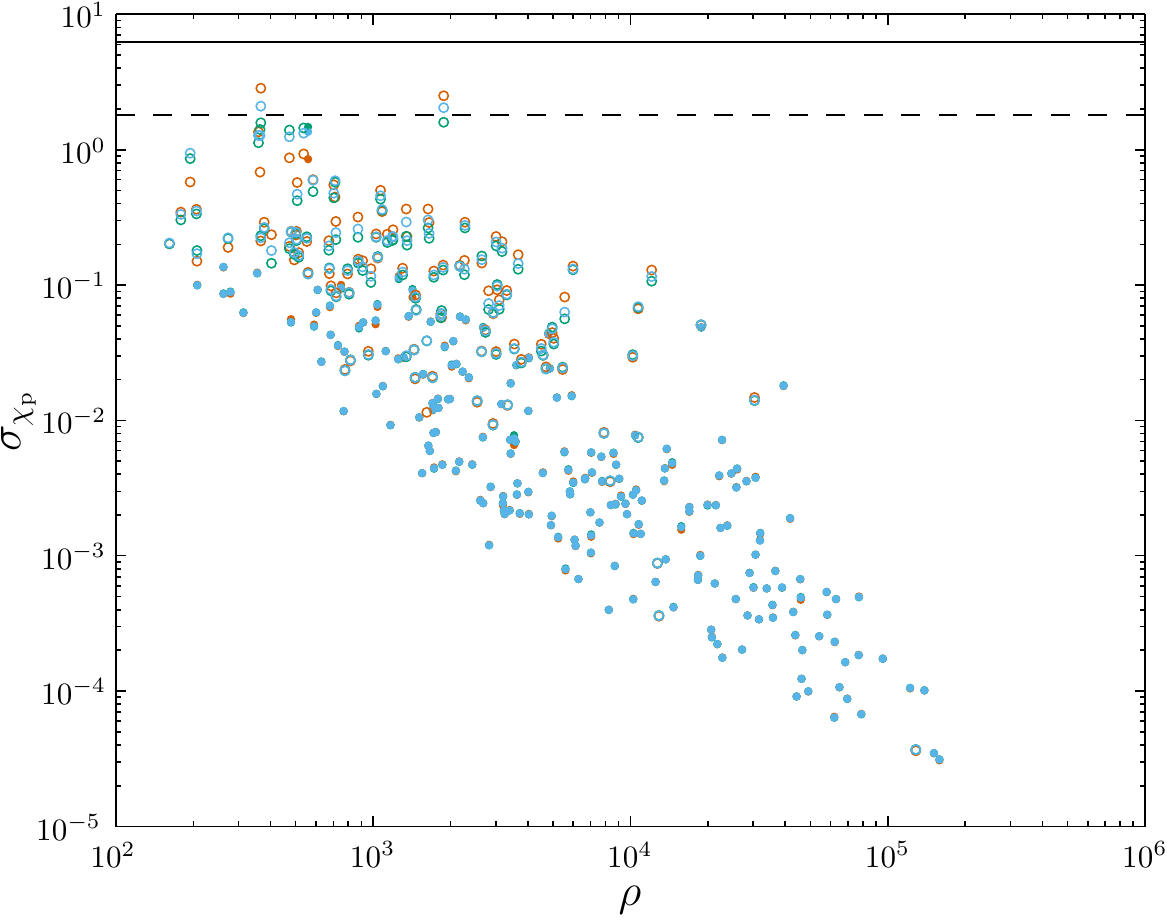}} \\
\contcaption{Distribution widths as functions of periapse $r\sub{p}$ and SNR $\rho$. Light blue is used for the standard deviation, red is the scaled $50$-percentile range and green is the scaled $95$-percentile range: all three coincide for a normal distribution. Filled circles are used for converged runs, open circles for those yet to converge. The dotted line indicates the current uncertainty for $M_\bullet$; the dashed lines the standard deviation for an uninformative prior, and the solid lines the total prior range.}
\end{center}
\end{figure*}
\begin{figure*}
\setcounter{subfigure}{18}
\begin{center}
\subfigure[Periapse time $t\sub{p}$ versus periapsis.]{\includegraphics[width=0.42\textwidth]{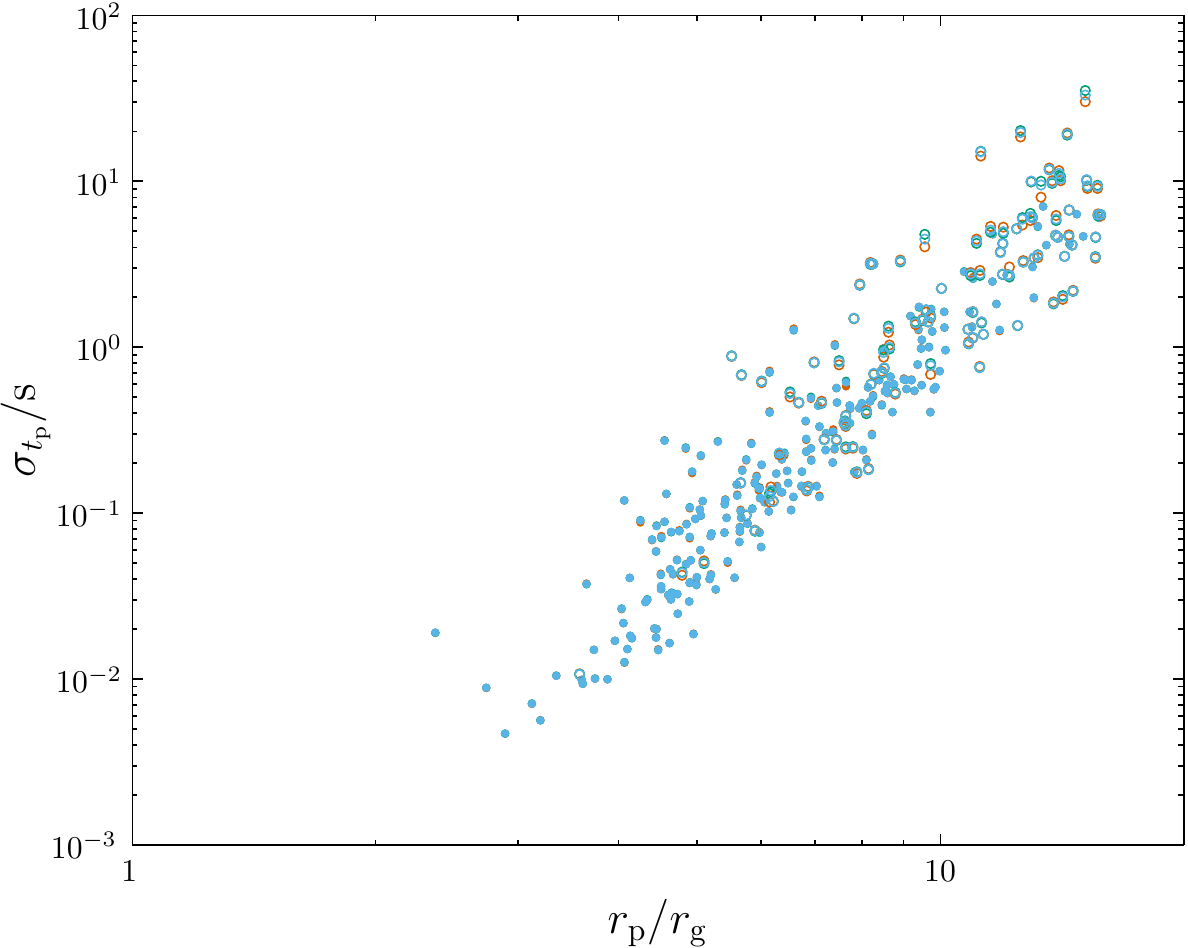}} \quad
\subfigure[Periapse time $t\sub{p}$ versus SNR.]{\includegraphics[width=0.43\textwidth]{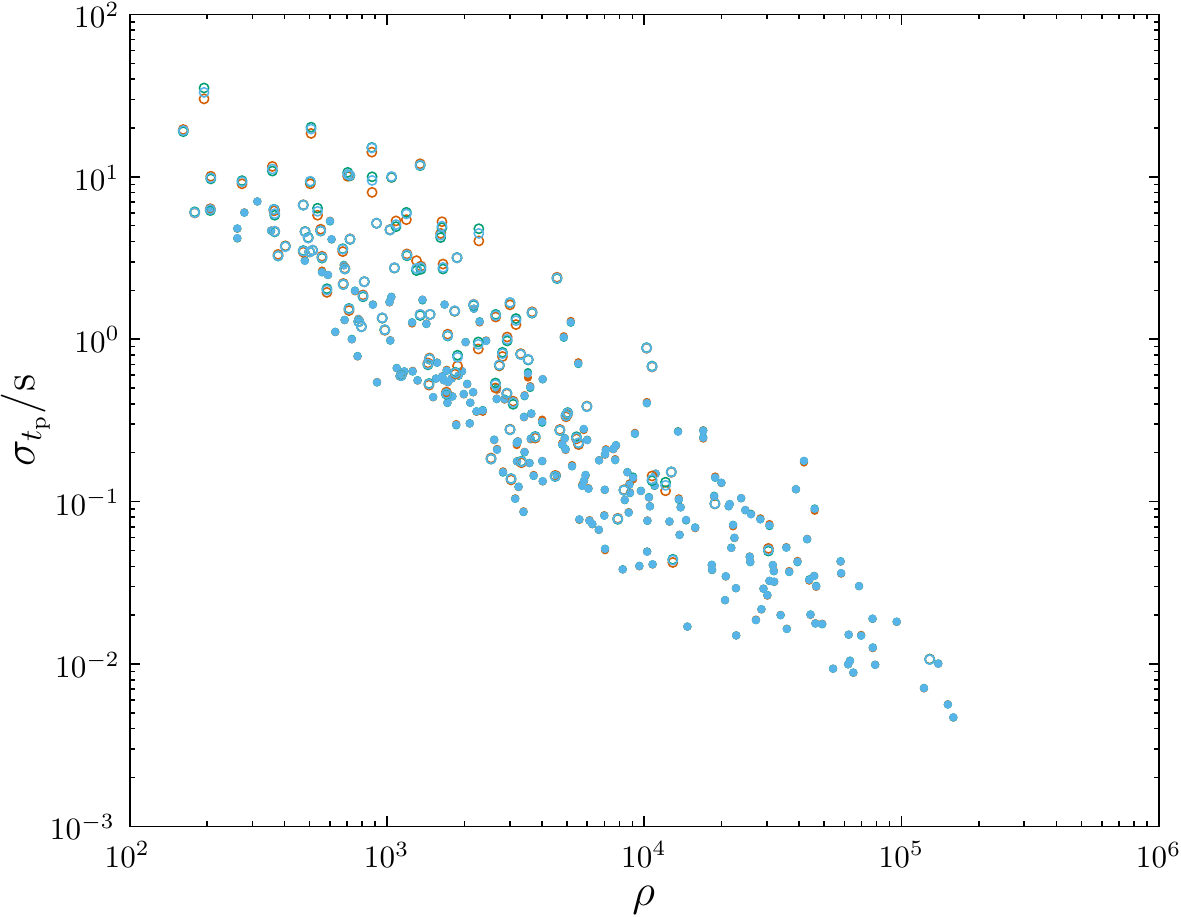}} \\
\contcaption{Distribution widths as functions of periapsis $r\sub{p}$ and SNR $\rho$. Light blue is used for the standard deviation, red is the scaled $50$-percentile range and green is the scaled $95$-percentile range: all three coincide for a normal distribution. Filled circles are used for converged runs, open circles for those yet to converge. The dotted line indicates the current uncertainty for $M_\bullet$; the dashed lines the standard deviation for an uninformative prior, and the solid line the total prior range.}
\end{center}
\setcounter{subfigure}{0}
\end{figure*}
For guidance, the dotted line corresponds to the current measurement uncertainty for $M_\bullet$; the dashed lines are from uniform priors for $a_\ast$, $\Phi\sub{K}$, $\phi\sub{p}$, $\chi\sub{p}$, $\cos\Theta\sub{K}$ and $\cos\iota$, and, for completeness, the solid line indicates the total prior range. We have no expectations for the width of the MBH mass distribution with respect to the current value; however, we would expect that the recovered distributions for the other parameters are narrower than for the case of complete ignorance. This may not be the case if the distribution is multimodal: in this event using the width is an inadequate description of the distribution. Only a few unconverged runs exceed these limits, and some appear to be multimodal.

\begin{figure*}
\begin{center}
\subfigure[MBH spin $a_\ast$.]{\includegraphics[width=0.45\textwidth]{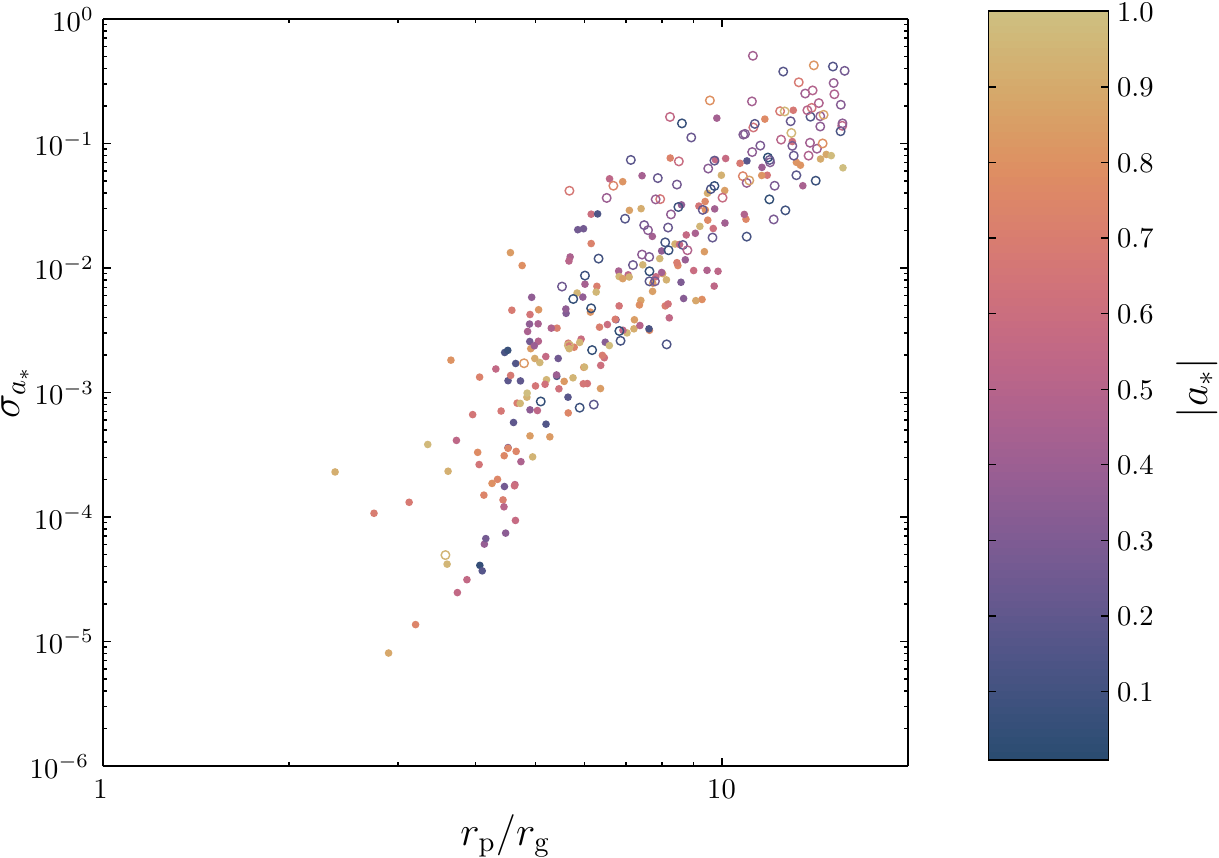}} \quad
\subfigure[Orientation angle $\Theta\sub{K}$.]{\includegraphics[width=0.45\textwidth]{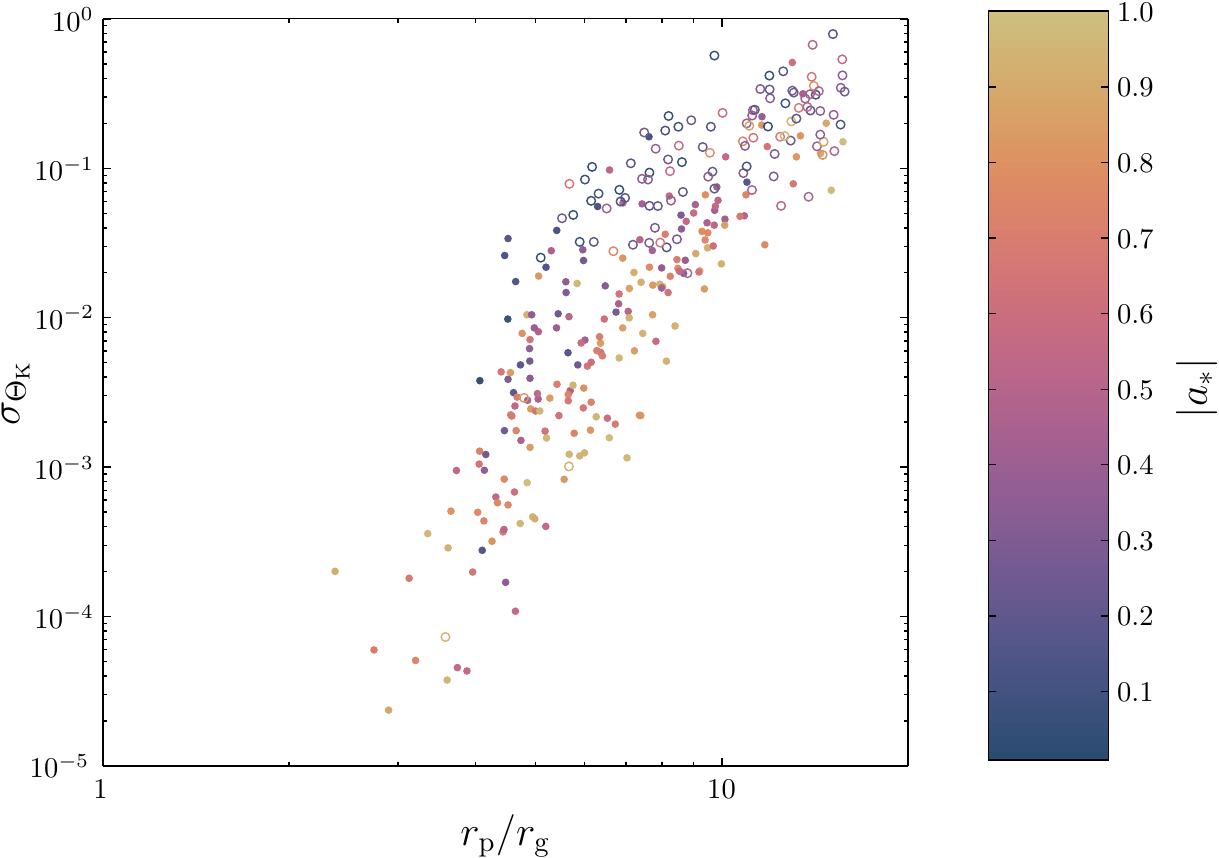}} \\
\subfigure[Orientation angle $\Phi\sub{K}$.]{\includegraphics[width=0.45\textwidth]{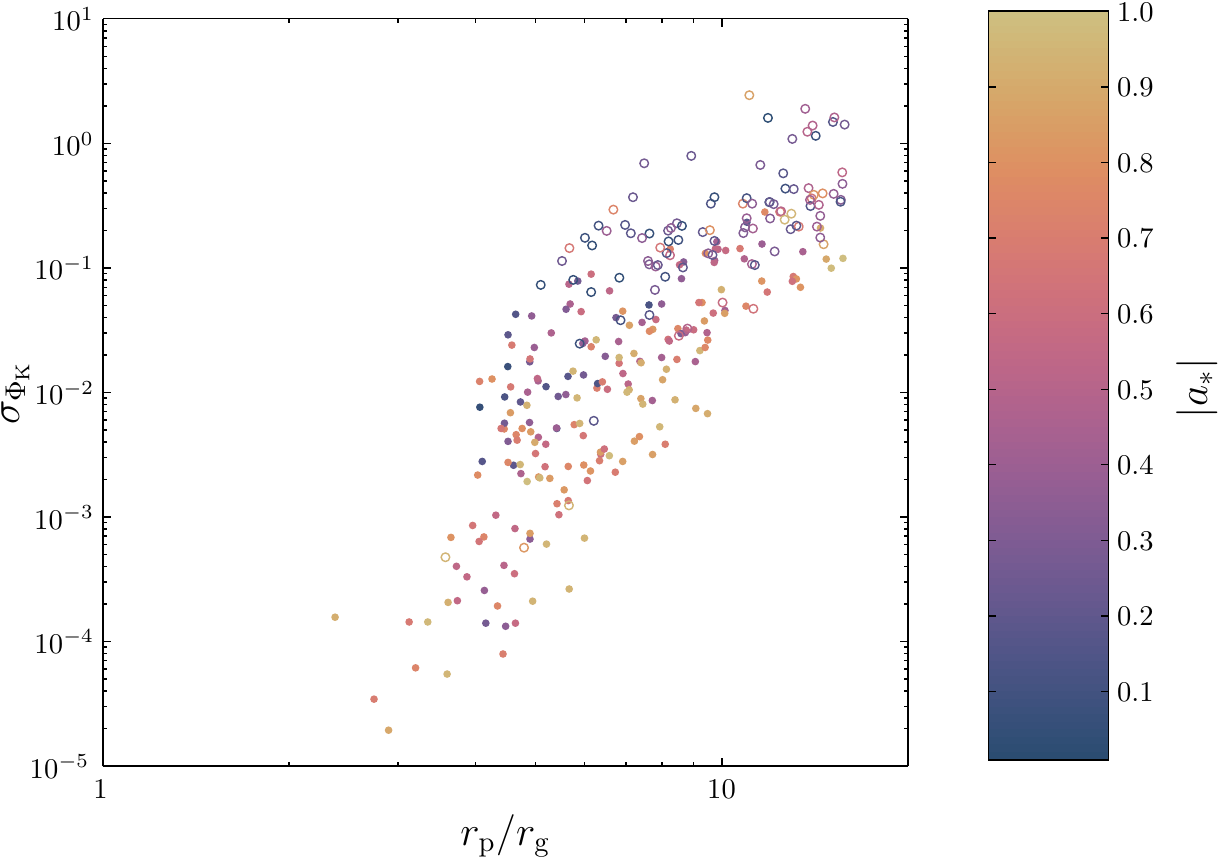}} \quad
\subfigure[Orbital inclination $\iota$.]{\includegraphics[width=0.45\textwidth]{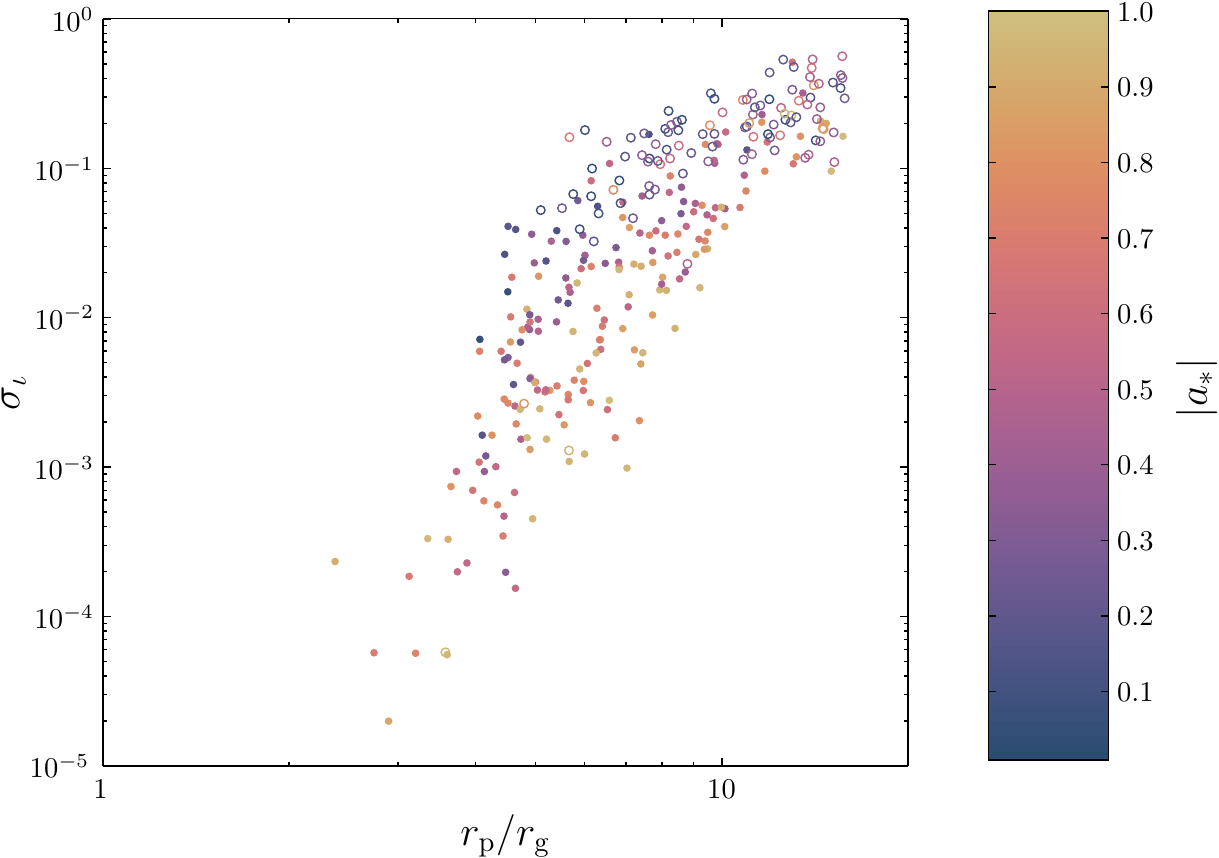}} \\
\subfigure[Periapse azimuthal phase $\phi\sub{p}$]{\includegraphics[width=0.45\textwidth]{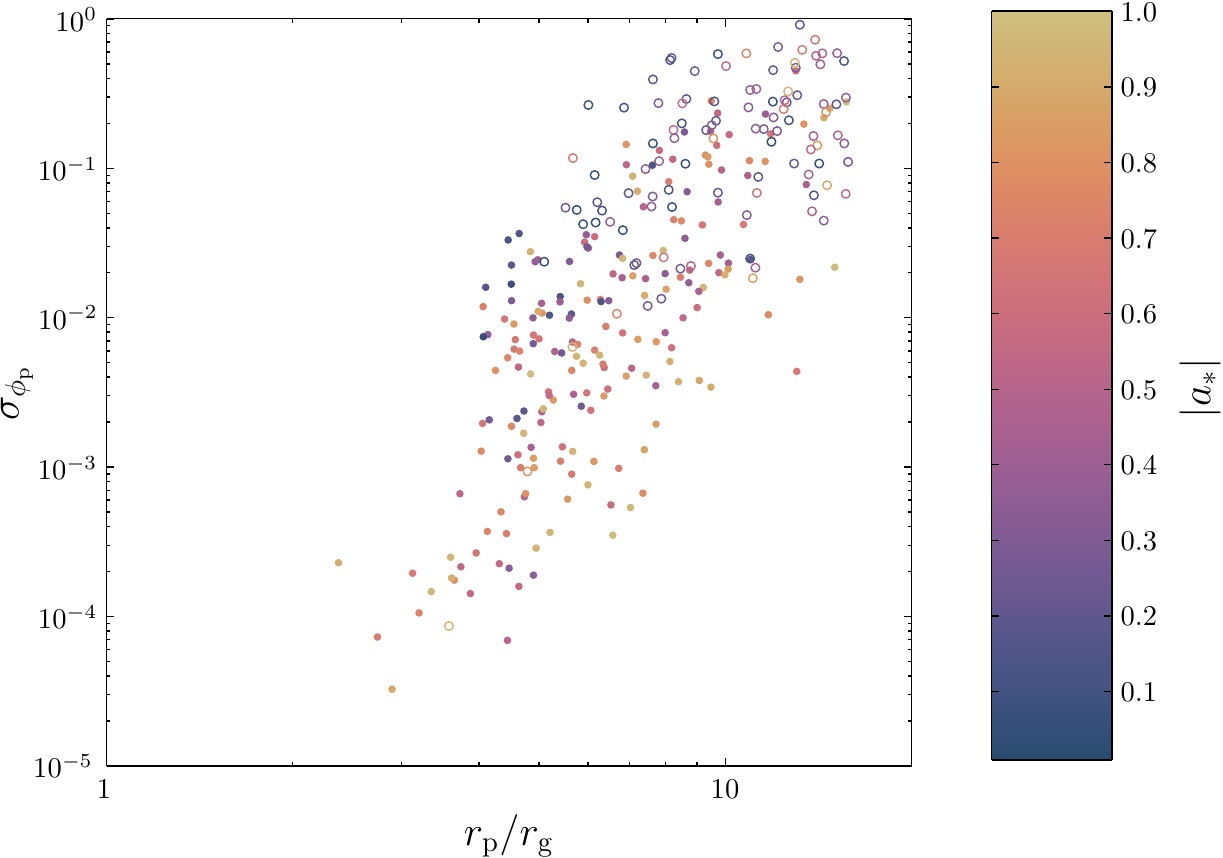}} \quad
\subfigure[Periapse polar phase $\chi\sub{p}$.]{\includegraphics[width=0.45\textwidth]{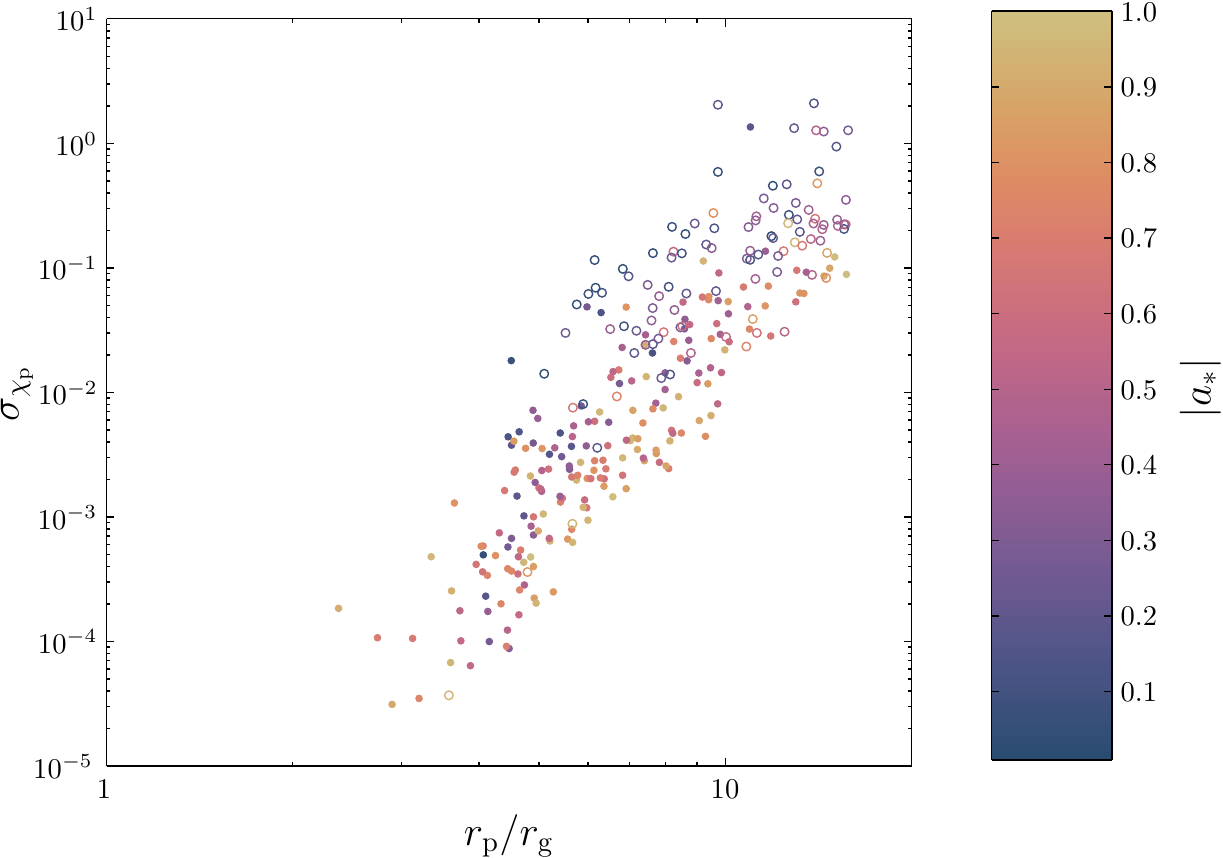}}
\caption{Parameter standard deviations versus periapsis $r\sub{p}$, showing dependence (or lack thereof) upon the spin magnitude $|a_\ast|$.\label{fig:sigmas-spin}}
\end{center}
\end{figure*}

The widths show a trend of decreasing with decreasing periapsis or increasing SNR, but there is a large degree of scatter. There does not appear to be a strong dependence upon any single input parameter, with the exception of the spin. The widths for $\iota$, $\Theta\sub{K}$, $\Phi\sub{K}$, $\phi\sub{p}$ and $\chi\sub{p}$ increase for smaller spin magnitudes. The dependence is shown in \figref{sigmas-spin}.
These parameters are defined with reference to the coordinate system established by the spin axis: for $a_\ast = 0$ we have spherical symmetry and there would be ambiguity in defining them. Therefore, it makes sense that they can be more accurately determined for larger spin magnitudes. The width for $a_\ast$, however, shows no clear correlation.

Comparing our MCMC and FIM results, we see there can be significant differences. Most parameters give results consistent to within an order (or two) of magnitude. The best agreement is for $t\sub{p}$, which is largely uncorrelated with the other parameters. The widths for $M_\bullet$, $a_\ast$, $L_\infty$ and $\iota$ show more severe differences; these parameters show the tightest degeneracies. The two methods do show signs of slowly converging with increasing SNR, as expected.

As a consistency check, to verify that the mismatch between the FIM and MCMC results is a consequence of parameter correlations, we calculated one-dimensional FIMs, only varying the MBH mass, and compared these to widths computed from MCMCs only sampling in mass. These were found to be in good agreement. The majority ($\sim 87\%$) have standard deviations consistent to within a factor of two; the rest within an order of magnitude.\footnote{One differed by more than an order of magnitude, and also failed to fulfil the (one dimensional) MM criterion; this was a numerical problem in calculating the FIM.} Some small difference is expected because of numerical error from calculating derivatives for the FIM by finite differencing.

\subsection{Scientific potential}

Having quantified the precision with which we could infer parameters from an EMRB waveform, we can now consider if it is possible to learn anything new.

Of paramount interest are the MBH mass and spin. The current uncertainty in the mass is $\sigma_{M_\bullet} = 0.36 \times 10^6 M_\odot$ ($\sim 8\%$). There are few runs amongst our data set that are not better than this: it appears that orbits of a $\mu = 10 M_\odot$ CO with periapses $r\sub{p} \lesssim 13 r\sub{g}$ should be able to match our current observational constraints. However, the EMRB is an independent measurement, and so a measurement of comparable precision to the current bound can still be informative. Accuracy of $1\%$ could be possible if $r\sub{p} \lesssim 8 r\sub{g}$.

The spin is less well constrained. To obtain an uncertainty for the magnitude of $0.1$, comparable to that achieved in X-ray measurements of active galactic nuclei, it appears that the periapsis needs to be $r\sub{p} \lesssim 11 r\sub{g}$. For smaller periapses, the uncertainty can be much less, indicating that an EMRB could be an excellent probe. The orientation angles for the spin axis may be constrained to better than $0.1$ for $r\sub{p} \lesssim 11 r\sub{g}$. It may well be possible to learn both the direction and the magnitude of the spin. This could illuminate the MBH's formation.

We have no {\it a priori} knowledge about the CO or its orbit, so anything we learn would be new. However, this is not particularly useful information, unless we observe multiple bursts, and can start to build up statistics for the dynamics of the GC. Using current observations for the distance to the GC, which could be further improved by the mass measurement from the EMRB, it is possible to infer a value for the mass $\mu$ from $\zeta$. This could inform us of the nature of the object (BH, NS or WD) and be a useful consistency check. A small value of $\zeta$, indicating a massive CO, would be unambiguous evidence for the existence of a stellar mass BH.

\section{Extra-galactic sources}\label{sec:Extragal}

We have so far only been concerned with properties of bursts from our own galaxy. This is the best source for bursts because of its proximity. A natural continuation is to consider EMRBs from other MBHs. \citet{Rubbo2006} suggested that \textit{LISA} should be able to detect EMRBs originating from the Virgo cluster, although the detectable rate may be only $10^{-4}\units{yr^{-1}}$ per galaxy \citep{Hopman2007}. Detectability depends upon the mass of the MBH; higher masses correspond to lower frequency bursts, which are harder to detect.

\begin{figure}
  \begin{center}
  \includegraphics[width=0.43\textwidth]{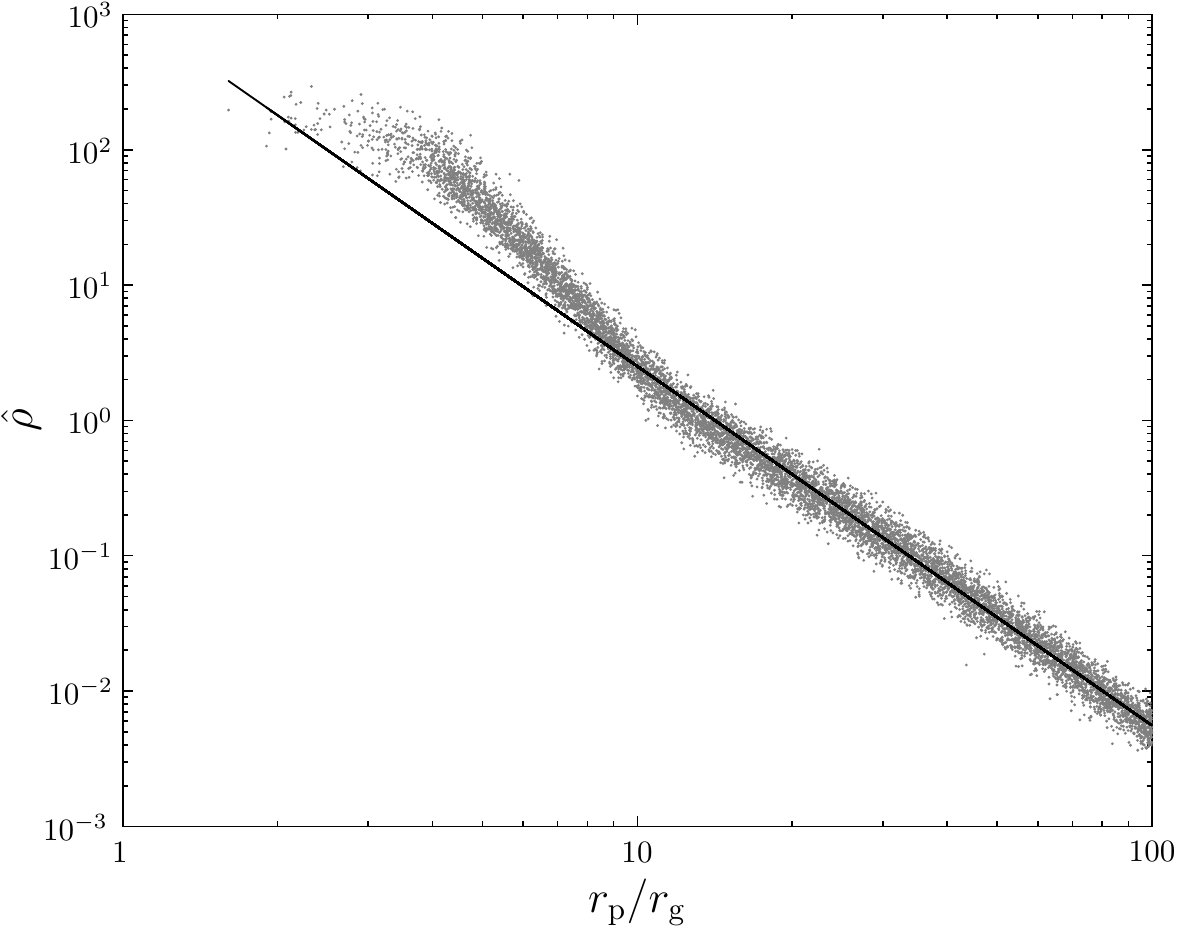}
    \caption{Signal-to-noise ratio as a function of periapse radius for a $\mu = 1 M_\odot$ CO about the MBH of M32. The plotted points are the values obtained by averaging over each set of intrinsic parameters. The best fit line is $\log\left(\rho\right) = -2.65\log(r\sub{p}/r\sub{g}) + 3.05$. This is fitted to orbits with $r\sub{p} > 18.8 r\sub{g}$ and has a reduced chi-squared value of $\chi^2/\nu = 1.26$.\label{fig:SNR-M32}}
  \end{center}
\end{figure}

Checking our nearest neighbours, we find bursts from Andromeda (M31) would not be detectable. This is because of the large mass of the MBH $M\sub{M31} = (1.4^{+0.9}_{-0.3}) \times 10^8 M_\odot$ \citep{Bender2005}. However, its companion M32 is more promising. It has a lighter MBH $M\sub{M32} = (2.5 \pm 0.5) \times 10^6 M_\odot$ \citep{Verolme2002}. The trend between the periapse radius and SNR is shown in \figref{SNR-M32}.
The fit is again for orbits with $f_\ast = \sqrt{GM\sub{M32}/r\sub{p}} < 1 \times 10^{-3}\units{Hz}$ to avoid the bucket of the noise curve. Bursts for a $1 M_\odot$ ($10 M_\odot$) can be detected with $\rho > 10$ if the periapse is smaller than $7 r\sub{g}$ ($14 r\sub{g}$).

\begin{figure}
  \begin{center}
  \includegraphics[width=0.43\textwidth]{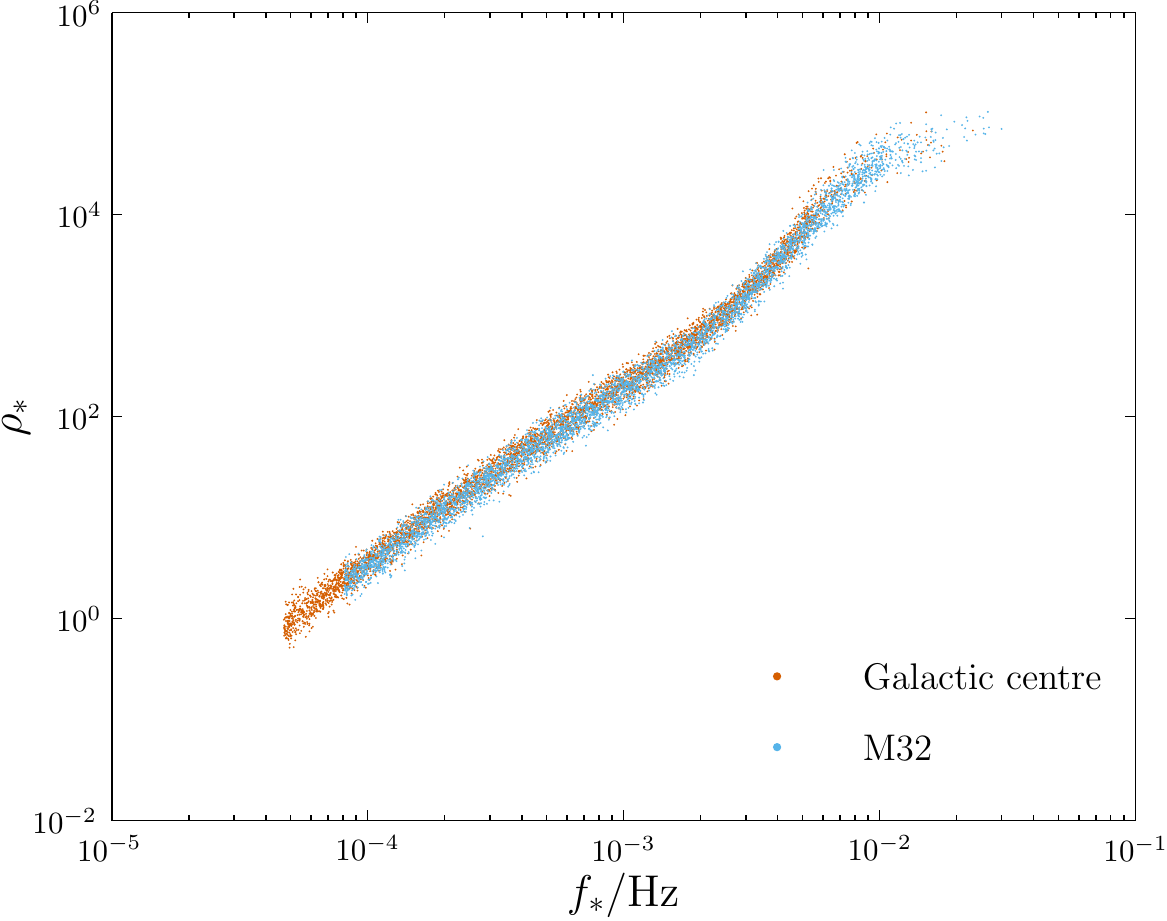}
    \caption{Scaled signal-to-noise ratio as a function of characteristic frequency.\label{fig:SNR-scaling}}
  \end{center}
\end{figure}

The general behaviour is the same as for the GC. Bursts from the two MBHs can be compared using their characteristic frequencies $f_\ast$ and scaled SNR
\begin{equation}
\rho_\ast[\boldsymbol{h}] = \left(\frac{\mu}{M_\odot}\right)^{-1}\left(\frac{R}{\mathrm{kpc}}\right)\left(\frac{M}{10^6 M_\odot}\right)^{-2/3}\rho[\boldsymbol{h}],
\end{equation}
where $R$ and $M$ are the appropriate distances and masses for the two MBHs. These scalings can be determined from the quadrupole piece of \eqnref{octupole} assuming a characteristic length scale set by $r\sub{p}$. \Figref{SNR-scaling} shows the trend for both galaxies.
The difference in sky position is largely washed out through the motion of the detector.

M31 and M32 are at a distance of $770\units{kpc}$ \citep{Karachentsev2004}. It therefore seems unlikely that bursts could be observed from the Virgo cluster at a distance of $R\sub{Virgo} \approx 16.5\units{Mpc}$ \citep{Mei2007}.

Triangulum (M33) is believed not to have an MBH. \citet{Merritt2001} use dynamical constraints to place an upper bound on the mass of a central BH of $M\sub{M33} < 3 \times 10^3 M_\odot$, \citet{Gebhardt2001} find a bound of $M\sub{M33} < 1.5 \times 10^3 M_\odot$. Observations of the ultra-luminous nuclear X-ray source (ULX) closest to the centre of M33 yield a best estimate of $M\sub{ULX} \sim \order{10} M_\odot$ for the source object's mass \citep{Foschini2004, Weng2009}. This is consistent with there being no MBH; the ULX originates from a stellar mass BH that is coincidentally located close to the core of the galaxy. Consequently, we do not expect to see any bursts from M33: to detect one would confirm the existence of a previously invisible MBH.

\section{Discussion}\label{sec:End}

We have outlined an approximate method of generating gravitational waveforms for EMRBs originating at the GC. This assumes that the orbits are parabolic and employs a numerical kludge approximation. The two coordinate schemes for a NK presented here yield almost indistinguishable results. We conclude that either is a valid choice for this purpose. There may be differences when the spin is large and the periapse is small: $\sim 10\%$ for $r\sub{p} \simeq 4 r\sub{g}$, $\sim 20\%$ for $r\sub{p} \simeq 2 r\sub{g}$.

The waveforms created appear to be consistent with results obtained using Peters and Mathews waveforms for large periapses, indicating that they have the correct weak-field form. The NK approach should be superior to that of Peters and Mathews in the strong-field regime as it uses the exact geodesics of the Kerr spacetime. Comparisons with energy fluxes from black hole perturbation theory indicate that typical waveform accuracy may be of order $5\%$, but this is worse for orbits with small periapses and may be $\sim 20\%$. These errors are greater than the differences resulting from the use of the alternative coordinate systems.

The signal-to-noise ratio of bursts is well correlated with the periapsis. Except for the closest orbits ($r\sub{p} \lesssim 7 r\sub{g}$), the SNR (per unit mass) may be reasonably described as having a power-law dependence of
\begin{equation}
\log\left(\hat{\rho}\right) \simeq -2.7\log\left(\frac{r\sub{p}}{r\sub{g}}\right) + 4.9.
\end{equation}
Signals should be detectable for a $1 M_\odot$ ($10 M_\odot$) object if the periapse is $r\sub{p} < 27 r\sub{g}$ ($r\sub{p} < 65 r\sub{g}$), corresponding to a physical scale of $1.7 \times 10^{11}\units{m}$ ($4.1 \times 10^{11}\units{m}$) or $5.6 \times 10^{-6}\units{pc}$ ($1.3 \times 10^{-5}\units{pc}$).

Using the NK waveforms we conducted an investigation, using Fisher matrix analysis, into how precisely we could infer parameters of the GC's MBH should such an EMRB be observed. However, we found that the linearised-signal approximation does not hold for these bursts over a wide range of SNR. This demonstrates the necessity of checking the approximation before quoting the results of an analysis~\citep{Vallisneri2008}.

We used MCMC results as a more robust measure of parameter estimation accuracy. Potentially, it is possible to determine very precisely the key parameters defining the MBH's mass and spin, if the orbit gets close enough to the MBH. It appears that we can achieve good results from a single EMRB with periapsis of $r\sub{p} \simeq 10 r\sub{g}$ for a $10 M_\odot$ CO. This translates to a distance of $6 \times 10^{10}\units{m}$ or $2 \times 10^{-6}\units{pc}$. Orbits closer than this would place stricter constraints. The best orbits yield uncertainties of almost one part in $10^5$ for the MBH mass and spin, far exceeding existing techniques. Conversely, orbits with $r\sub{p} \gtrsim 20 r\sub{g}$ are unlikely to provide any useful information.

Before we can quote results for how accurately we can determine the various parameters, we must consider the probability of each orbit. This will be the subject of a companion paper, building upon the earlier results of \citet{Rubbo2006} and \citet{Hopman2007}, who only considered approximate forms for the SNR, rather than using waveforms. Using a model for the nuclear star cluster of the GC it is possible to define distributions for angular momenta $L_\infty$, for a species of mass $\mu$. With these we can estimate the event rate and how much information, on average, we could hope to obtain from EMRB observations. If it is likely that we would observe multiple EMRBs, it may be possible to combine results to tighten uncertainties.

Some consideration should also be given to methods of fitting a waveform to an observed signal. Given a noisy data stream, how could EMRBs be extracted? The parabolic spectrum has a characteristic profile, suggesting that matched filtering could be possible. Complications could arise in fitting parameters to a waveform: we have seen that there exist complicated degeneracies between parameters. These issues would warrant further investigation should the event rate be high enough.

While we have only considered bursts from our own galaxy in detail, it should be possible to observe bursts from other nearby galaxies if their MBH is of the appropriate mass. This makes M32 a viable candidate. The SNR shows a similar dependence upon periapsis as for the GC, and may be described by a power-law of
\begin{equation}
\log\left(\hat{\rho}\right) \simeq -2.7\log\left(\frac{r\sub{p}}{r\sub{g}}\right) + 3.1,
\end{equation}
for orbits with $r\sub{p} \gtrsim 10 r\sub{g}$. For a $1 M_\odot$ ($10 M_\odot$) object, bursts should be detectable for periapses $r\sub{p} \lesssim 7 r\sub{g}$ ($r\sub{p} \lesssim 14 r\sub{g}$), corresponding to $2.6 \times 10^{10}\units{m}$ ($4.9 \times 10^{10}\units{m}$) or $8.4 \times 10^{-7}\units{pc}$ ($1.6 \times 10^{-6}\units{pc}$). This is a small region of parameter space, so we conclude that extra-galactic bursts are likely to be rare.

\section*{Acknowledgments}

The authors are indebted to Michele Vallisneri for discussions on the (im)proper use of Fisher matrices; they thank Stephen Taylor for advice on adaptive MCMC methods and Dave Green for helpful suggestions regarding apodization. They are grateful to Donald Lynden-Bell for his suggestions, and would like to thank Pau Amaro-Seoane for useful comments. CPLB is supported by STFC. JRG is supported by the Royal Society. The MCMC simulations were performed using the Darwin Supercomputer of the University of Cambridge High Performance Computing Service (\url{http://www.hpc.cam.ac.uk/}), provided by Dell Inc.\ using Strategic Research Infrastructure Funding from the Higher Education Funding Council for England. Figures \ref{fig:MCMC-1}, \ref{fig:MCMC-2} and \ref{fig:MCMC-3} were produced using the colour scheme of \citet{Green2011}.

\bibliographystyle{mn3e}
\bibliography{Parabolic}

\appendix

\section{Window functions}\label{ap:window}

When performing a Fourier transform using a computer we must necessarily only transform a finite time-span $\tau$. The effect of this is the same as transforming the true, infinite signal multiplied by a unit top-hat function of width $\tau$. Transforming yields the true waveform convolved with a $\sinc$. If $\tilde{h}'(f)$ is the computed Fourier transform then
\begin{equation}
\tilde{h}'(f) = \intd{-\tau/2}{\tau/2}{h(t)\exp({2\pi i ft})}{t} = \left[\tilde{h}(f) \ast \tau \sinc(\pi f\tau)\right],
\end{equation}
where $\tilde{h}(f) = \mathscr{F}\left\{h(t)\right\}$ is the unwindowed Fourier transform of the infinite signal. This windowing of the data is a problem innate in the method, and results in spectral leakage.

\begin{figure*}
  \begin{center}
   \subfigure[Spectrum using no window. The calculated SNR is $\rho \simeq 12.5$.]{\label{fig:Windowing_Rectangular} \includegraphics[width=0.43\textwidth]{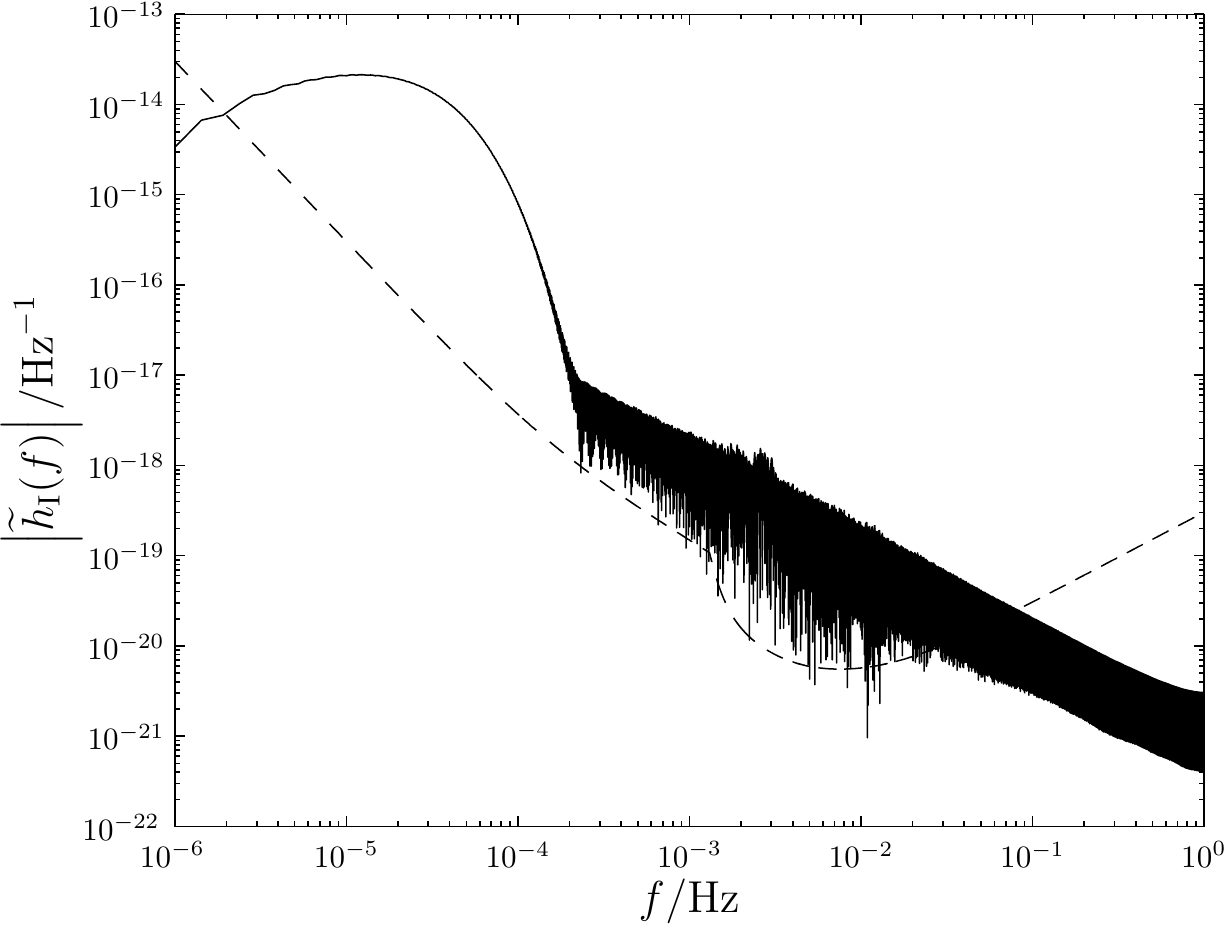}} \quad
   \subfigure[Spectrum using a Nuttall window. The calculated SNR is $\rho \simeq 8.5$.]{\label{fig:Windowing_Nuttall} \includegraphics[width=0.43\textwidth]{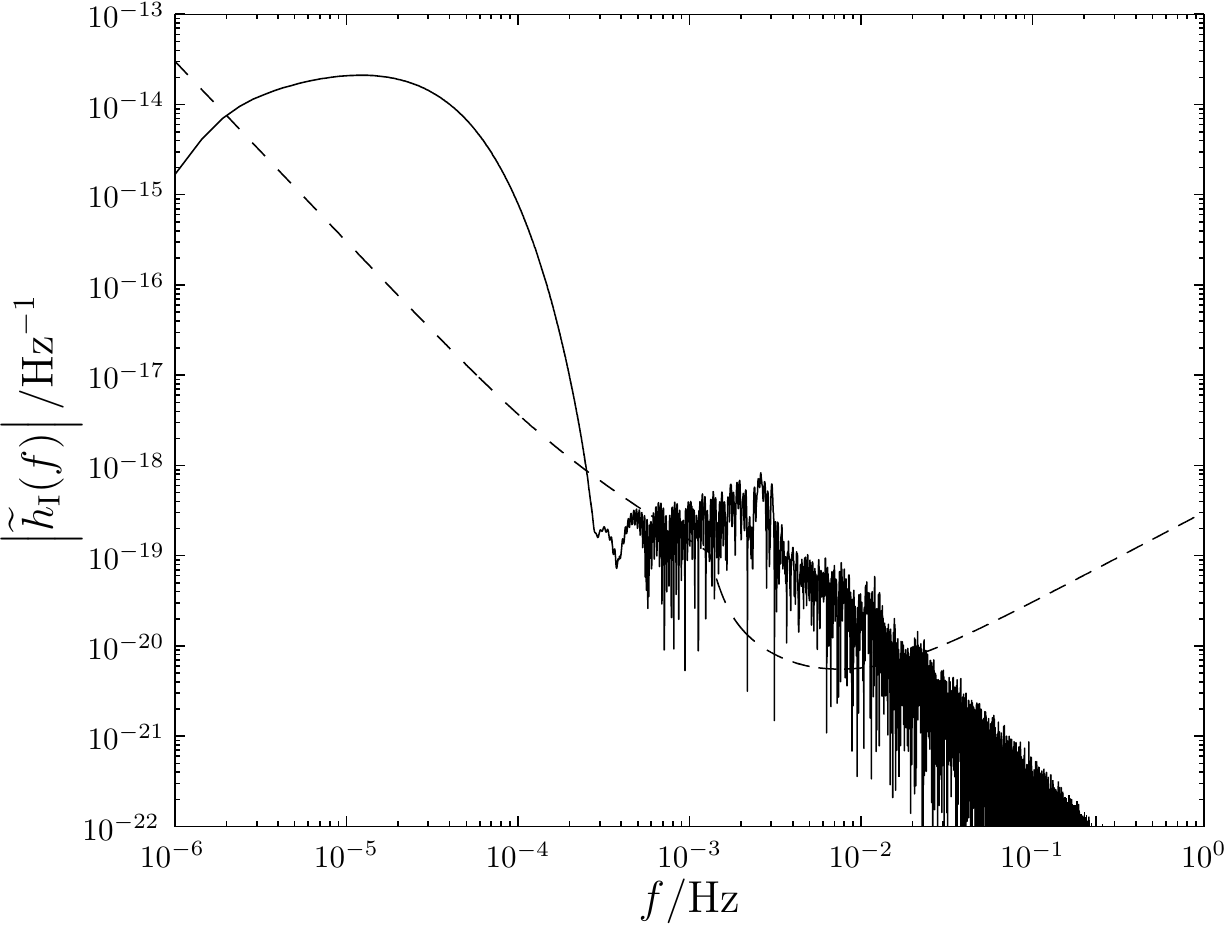}}
    \caption{Example spectra calculated using (a) a rectangular window and (b) Nuttall's four-term window with continuous first derivative \citep{Nuttall1981}. The spin of the MBH is $a_\ast = 0.5$, the mass of the orbiting CO is $\mu = 10 M_\odot$, the periapsis is $r\sub{p} = 50 r\sub{g}$ and the inclination is $\iota = 0.1$. The high-frequency tail is the result of spectral leakage. The level of the \textit{LISA} noise curve is indicated by the dashed line. The spectra are from detector I, but the detector II spectra look similar.\label{fig:Windowing}}
  \end{center}
\end{figure*}

Fig.\ A\ref{fig:Windowing_Rectangular} shows the computed Fourier transform for an example EMRB. The waveform has two distinct regions: a low-frequency curve, and a high-frequency tail. The low-frequency signal is the spectrum we are interested in; the high-frequency components are a combination of spectral leakage and numerical noise. The $\order{1/{f}}$ behaviour of the $\sinc$ gives the shape of the tail. This has possibly been misidentified in figure 8 of \citet{Burko2007} as the characteristic strain for parabolic encounters.

Despite being many orders of magnitude below the peak level, the high-frequency tail is still well above the noise curve for a wide range of frequencies. It therefore contributes to the evaluation of any inner products, and could mask interesting features. It is possible to reduce the leakage using apodization: to improve the frequency response of a finite time series one can use a weighting window function $w(t)$ which modifies the impulse response in a prescribed way.

The simplest window function is the rectangular (or Dirichlet) window $w\sub{R}(t)$; this is the top-hat described above. Other window functions are generally tapered.\footnote{When using a tapered window function it is important to ensure that the window is centred upon the signal; otherwise the calculated transform has a reduced amplitude.} There is a wide range of window functions described in the literature \citep*{Harris1978,Kaiser1980,Nuttall1981,McKechan2010}. The introduction of a window function influences the spectrum in a manner dependent upon its precise shape. There are two distinct distortions: local smearing due to the finite width of the centre lobe, and distant leakage due to finite amplitude sidelobes. The window function may be optimised such that the peak sidelobe has a small amplitude, or such that the sidelobes decay away rapidly with frequency. Choosing a window function is a trade-off between these various properties, and depends upon the particular application.

For use with the parabolic spectra, the primary concern is to suppress the sidelobes. Many windows with good sidelobe behaviour exist; we consider three: the Blackman-Harris minimum four-term window \citep{Harris1978, Nuttall1981}
\begin{equation}
w\sub{BH}(t) = \sum_{n=0}^{3} a\super{BH}_n\cos\left(\frac{2n\pi t}{\tau}\right),
\end{equation}
where
\begin{equation}
\begin{split}
a\super{BH}_0 = 0.35875, \quad a\super{BH}_1 = 0.48829,\\
a\super{BH}_2 = 0.14128, \quad a\super{BH}_3 = 0.01168;
\end{split}
\end{equation}
the Nuttall four-term window with continuous first derivative \citep{Nuttall1981}
\begin{equation}
w\sub{N}(t) = \sum_{n=0}^{3} a\super{N}_n\cos\left(\frac{2n\pi t}{\tau}\right),
\end{equation}
where
\begin{equation}
\begin{split}
a\super{N}_0 = 0.355768, \quad a\super{N}_1 = 0.487396,\\
a\super{N}_2 = 0.144232, \quad a\super{N}_3 = 0.012604,
\end{split}
\end{equation}
and the Kaiser-Bessel window \citep{Harris1978, Kaiser1980}
\begin{equation}
w\sub{KB}(t;\beta) = \frac{I_0\left[\beta\sqrt{1 - (2 t/\tau)^2}\right]}{I_0(\beta)},
\end{equation}
where $I_\nu(z)$ is the modified Bessel function of the first kind, and $\beta$ is an adjustable parameter. Increasing $\beta$ reduces the peak sidelobe, but also widens the central lobe.

The Kaiser-Bessel window has the smallest peak sidelobe, but the worst decay ($1/f$); the Nuttall window has the best asymptotic behaviour ($1/f^3$); the Blackman-Harris window has a peak sidelobe similar to the Nuttall window, and decays asymptotically as fast (slow) as the Kaiser-Bessel window, but has the advantage of having suppressed sidelobes next to the central lobe.

\begin{figure}
  \begin{center}
  \includegraphics[width=0.43\textwidth]{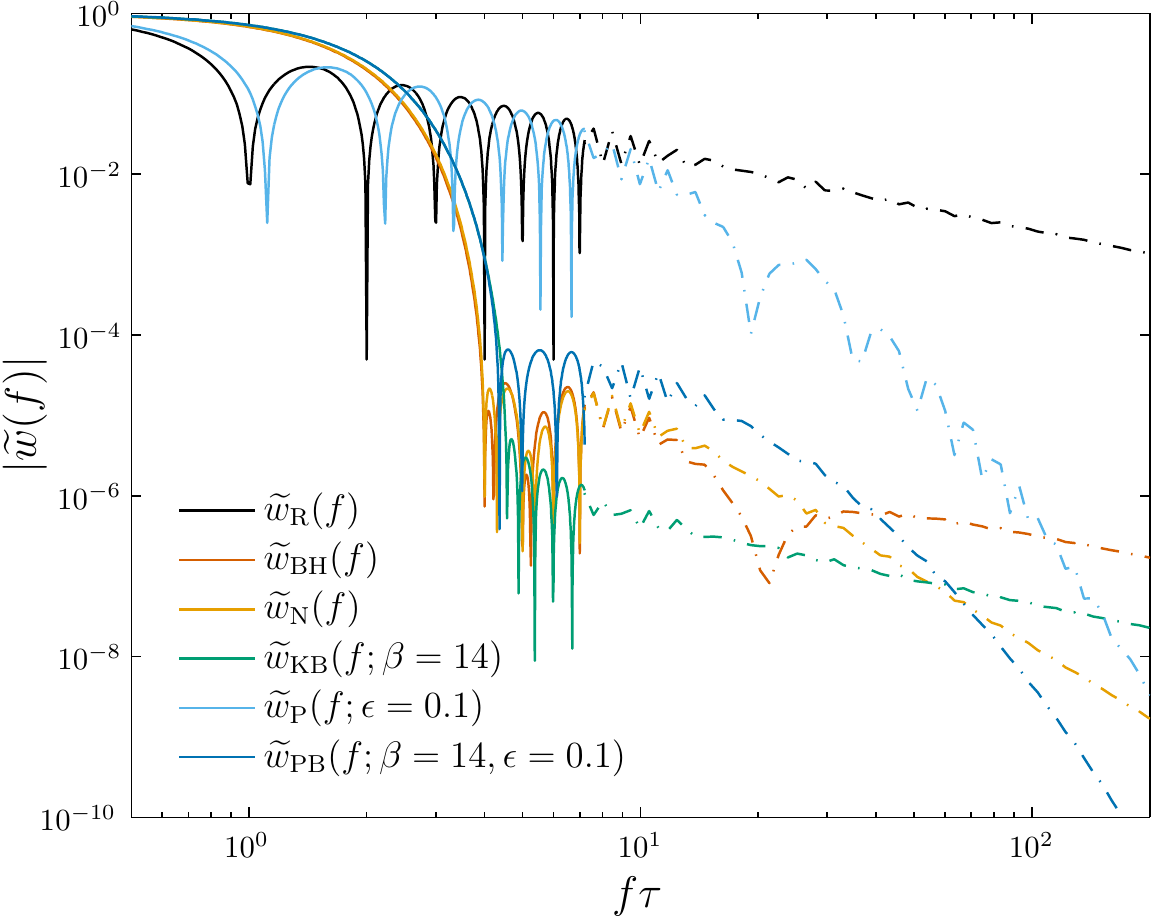}
    \caption{Window function frequency response. To avoid clutter, the response function is only plotted in detail until $f\tau = 8$, above this a smoothed value is used, as indicated by the dot-dashed line. As well as having good asymptotic behaviour, the Planck-taper window has the narrowest main lobe, except for the rectangular window.}
    \label{fig:Response}
  \end{center}
\end{figure}

Another window has been recently suggested for use with gravitational waveforms: the Planck-taper window (\citealt*{Damour2000}; \citealt{McKechan2010})
\begin{equation}
w\sub{P}(t; \epsilon) = \begin{cases}
 {\displaystyle \recip{\exp(Z_+)+1}} & {\displaystyle \hphantom{\left(\recip{2} - \epsilon\right)}-\frac{\tau}{2} \leq t < -\tau\left(\recip{2} - \epsilon\right)} \\
 1 & {\displaystyle -\tau\left(\recip{2} - \epsilon\right) < t < \tau\left(\recip{2} - \epsilon\right)} \\
 {\displaystyle \recip{\exp(Z_-)+1}} & {\displaystyle \hphantom{-}\tau\left(\recip{2} - \epsilon\right) < t \leq \frac{\tau}{2}}
\end{cases},
\end{equation}
with
\begin{equation}
Z_\pm(t; \epsilon) = 2\epsilon\left[\recip{1 \pm 2(t/\tau)} + \recip{1 - 2\epsilon \pm 2(t/\tau)}\right].
\end{equation}
This was put forward for use with binary coalescences, and has superb asymptotic decay. However, the peak sidelobe is high, which is disadvantageous here. We therefore propose a new window function: the Planck-Bessel window which combines the Kaiser-Bessel and Planck-taper windows to produce a window which inherits the best features of both, albeit in a diluted form,
\begin{equation}
w\sub{PB}(t;\beta,\epsilon) = w\sub{P}(t; \epsilon)w\sub{KB}(t;\beta).
\end{equation}
The window functions' frequency responses are plotted in \figref{Response}.
There is no window that performs best everywhere.

\Figref{Windowing} shows the computed Fourier transforms for an example EMRB using no window (alternatively a rectangular window), and the Nuttall window.\footnote{The Blackman-Harris, Kaiser-Bessel and Planck-Bessel windows give almost identical results.} Using the Nuttall window, the spectral leakage is greatly reduced; the peak sidelobe is lower, and the tail decays away as $1/{f^3}$ instead of $1/{f}$. The low frequency signal is not appreciably changed.

The choice of window function influences the results as it changes the form of $\widetilde{h}(f)$. The variation in results between windows depends upon the signal: variation is greatest for low frequency bursts, as then there is greatest scope for leakage into the detector band; variation is least significant for zoom-whirl orbits as then there are strong signals to relatively high frequencies, and spectral leakage is confined mostly to below the noise level. To quantify the influence of window functions, we studied the diagonal elements of the Fisher matrix from a selection of orbits with periapses ranging from $\sim 10 r\sub{g}$--$300 r\sub{g}$. For orbits with small periapses all five windows (excluding the rectangular window) produced very similar results: the Planck-taper window differed by a maximum of $\sim 0.5\%$ from the others, which all agreed to better than $0.1\%$. The worst case results came from the lowest frequency orbits, then the Planck-taper window deviated by a maximum of $\sim 30\%$ in the value for the Fisher matrix elements, the Blackman-Harris deviated by $\sim 20\%$ and the others agreed to better than $\sim 5\%$. The Planck-taper window's performance is limited by its poor sidelobe behaviour; the Blackman-Harris has the worst performance at high frequency.

For this work we have used the Nuttall window. Its performance is comparable to the Kaiser-Bessel and Planck-Bessel windows, but it is computationally less expensive as it does not contain Bessel functions. Results should be accurate to a few percent at worst, and results from closer orbits, which provide better constraints, should be less affected by the choice of window function. Therefore, we are confident that none of our conclusions are sensitive to the particular windowing method implemented.

\bsp

\label{lastpage}

\end{document}